\documentclass[11pt,a4paper]{article}
\usepackage{jheppub}
\usepackage[utf8]{inputenc}
\usepackage{graphicx,fancybox,float,comment}

\usepackage{units}
\usepackage{subcaption}
\usepackage{multirow,array,arydshln}
\usepackage[dvipsnames]{xcolor}
\usepackage{braket,tensor}
\usepackage{amsmath,amssymb,amsfonts}
\usepackage{tikz-cd}

\usepackage{ulem}

\global\interfootnotelinepenalty=1000

\graphicspath{{graphics}}

\title{Dissipative hydrodynamic actions and horizon symmetries in gravity}

\author[1]{Mike Blake,}
\author[2,3]{Arpit Das}
\author[4]{and Richard A. Davison}

\affiliation[1]{School of Mathematics, University of Bristol, Woodland Road, Bristol BS8 1UG, U.K.} 
\affiliation[2]{School of Mathematics and Maxwell Institute for Mathematical Sciences, University of Edinburgh,\\ Edinburgh EH9 3FD, U.K.}
\affiliation[3]{Higgs Centre for Theoretical Physics,
University of Edinburgh,\\ Edinburgh EH8 9YL, U.K.}
\affiliation[4]{Department of Mathematics and Maxwell Institute for Mathematical Sciences,\\ Heriot-Watt University, Edinburgh EH14 4AS, U.K.}

\emailAdd{mike.blake@bristol.ac.uk}
\emailAdd{arpit.das@ed.ac.uk}
\emailAdd{r.davison@hw.ac.uk}

\abstract{We give a prescription to compute a dissipative action describing the large-scale thermal stress tensor dynamics of a holographic quantum field theory dual to AdS$_4$ gravity, in the context of the Schwinger-Keldysh formalism. Our prescription is valid to quadratic order in perturbations about the thermal equilibrium state. The hydrodynamical degrees of freedom of this action are realised in gravity as relative diffeomorphisms between the black hole horizon and the two asymptotic boundaries of the Crossley-Glorioso-Liu contour. We explicitly compute the action to first order in derivatives, and confirm it correctly reproduces the known hydrodynamic Green's functions. Our prescription requires a choice of horizon boundary conditions for the metric. We study the horizon symmetries that preserve these, and their relation to conjectured hydrodynamic symmetries responsible for many-body quantum chaos.}

\begin{document}

\maketitle
\section{Introduction}

\paragraph{} Hydrodynamics provides a universal description of the long distance, late time dynamics of conserved quantities in many-body quantum systems and quantum field theory. Traditionally, hydrodynamics is formulated at a classical level, in which the expectation values of conserved currents are expanded in terms of hydrodynamical variables such as the local temperature and fluid velocity (see \cite{Kovtun:2012rj} for a pedagogical review). It had been a long standing question to construct an action principle from which the classical hydrodynamical equations emerge. The formulation of such an action principle was non-trivial due to the fact that hydrodynamics is, in general, dissipative. As a result, the construction of such actions must be formulated in the context of the Schwinger-Keldysh formalism, requiring a doubling of the hydrodynamical degrees of freedom.

\paragraph{}In recent years, there have been significant developments from a number of authors in formulating hydrodynamical actions.  In particular, a systematic treatment of actions for dissipative hydrodynamics was eventually provided in~\cite{Crossley:2015evo}, both for the case of a single conserved U(1) charge and for energy-momentum conservation (see \cite{Grozdanov:2013dba,Haehl:2015foa,Haehl:2017zac,Jensen:2017kzi,Liu:2018kfw} for previous and related work). The actions constructed in~\cite{Crossley:2015evo} have several advantages over the classical formulation of hydrodynamics. These include constraints on terms in the constitutive relations from first principles, and the incorporation of quantum and statistical fluctuations in a systematic manner. The study of such Schwinger-Keldysh hydrodynamic actions is now a substantial field of research -- see~\cite{Delacretaz:2026owo} for a recent review. 

\paragraph{} As in classical hydrodynamics, the Schwinger-Keldysh actions of~\cite{Crossley:2015evo} operate at the level of an effective theory: they are constrained by underlying symmetry principles, and are conventionally constructed in a derivative expansions with coefficients determined by the microscopic theory. In practice, explicitly constructing the effective action for a given microscopic system is challenging. One class of systems for which one might hope to be able to perform such a construction are quantum field theories described by the AdS/CFT correspondence.\footnote{Another example are SYK chains, for which an action for non-linear fluctuating hydrodynamics has recently been derived (in the low temperature limit) in~\cite{Bucca:2026xmh}.} For such theories there is a well-established procedure for deriving the classical description of hydrodynamics from the gravitational equations of motion (the fluid/gravity correspondence \cite{Bhattacharyya:2007vjd, Bhattacharyya:2008ji, Banerjee:2008th, Erdmenger:2008rm, Hubeny:2011hd}). Furthermore, it has been possible to derive Schwinger-Keldysh actions describing the dissipative hydrodynamics of a single U(1) conserved charge (as well as several generalisations) for these theories~\cite{CGL, deBoer:2018qqm, Haehl:2025zfn} (see \cite{Landry:2019iel, Iqbal:2020lrt, Loganayagam:2020eue, Loganayagam:2020iol, Bu:2021clf, Bu:2021jlp, Bu:2022esd, Ghosh:2022fyo, Baggioli:2023tlc, Rangamani:2023mok, Akyuz:2023lsm, Martin:2024mdm, Baggioli:2024zfq, Bu:2024fhz, Ammon:2025vod, Bu:2026mxr} for various related constructions). For this paper, the most directly relevant calculation is that of~\cite{CGL}, for which the hydrodynamic action is obtained by placing the bulk Maxwell equations partially on-shell, in a suitably analytically continued black hole geometry. In particular, in the CGL geometry studied in~\cite{CGL}, the holographic radial coordinate has two branches which give rise to the two copies of the hydrodynamical fields that appear in the Schwinger-Keldysh action. 

\paragraph{} It remains, however, an important open question to obtain a dissipative hydrodynamic action describing the stress tensor dynamics using this approach. There have been several previous attempts to derive such hydrodynamical actions from gravity, including \cite{Nickel:2010pr}, \cite{Crossley:2015tka} and \cite{deBoer:2015ija}. These works constructed actions for hydrodynamic degrees of freedom corresponding to relative diffeomorphisms between the horizon and the asymptotic boundary. However they worked solely in a one-sided geometry, and as such were unable to systematically include dissipative effects.\footnote{JT \cite{Jensen:2016pah,Maldacena:2016upp,Engelsoy:2016xyb} and AdS$_3$ gravity \cite{Cotler:2018zff} can be described by actions for reparameterisation fields, which can be considered as a special case of hydrodynamical actions with no dissipation. In~\cite{Blake:2026axf} we revisit the construction of hydrodynamic actions in these simpler settings, using the same methodology of this paper.}

\paragraph{}  An additional motivation for constructing such hydrodynamic actions in holographic theories comes from conjectured relationships between hydrodynamics and many-body quantum chaos. In particular, it was proposed in~\cite{Blake:2017ris, Blake:2021wqj} that in maximally chaotic systems hydrodynamic actions should contain exponentially growing modes that are responsible for the exponential growth of out-of-time-ordered correlation functions. Furthermore, it has recently been argued in~\cite{Knysh:2024asf} that such modes must have a precise form related to geometric symmetries of black hole horizons. By constructing such hydrodynamic actions, we can test these important conjectures.

\paragraph{} The first main result of this paper is to present a consistent derivation of a Schwinger-Keldysh action describing the stress-tensor dynamics of the holographic QFT dual to the AdS$_4$-Schwarzschild black hole. Our calculation will be performed to quadratic order in the amplitude of perturbations about equilibrium, and to first order in the hydrodynamical derivative expansion. Our approach will be similar to that of \cite{Nickel:2010pr}, \cite{Crossley:2015tka} and \cite{deBoer:2015ija}, in that we will construct a partially on-shell action for relative diffeomorphisms, obtained by imposing a subset of the Einstein equations. The key novel ingredient, compared to the aforementioned approaches, is that we will study the Einstein equations on the analytically continued CGL contour \cite{CGL} with two asymptotic boundaries. This will allow us to systematically include the effects of dissipation. 

\paragraph{} A crucial aspect of our calculation is identifying appropriate boundary conditions to impose at the black hole horizon. In particular, in the CGL computation of the hydrodynamic action for a U(1) charge~\cite{CGL} one imposes two types of boundary conditions. Firstly, the spatial components of the bulk gauge field are made continuous after analytic continuation near the horizon of the CGL contour. For the time component a qualitatively different boundary condition is imposed: $A_t^{\sigma}(r_0) = 0$, where $\sigma = 1,2$ refers to the branch of the CGL contour, $t$ is the ingoing bulk coordinate and $r_0$ the horizon radius. In the stress tensor case we find that, working in radial gauge near the horizon, we can relate the metric perturbations $\delta g^{\sigma}_{xy}$ and $\delta g^{\sigma}_{m} = \delta g^{\sigma}_{xx} - \delta g^{\sigma}_{yy}$ by analytic continuation near the horizon. However we are then required to impose an additional four boundary conditions at the horizon for each branch of the CGL contour. We will primarily study the following set of boundary conditions
\begin{equation}
\delta g^{\sigma}_{t \mu}(r_0) = 0,  \hspace{2.0cm} 2 r_0^2 \partial_{r} \delta g^{\sigma}_{tt}(r_0) = \partial_{t}\delta g^{\sigma}_{xx}(r_0) + \partial_{t} \delta g^{\sigma}_{yy}(r_0),
\label{bcsintro}
\end{equation}
where $\mu = (t, x, y)$ and $r$ is the bulk radial coordinate. With these boundary conditions, it is manifest that integrating out the hydrodynamical fields corresponds to putting the gravitational action fully on-shell.

\paragraph{} The output of our calculation is a hydrodynamic Schwinger-Keldysh action $S_{\mathrm{hydro}}[\delta g_{\mu \nu}^{\sigma (s)}, \xi_{\mu}^{\sigma}]$. This action has, in particular, the following properties. 

\begin{enumerate}
\item The action depends only on hydrodynamic fields $\xi_{\mu}^{\sigma}$ and external sources for the stress tensor (i.e.~boundary metric) $\delta g_{\mu \nu}^{\sigma (s)}$.  Here $\sigma = 1,2$ is an index labelling the branch of the Schwinger-Keldysh contour and $\mu=(t,x,y)$ a boundary spacetime index. The hydrodynamical fields $\xi_{\mu}^{\sigma}$ are defined as relative diffeomorphisms between the black hole horizon and the (two) asymptotic boundaries of the CGL contour. 
\item The action depends on $\xi_{\mu}^{\sigma}$ and $\delta g_{\mu \nu}^{\sigma (s)}$ only through the combinations ${\mathcal B}_{\mu \nu}^{\sigma}$, defined to be the perturbations of the induced boundary metric under a diffeomorphism. Momentarily suppressing the branch index, we define 
\begin{equation}
{\mathcal B}_{\mu \nu}(x) = \frac{\partial y^{\alpha}}{\partial x^{\mu} } \frac{\partial y^{\beta}}{\partial x^{\nu}}  g^{(s)}_{\alpha \beta}(y(x)) - \eta_{\mu \nu}, 
\end{equation}
where $y^{\mu}= x^{\mu} - \xi^{\mu}$ and $g_{\mu \nu}^{(s)} = \eta_{\mu \nu} + \delta g_{\mu \nu}^{(s)}$.
Note that ${\mathcal B}_{\mu \nu}$, and hence the action, are invariant under the linearised (in $\epsilon$) diffeomorphisms
\begin{equation}
\delta \xi^{\mu}(x) = \epsilon^{\mu}(y(x)), \qquad \delta g^{(s)}_{\mu \nu}(y(x)) \to \delta g^{(s)}_{\mu \nu}(y(x)) + \nabla_{\mu}( g^{(s)}_{\nu \rho} \epsilon^{\rho}) + \nabla_{\nu} (g^{(s)}_{\mu \rho}\epsilon^{\rho}),
\label{diffeointro}
\end{equation} 
where the covariant derivatives are defined with respect to $y^{\mu}$. 
\item Defining the stress tensor on each branch of the Schwinger-Keldysh contour by
\begin{equation}
T^{\mu \nu, 1} = \frac{2}{\sqrt{-g}} \frac{\delta S_{\mathrm{hydro}}}{\delta g^{1(s)}_{\mu \nu}}, \hspace{1.0cm} T^{\mu \nu, 2} = -\frac{2}{\sqrt{-g}} \frac{\delta S_{\mathrm{hydro}}}{\delta g^{2 (s)}_{\mu \nu}},
\label{tmunudef}
\end{equation}
the equations of motion of the hydrodynamical action are equivalent to imposing covariant conservation of the stress tensor, and are furthermore equivalent to the unimposed Einstein equations. 
\item Integrating out the hydrodynamical fields gives the generating functional for real-time two-point functions of the stress tensor. These two point functions are consistent with the known results computed using variational hydrodynamics and the fluctuation-dissipation relation.  
\end{enumerate}

\paragraph{} As we have highlighted, a key motivation for computing a hydrodynamic action -- rather than the correlators directly -- is to understand if it possesses conjectured symmetries that are related to quantum chaos~\cite{Knysh:2024asf, Blake:2017ris, Blake:2021wqj}. In particular it was argued in \cite{Blake:2017ris} that the hydrodynamic actions of maximally chaotic quantum field theories possess a point-wise shift symmetry that gives rise to exponentially growing zero modes of the hydrodynamic fields. Such symmetries can be thought of as redundancies (i.e.~gauge symmetries) of the hydrodynamic theory defined on a two-fold contour. However, for correlation functions defined on a four-fold contour, these transformations are not redundancies and give rise to the exponential growth that appears in OTOCs (see \cite{Bucca:2026enl} for recent work extending this). Additionally, the fact that the hydrodynamic action can be written with such redundancies makes manifest that the two-point functions of the stress tensor will exhibit pole-skipping \cite{Blake:2017ris,Grozdanov:2017ajz,Blake:2018leo}. For non-maximally chaotic theories, the hydrodynamic action is not expected to realise these symmetries: see \cite{Choi:2020tdj,Gao:2023wun,Choi:2023mab,Gao:2025ohk} for discussions of effective theories of scrambling in such cases.

\paragraph{} In the context of our prescription, we argue that such symmetries can be identified as diffeomorphisms preserving our horizon boundary conditions. We discuss the expected symmetries for two distinct choices of horizon boundary conditions. Firstly, we consider diffeomorphisms preserving the boundary conditions~\eqref{bcsintro}. The resulting symmetries are reminiscent, but not identical, to those argued for in~\cite{Knysh:2024asf}. They include certain time-independent shifts of the hydrodynamic fields, and we perform a non-trivial check that our action is indeed invariant under such symmetries. Additionally, they include exponentially growing and decaying (in time) symmetries reminiscent of those conjectured via quantum chaos~\cite{Knysh:2024asf, Blake:2017ris, Blake:2021wqj}.\footnote{We cannot explicitly test if such time-dependent modes are symmetries of the action at present, since this would require the action to all orders in derivatives.} However the profile of the exponentially growing mode is distinct from that identified in~\cite{Knysh:2024asf}.

\paragraph{} With this distinction in mind, we also consider the hydrodynamic action obtained using a second set of horizon boundary conditions, in which we replace~\eqref{bcsintro} by 
\begin{equation}
\delta g^{\sigma}_{t \mu}(r_0) = 0,  \hspace{2.0cm} \partial_{r} \delta g^{\sigma}_{tt}(r_0) = 0,
\label{bcsintro2}
\end{equation}
keeping the treatment of $\delta g_{xy}^{\sigma}$ and $\delta g_{m}^{\sigma}$ unchanged. The near-horizon diffeomorphisms preserving the boundary conditions~\eqref{bcsintro2} are precisely equivalent to the horizon symmetries argued for in~\cite{Knysh:2024asf}. Furthermore, to first order in boundary derivatives, the hydrodynamic action is identical to that obtained using the boundary conditions~\eqref{bcsintro}. However, there is a subtlety that prevents us immediately interpreting the diffeomorphisms of~\cite{Knysh:2024asf} as symmetries of the full action obtained using the boundary conditions~\eqref{bcsintro2}. Unlike the boundary conditions~\eqref{bcsintro}, we find that applying~\eqref{bcsintro2} (alongside the conditions on $\delta g_{xy}^\sigma$ and $\delta g_m^\sigma$) is not manifestly consistent with the variational principle at the horizon. As a result, solutions of the bulk Einstein equations are not guaranteed to be solutions of the equations of motion of the hydrodynamic action obtained using these boundary conditions.

\paragraph{}Before presenting our calculation, let us further contextualise our work by reviewing two alternative approaches to constructing dissipative hydrodynamical actions that have appeared in the literature. In parallel to the developments highlighted above, there has been an independent line of works which derive dissipative hydrodynamic actions \cite{Jana:2020vyx,Ghosh:2020lel,He:2021jna,He:2022jnc,He:2022deg,Loganayagam:2022zmq,Loganayagam:2022teq}, even beyond quadratic order.  These works do not directly work in terms of the metric, but package gravitational perturbations into gauge-invariant fields. Whilst successful, it is not directly clear how such approaches are related to the hydrodynamical dissipative action formalism developed in \cite{CGL}. We therefore view these results as an important but independent line of research to the approach we will outline here. Another approach for computing a hydrodynamic Schwinger-Keldysh action (for AdS$_5$ gravity) was proposed in~\cite{Bu:2025zad}, and looks superficially similar to our prescription. As discussed further in Appendix~\ref{app:bu}, we have been unable to reproduce these results. As such we believe that the results presented in this paper constitute the first consistent derivation of a dissipative hydrodynamical action, in the spirit of \cite{CGL}, from gravitational dynamics. 

\paragraph{} The paper is organised as follows. After reviewing the equilibrium state in Section~\ref{sec:background}, in Section~\ref{sec:transverse} we construct the hydrodynamical action for transverse metric perturbations -- this is particularly instructive because this is structurally similar to the CGL study of a U(1) charge presented in~\cite{CGL}. In Section~\ref{sec:longitudinal} we do the same for the longitudinal perturbations with the boundary conditions \eqref{bcsintro}. In Section~\ref{sec:tests} we present the full action and perform various consistency checks, expanding on the enumerated statements above. In Section~\ref{sec:horizonsyms} we discuss the symmetries of the hydrodynamic action that we expect to arise due to diffeomorphisms that preserve our horizon boundary conditions. Finally in Section~\ref{sec:altBCs} we discuss the hydrodynamic action and horizon symmetries for the alternative set of boundary conditions~\eqref{bcsintro2}. Several technical details are covered in the Appendices, including a more elementary presentation of the horizon symmetries argued for in~\cite{Knysh:2024asf} that is sufficient for the simple state we are studying (Appendix~\ref{app:knysh}). In particular, this Appendix relates the boundary conditions~\eqref{bcsintro2} to the geometric presentation of horizon symmetries in~\cite{Knysh:2024asf}.

\section{Background geometry }
\label{sec:background}
\paragraph{} We now turn to our main goal, which is to construct a hydrodynamic action describing the stress tensor dynamics for the dual of AdS$_4$-Schwarzschild. Such a hydrodynamical theory is dissipative beyond ideal order, and hence to construct an effective action it is necessary to work in Schwinger-Keldysh formalism. To do this we will work in the framework of Crossley, Glorioso and Liu (CGL)~\cite{CGL}, who formulated a prescription for deriving Schwinger-Keldysh actions from solving bulk dynamics on a contour in which the bulk radial direction is suitably analytically continued.

\begin{figure}[h!]
    \centering
    \includegraphics[width=0.8\linewidth]{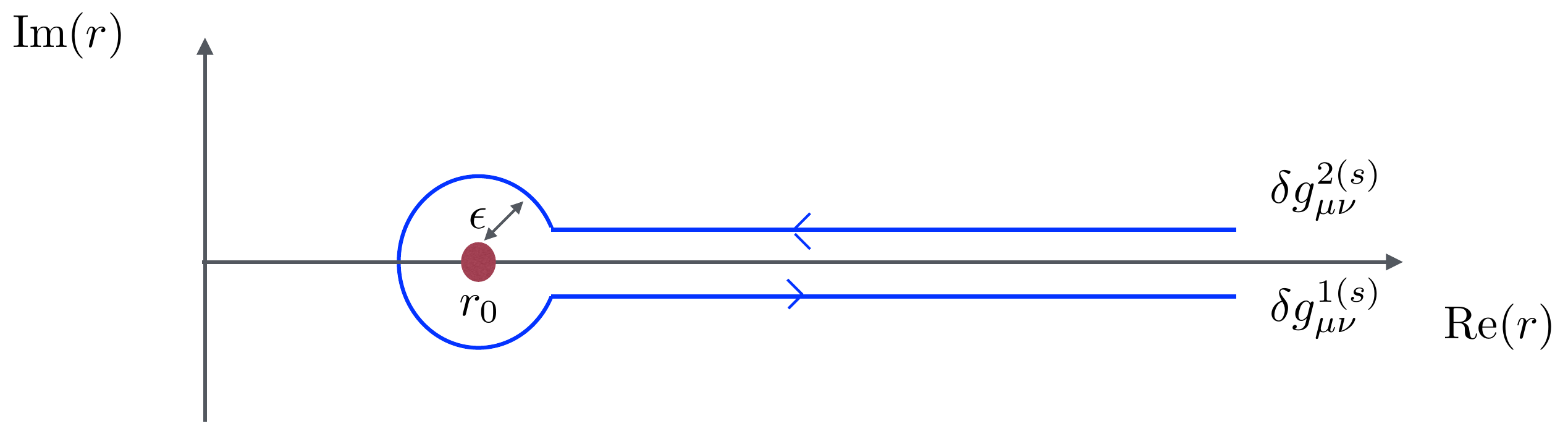}
    \caption{The CGL contour on which the linearised Einstein equations are studied. UV boundary conditions are imposed in Fefferman-Graham gauge near the asymptotic boundaries. The arrows represent the orientation of radial integration in the Einstein-Hilbert action.}
    \label{cglcontour}
\end{figure}

\paragraph{} Our starting point is the action $S = S_{\mathrm{EH}}  + S_{\mathrm{GH}}$, where for $3+1$ bulk dimensions we have
\begin{equation}
S_{\mathrm{EH}} = \int d^4x \sqrt{-g} (R + 6),   \hspace{2.0cm}  S_{\mathrm{GH}} = \int d^3x  \sqrt{-\gamma}(2 K - 4- R[\gamma]).
\label{EHads4}
\end{equation}
$S_{\mathrm{GH}}$ is a boundary term that we will evaluate at the conformal boundary of the spacetime, as explained in more detail below. The first term in $S_{\mathrm{GH}}$ is the Gibbons-Hawking term while the others are the usual holographic counterterms \cite{deHaro:2000vlm}.
The equations of motion are 
\begin{equation}
E_{M N} =  R_{M N} - \frac{g_{M N}}{2}(R + 6) = 0.
\end{equation}
We are interested in the planar AdS$_4$-Schwarzschild black hole solution, which can be written in Schwarzschild coordinates $(\tau, r, x, y)$ as
\begin{equation}
\label{eq:BGSchwarzCoords}
ds^2 = -D(r) d\tau^2 + \frac{dr^2}{D(r)} + r^2(dx^2 + dy^2), \hspace{2.0cm} D(r) = r^2\bigg(1- \frac{r_0^3}{r^3} \bigg),
\end{equation}
where the conformal boundary is at $r\rightarrow\infty$.
We introduce the ingoing coordinate $t$ through
\begin{equation}
t = \tau + r_*(r),\quad\quad\quad\quad\quad\quad r_*(r)=\int_{\infty}^{r} \frac{d\tilde{r}}{D(\tilde{r})}.
\label{ingoing}
\end{equation}
The background metric \eqref{eq:BGSchwarzCoords} in ingoing coordinates $(t, r, x, y)$ is then
\begin{equation}
ds^2 = -r^2f(r) dt^2 + 2 dt dr + r^2(dx^2 + dy^2), \hspace{2.0cm} f(r) = \bigg(1- \frac{r_0^3}{r^3} \bigg).
\end{equation}
This solution is dual to a thermal state with temperature $4 \pi T = 3 r_0$, entropy density $s = 4 \pi r_0^2$, energy density $\varepsilon\equiv \langle T^{tt} \rangle = 2 r_0^3$ and pressure $p\equiv \langle T^{xx} \rangle = \langle T^{yy} \rangle = r_0^3$.
\paragraph{} Our goal is to derive a hydrodynamic action describing perturbations about this equilibrium solution. To do this, we study perturbations defined in the bulk on the CGL contour defined by the analytic continuation of the radial coordinate shown in Fig~\ref{cglcontour}. In practice, we will construct partially on-shell solutions to the equations of motion, piecewise on the two branches of the contour, and then relate these two solutions by imposing appropriate boundary conditions near the horizon $r=r_0$. Schematically, we can think of the two straight branches of the contour as capturing the two exteriors of the maximally extended black hole spacetime.

\paragraph{}Due to the rotational symmetry of the background spacetime in the $(x,y)$ plane, linearised perturbations of the metric can be decomposed into two decoupled sets. We call these transverse and longitudinal perturbations and now treat them independently in turn, before combining them to obtain a rotationally invariant action.

\section{Transverse perturbations}
\label{sec:transverse}

\paragraph{} We first study the linearised metric perturbations in the transverse channel
\begin{equation}
\delta ds^2 = 2 \delta g^{\sigma}_{ty}(t,r,x) dt dy + 2 \delta g^{\sigma}_{yr}(t,r,x) dr dy + 2 \delta g^{\sigma}_{xy}(t,r,x) dx dy,
\label{transversepert}
\end{equation}
where we have used rotational symmetry to align the spatial dependence of the fields with the $x$ direction, without loss of generality. We will use $z$ to denote the complexified radial coordinate, and reserve $r$ for the two real segments (which is what we will be mostly concerned with). We distinguish between the two real segments by a superscript index $\sigma = 1,2$ on the fields, rather than by a different coordinate. On the upper real segment we denote the solution $\delta g^2_{\mu\nu}(r-r_0)$ and on the lower real segment we denote it $\delta g^1_{\mu\nu}(r-r_0)$. We will describe later how we match these solutions around the branch point at $r=r_0$.

\paragraph{} We decompose the metric perturbations in the following way 
\begin{eqnarray}
\delta g^{\sigma}_{ty}(t,r,x)&=& r^2 h^{\sigma}_{ty}(t,r,x)+ \partial_{t} \zeta^{\sigma}_{y}(t,r,x), \nonumber \\
\delta g^{\sigma}_{yr}(t, r, x)&=& - \frac{2}{r} \zeta^{\sigma}_{y}(t,r,x) + \partial_{r} \zeta^{\sigma}_{y}(t,r,x), \nonumber \\
\delta g^{\sigma}_{xy}(t,r,x)  &=& r^2 h^{\sigma}_{xy}(t,r,x) + \partial_{x} \zeta^{\sigma}_{y}(t,r,x).
\label{transverseansatz} 
\end{eqnarray}
This decomposition has a natural interpretation -- the fields $h^{\sigma}_{\mu \nu}$ correspond to a metric perturbation in radial gauge in ingoing coordinates. The fields $\delta g^{\sigma}_{MN}$ then correspond to acting on this perturbation with a diffeomorphism generated by the field $\zeta^{\sigma}_{y}$. As we will explicitly see soon, $\zeta^{\sigma}_{y}$ identically cancels in the Einstein equations, which give relations only between the $h^{\sigma}_{\mu\nu}$. Throughout, we raise and lower indices on the bulk fields $\zeta_\mu^\sigma(t,r,x)$ with the AdS$_4$-Schwarzschild metric.

\paragraph{}Although not constrained by the Einstein equation, we will not allow $\zeta^{\sigma}_{y}(t,r,x)$ to be completely arbitrary bulk diffeomorphisms. In particular we will perform a partial gauge fixing to control their asymptotics near the AdS boundaries and the black hole horizon. In the UV we will constrain the expansions of $\zeta^{\sigma}_{y}(t,r,x)$ at large $r$ such that the spacetime is asymptotically AdS$_4$ (this requires transforming to Fefferman-Graham coordinates near the boundary as is performed explicitly in Appendix~\ref{app:uvbcs}). Near the horizon, we will impose radial gauge in ingoing coordinates (i.e. $\delta g^{\sigma}_{yr} = 0$). This can be done order-by-order in an expansion in $(r-r_0)$. At leading order it requires
\begin{equation}
\zeta^{y,\sigma}(t, r\rightarrow r_0, x) = Y^{\sigma}(t,x)+O((r-r_0)^2), 
\label{irdiffeotransverse}
\end{equation}
near the horizon. $Y^{\sigma}(t,x)$ generates a residual gauge transformation near the horizon: it preserves radial gauge $\delta g^\sigma_{yr}(r_0)=0$ while acting non-trivially on the other components of the metric near the horizon.

\subsubsection*{Hydrodynamic variables}

\paragraph{} Starting from the (3+1)-dimensional bulk action on the CGL contour, we now seek to construct a hydrodynamic action for transverse perturbations. This is a (2+1)-dimensional action of the form 
\begin{equation}
\label{eq:hydroactionstructure}
S_{\mathrm{hydro}} [\xi_y^{\sigma}, \delta g_{ty}^{\sigma (s)}, \delta g_{xy}^{\sigma (s)}],
\end{equation}
where $\xi_y^{\sigma}$ are the dynamical fields and $\delta g_{ty}^{\sigma (s)}, \delta g_{xy}^{\sigma (s)}$ are external sources. These fields and sources are functions of only the boundary coordinates $(t, x)$. Throughout, we will raise and lower indices on $\xi_\mu^{\sigma}$ and $\delta g_{\mu\nu}^{\sigma (s)}$ using the Minkowski metric.

\paragraph{}The hydrodynamic variables that will arise naturally from our analysis are
\begin{equation}
\xi_{y}^{\sigma}(t,x) = [\zeta^{y}]^{\infty}_{r_0} = \zeta^{y, \sigma}(t, r=\infty, x) - Y^{\sigma}(t,x).
\label{transversehydrofield}
\end{equation}
These are relative diffeomorphisms in the $y$-direction between the asymptotic boundaries of the CGL contour and the black hole horizon. They can equivalently be expressed as Wilson line-like integrals of the metric perturbations
\begin{equation}
\xi_y^\sigma(t,x)=\int^\infty_{r_0}\frac{\delta g_{yr}^\sigma(t,r,x)}{r^2} dr,
\end{equation}
after using equation \eqref{transverseansatz} to express $\zeta_y$ in terms of the metric. They vanish identically if one works in radial gauge.

\paragraph{} It is common in computations of Schwinger-Keldysh actions to arrange the fields into symmetric and antisymmetric combinations defined by 
\begin{equation}
\xi_{\mu}^{+} = \frac{\xi_{\mu}^1 + \xi_{\mu}^2}{2}, \hspace{2.0cm} \xi_{\mu}^{-} =  \xi_{\mu}^1 - \xi_{\mu}^2,
\label{hydrora}
\end{equation}
and likewise for the sources
\begin{equation}
\delta g_{\mu \nu}^{+ (s)} = \frac{\delta g_{\mu \nu}^{1 (s)}  + \delta g_{\mu \nu}^{2 (s)} }{2},  \hspace{2.0cm} \delta g_{\mu \nu}^{- (s)} =  \delta g_{\mu \nu}^{1 (s)}  -  \delta g_{\mu \nu}^{2 (s)}. 
\label{metricra}
\end{equation}
It will be convenient to introduce the following combinations
\begin{eqnarray}
B_{t y}^{\sigma} = \delta g_{t y}^{\sigma (s)} - \partial_{t} \xi_{y}^{\sigma}, \hspace{2.0cm} B_{x y}^{\sigma} = \delta g_{x y}^{\sigma (s)} - \partial_{x} \xi_{y}^{\sigma}, 
\label{transversebmunu}
\end{eqnarray}
together with their symmetric and antisymmetric combinations $B_{t y}^{\pm}, B_{xy}^{\pm}$.\footnote{We always define symmetric and antisymmetric combinations of fields as in \eqref{hydrora} and \eqref{metricra}, with a factor of 2 in the denominator of the symmetric combination.} The combinations~\eqref{transversebmunu} are invariant under the linearised boundary diffeomorphisms
\begin{equation}    \xi_y^\sigma\rightarrow\xi_y^\sigma+\epsilon_y^\sigma,\quad\quad\quad\quad \delta g_{t y}^{\sigma (s)}\rightarrow\delta g_{t y}^{\sigma (s)}+\partial_t\epsilon_y^\sigma,\quad\quad\quad\quad \delta g_{x y}^{\sigma (s)}\rightarrow\delta g_{x y}^{\sigma (s)}+\partial_x\epsilon_y^\sigma,
\end{equation}
and will naturally arise in our construction below.

\paragraph{} Our starting point for computing this hydrodynamic action is the bulk action~\eqref{EHads4}, now defined on the full CGL contour with the boundary term $S_{\mathrm{GH}}$ at the asymptotic boundary of each branch. Note the orientation of the radial integrals and boundary terms is indicated by the arrows in Fig.~\ref{cglcontour}. In practice the gravitational action will essentially amount to taking the difference between~\eqref{EHads4} on the two branches, up to subtleties regarding certain horizon contributions that will be explained later. We expand the action evaluated on the metric perturbations~\eqref{transversepert} as
\begin{equation}
\label{eq:schematicbulkexp}
S = S_0 + S_1[\delta g^{\sigma}_{\mu \nu}] + S_2[\delta g^{\sigma}_{\mu \nu}] + \dots .
\end{equation}
Here $S_0$ is the action evaluated on the background solution, $S_1$ the action that is linear in the amplitude of the perturbations and $S_2$ quadratic. The dots indicate cubic and higher terms that we will neglect. To convert the action \eqref{eq:schematicbulkexp} for bulk metric perturbations into a boundary action for hydrodynamic fields of the form \eqref{eq:hydroactionstructure}, we will evaluate $S_1$ and $S_2$ for partially on-shell solutions: those for which we have imposed a subset of the Einstein equations. 

\subsection*{Equations of motion}

\paragraph{} On each branch of the contour there are 3 Einstein equations $\delta E^{\sigma}_{ty}, \delta E^{\sigma}_{ry}, \delta E^{\sigma}_{xy}$ governing the bulk metric perturbations \eqref{transversepert}. These can be arranged into two dynamical equations (second order in radial derivatives) and one constraint equation (first order). With the metric decomposed as in \eqref{transverseansatz}, the second order equations are
\begin{eqnarray}
\label{ads4transversedynamical}
\delta E^{\sigma}_{ry} &=& 2 r \partial_{r} h^{\sigma}_{ty} + \frac{1}{2} \partial_{r} \partial_{x} h^{\sigma}_{xy} + \frac{1}{2} r^2 \partial_{r}^2 h^{\sigma}_{ty},\\
\delta E^{\sigma}_{xy} &=& \frac{1}{2}\bigg[ 2 r \partial_{x} h^{\sigma}_{ty} + (-4 r^3 + r_0^3) \partial_{r} h^{\sigma}_{xy} + r^2 \partial_{r} \partial_{x} h^{\sigma}_{ty} - r^4f(r) \partial_{r}^2 h^{\sigma}_{xy} - 2 r \partial_{r} (r \partial_{t} h^{\sigma}_{xy}) \bigg], \nonumber 
\end{eqnarray} 
while the constraint, which we label $\text{momentum}_y$ for reasons that will be apparent shortly, is 
\begin{equation}
\begin{aligned}
\label{constrainttransverse}
\mathrm{momentum}^{\sigma}_y &\,= 2\left(r^2  D(r) \delta E^{\sigma}_{ry} + r^2 \delta E^{\sigma}_{ty}\right) \\
&\,=-\partial_t\left(r^4\partial_rh_{ty}^\sigma-r^2\partial_xh_{xy}^\sigma\right)-\partial_x\left(-r^4f\partial_rh_{xy}^\sigma+r^2\partial_xh_{ty}^\sigma\right).
\end{aligned}
\end{equation}

Notice that the diffeomorphism field $\zeta^\sigma_y$ drops out of the Einstein equations \eqref{ads4transversedynamical} and \eqref{constrainttransverse} and so these are three equations for two fields (on each branch). However, the system is not overdetermined: adding radial derivatives of the constraint to a dynamical equation generates the other dynamical equation:
\begin{equation}
    \partial_r\left(\text{momentum}^\sigma_y\right)+2\partial_x \delta E_{xy}^\sigma+2r^2\partial_t \delta E_{ry}^\sigma=0.
\end{equation}
Equivalently, imposing both dynamical equations requires the constraint to be radially independent. The constraint will ultimately be the $y$ component of the momentum conservation equation arising from our hydrodynamic action, which is why we label it $\text{momentum}^\sigma_y$.

\subsubsection*{Partially on-shell solution}
\paragraph{} Our strategy for computing the hydrodynamical action is then as follows. We will solve the dynamical equations~\eqref{ads4transversedynamical} but not the constraint equations~\eqref{constrainttransverse}. We call these partially on-shell solutions, and then substitute these into the action to obtain the hydrodynamic action. The constraint equations (i.e.~the local conservation law for the $y$-component of the momentum) will end up being equivalent to the equations of motion obtained by varying the hydrodynamical action with respect to the fields $\xi^{\sigma}_y$. The dynamical equations for $h^{\sigma}_{xy}, h^{\sigma}_{ty}$ cannot be solved in closed form exactly, and so we will solve them in an expansion with respect to boundary derivatives $\partial_{t}, \partial_{x}$. We denote this expansion via
\begin{equation}
h^{\sigma}_{\mu \nu}(t,r,x) = h^{\sigma}_{\mu \nu, 0}(t,r,x) + h^{\sigma}_{\mu \nu, 1}(t,r,x)+ \dots.
\end{equation}
We will carry out the computation of the action by solving to first order in derivatives.  

\paragraph{} We begin by presenting the solutions to the partially on-shell equations of motion. We expand the equations~\eqref{ads4transversedynamical} to zeroth order in $\partial_{t}, \partial_{x}$ and solve for $h_{ty, 0}^{\sigma}, h_{xy, 0}^{\sigma} $. This yields  
\begin{eqnarray}
h^{\sigma}_{ty, 0}(t,r,x) &=& - \frac{\beta^{\sigma}_1(t,x)}{3 r^3} + \beta^{\sigma}_2(t, x), \nonumber \\
 h^{\sigma}_{xy, 0}(t,r,x) &=& \beta^{\sigma}_3(t,x) + \beta^{\sigma}_4(t,x) \mathrm{log}f(r).
 \label{hxyzeroorder}
\end{eqnarray}

At first order in derivatives the solutions are\footnote{The additional $1$ subscript on $\beta$s indicates quantities that are first order in the derivative expansion. While these quantities can be removed by appropriate redefinitions of the original $\beta_i^\sigma$, we will keep them explicit.}
\begin{eqnarray}
h^{\sigma}_{ty, 1}(t,r,x) &=& - \frac{\beta^{\sigma}_{11}(t,x)}{3 r^3} + \beta^{\sigma}_{21}(t,x) - \frac{\partial_{x} \beta_{4}^{\sigma}(t,x)}{r} - f(r) H(r) \partial_{x} \beta_{4}^{\sigma}(t,x),  \nonumber \\
h^{\sigma}_{xy,1}(t,r,x)&=& \beta^{\sigma}_{31}(t,x) + \beta^{\sigma}_{41}(t,x) \mathrm{log}f(r)   - H(r) \mathrm{log}f(r) \partial_t \beta^{\sigma}_{4}(t,x)  +  \bigg(\frac{r - r_0}{3 r r_0^4}\bigg)\partial_{x} \beta^{\sigma}_1(t,x) \nonumber \\ 
&-& H(r)\left(\partial_{t} \beta^{\sigma}_{3}(t,x)  + \frac{1}{3r_0^3}\partial_{x} \beta_1^{\sigma}(t,x)- \partial_{x} \beta_2^{\sigma}(t,x)\right),
\label{hxyfirstorder}
\end{eqnarray}
where
\begin{equation}
\label{eq:Hdefn}
H(r) = \frac{\pi}{2\sqrt{3}r_0}+r_*(r) = \frac{1}{6 r_0}\left( 2 \sqrt{3} \tan^{-1}\left(\frac{2 r + r_0}{\sqrt{3} r_0}\right) + \mathrm{log}\frac{(r - r_0)^2}{r^2 + r r_0 + r_0^2} \right).
\end{equation}
%
%
%
Using these solutions we can compute the constraint equations~\eqref{constrainttransverse} to second order in boundary derivatives. This yields 
\begin{equation}
\mathrm{momentum}^{\sigma}_y = 3 r_0^3 \partial_{x} \beta^{\sigma}_4  - \partial_{t} \beta^{\sigma}_1 + 3 r_0^3 \partial_{x} \beta^{\sigma}_{41} - \partial_{t} \beta^{\sigma}_{11}.  
\end{equation}
As previously indicated these are radially independent now that the dynamical equations have been imposed. 
\paragraph{} On each branch of the contour there are $4$ independent integration constants at each order in the derivative expansion (i.e. $\beta_1^{\sigma} \dots \beta_4^{\sigma}$ at zeroth order). We now fix these integration constants by applying boundary conditions near the asymptotic boundaries and horizon.

\subsubsection*{Asymptotic and horizon boundary conditions}

\paragraph{} Let us first discuss the boundary conditions we impose on our partially on-shell solutions near the asymptotic boundaries of the CGL contour. We start by expanding the bulk diffeomorphism fields $\zeta^{\sigma}_y$ for large $r$ as\footnote{With these asymptotics, the hydrodynamic field $\xi_y^\sigma(t,x)=\zeta^{\sigma (2)}_y(t,x)-Y^\sigma(t,x)$.}
\begin{equation}
\zeta^{\sigma}_y(t,r,x) = \zeta^{\sigma (2)}_y(t,x) r^2 + \zeta^{\sigma (1)}_y(t,x) r + \dots .
\label{zetayexpansion}
\end{equation}
We then restrict this expansion such that, when we switch to Schwarzschild coordinates, the partially on-shell solutions are in Fefferman-Graham gauge. Having done this, we then simply use the usual holographic dictionary to identify the field theory metric from the asymptotic bulk metric in this gauge. The details of this computation are in Appendix~\ref{app:uvbcs}. The result is that at zeroeth order in derivatives we can relate two of the integration constants to the field theory metric perturbations via
\begin{equation}
\label{eq:transverseUVBC1}
\beta_2^{\sigma} = \delta g_{ty}^{\sigma (s)} - \partial_t \zeta_y^{\sigma (2)},  \hspace{2.0cm} \beta_3^{\sigma} = \delta g_{xy}^{\sigma (s)} - \partial_x \zeta_y^{\sigma (2)}.
\end{equation}
Note that we view the bulk fields $\partial_{t} \zeta^{\sigma}_y, \partial_{x} \zeta^{\sigma}_y$ as zeroth order in our derivative expansion when imposing boundary conditions. At first order in derivatives we find
\begin{equation}
\begin{aligned}
\label{eq:transverseUVBC2}
\beta_{21}^\sigma=\frac{\pi}{2\sqrt{3}r_0}\partial_x\beta_4^\sigma,\quad\quad\quad
\beta_{31}^{\sigma}=-\frac{(2\sqrt{3}-\pi)}{6\sqrt{3}r_0^4}\partial_x\beta_1^\sigma+\frac{\pi}{2\sqrt{3}r_0}(\partial_t B_{xy}^{\sigma}-\partial_x B_{ty}^{\sigma }).
\end{aligned}
\end{equation}

\paragraph{}Now we turn to the boundary conditions near the horizon on each branch. These are motivated by the gauge field case studied in~\cite{CGL}, which is structurally similar, and in Appendix \ref{app:varprinciple} we discuss their relation to the variational principle. Our first horizon boundary condition is to fix $\delta g_{ty}^\sigma(r_0)=0$ on both branches. At the first two orders in the derivative expansion this  requires
\begin{equation}
\beta_1^{\sigma} =  3 r_0^3(\beta_2^{\sigma} + \partial_{t} Y^{\sigma}) =  3 r_0^3 B_{ty}^{\sigma},\quad\quad\quad\quad\quad \beta^{\sigma}_{11}= \frac{\sqrt{3} \pi - 6}{2} r_0^2 \partial_{x} \beta^{\sigma}_4,
\end{equation}
where we inserted \eqref{eq:transverseUVBC1} to get the second equality in the first expression.

\paragraph{}The second horizon boundary condition concerns the remaining field $\delta g_{xy}$. On both branches, this field has logarithmic divergences near the horizon, which can be understood as the derivative expansion of the solution $\sim(r-r_0)^{-i\omega/4\pi T}$ that is outgoing on the future horizon.\footnote{The contributions of such terms to $\delta g_{ty}$ are suppressed by powers of $(r-r_0)$ near the horizon and so they played no role above.} The boundary condition we impose is that $\delta g_{xy}$ should be continuous on the CGL contour, after suitably extending the solution on the lower branch around the circle at the horizon to where it joins the upper branch.\footnote{The explicit values of fields around the circle are unimportant in the limit $\epsilon\rightarrow0$ where the circle shrinks to zero, and so we don't present them.} In practice this means we expand the solution for $\delta g_{xy}^1(r)$ near the horizon,\footnote{Where it has the form $\delta g_{xy}^1(r\rightarrow r_0)=a(t,x)\log^2(r-r_0)+b(t,x)\log(r-r_0)+c(t,x)+\ldots$.} analytically continue $\log(r-r_0)\rightarrow\log (r-r_0)-2\pi i$, and then match the constant and $\log(r-r_0)$ terms in the subsequent expression to the near-horizon expansion of $\delta g_{xy}^2(r)$.\footnote{This automatically fixes the coefficients of the $\log^2(r-r_0)$ terms on each branch to be equal to the order in the derivative expansion that we are working.} At the first two orders in the derivative expansion this gives the conditions 
\begin{equation}
\beta_4^{\sigma} =  -\frac{i}{2 \pi} B_{xy}^{-},\quad\quad\quad\quad\quad \beta^{\sigma}_{41} = -\frac{\sqrt{3} i}{12 r_0} \partial_{t} B_{xy}^{-} +\frac{i}{2 \pi r_0} \partial_{x} B_{ty}^{-} + \frac{1}{3 r_0} \partial_{t} B_{xy}^{\bar{\sigma}},
\end{equation}
where $\bar{\sigma} = 1$ if $\sigma=2$ and $\bar{\sigma} = 2$ if $\sigma = 1$.

\paragraph{}With this, we have now fixed all of the $\beta^\sigma$ integration constants in the partially on-shell solution in terms of the sources $\delta g_{\mu y}^{\sigma (s)}$ and  gauge transformations $\zeta_y^{\sigma (2)}$, $Y^\sigma$. These results are summarised in Appendix \ref{app:ads4nearhorizon}. Our final step is to insert these partially on-shell solutions into the gravitational action, which will reduce it to a (2+1)-dimensional action dependent only on the hydrodynamic fields $\xi_y^\sigma$ and sources $\delta g_{\mu y}^{\sigma (s)}$.

\subsection*{Quadratic hydrodynamical action}

\paragraph{} The linear in amplitude action $S_1$ vanishes for transverse perturbations, and so the leading contribution is the term $S_2$ that is quadratic in amplitudes.\footnote{Here, and throughout, we will regularly integrate by parts and then neglect terms that are total derivatives in the $(t,x)$ coordinates.} To compute the partially on-shell action we note that the quadratic Einstein-Hilbert action on transverse perturbations can be written (up to a boundary term) on each branch as 
\begin{equation}
S_{\mathrm{EH}} = S_{\mathrm{bulk}} + S_{\mathrm{boundary}}, 
\end{equation}
where the bulk action is
\begin{eqnarray}
S_{\mathrm{bulk}} &=&  \int d^{4} x r^2 (\delta E_{ty} \delta g^{ty} + \delta E_{ry} \delta g^{ry} + \delta E_{xy} \delta g^{xy}) \nonumber \\
&=&-\frac{1}{2}\int d^4x \left( \partial_r \left( \zeta^y \mathrm{momentum}_y \right) + 2r^2 \delta E_{ry} h_{ty} + 2\delta E_{xy} h_{xy} \right),
\end{eqnarray}
and we will return to $S_{\mathrm{boundary}}$ shortly. When going partially on-shell, we imposed $\delta E_{ry} = \delta E_{xy} = 0$. Going partially on-shell therefore corresponds to integrating out the bulk spacetime, such that $S_{\mathrm{bulk}}$ also reduces to a boundary term.

\paragraph{}In order to evaluate the contributions of all of the boundary terms, we need to be careful as it is not clear, a priori, whether the action is continuous around the horizon on the CGL contour. To deal with this, we will split the contour integral into two: one along the upper branch, and one along the lower branch plus the segment that circles the horizon.

\paragraph{}We will deal first with the explicit boundary contributions from $S_{\mathrm{boundary}}$
\begin{eqnarray}
S_{\mathrm{boundary}} &=& -\int d^3x\frac{1}{2 r^4} \bigg( 2 r^3 \delta g_{ty}^2 + 3 r^4 \delta g_{ty} \partial_{r} \delta g_{ty} -  r^3f(r) \delta g_{xy} (- 4\delta g_{xy} + 3 r \partial_{r} \delta g_{xy}) \nonumber \\
&+& 3 r (2 r^6 - 3 r^3 r_0^3 + r_0^6) \delta g_{yr}^2 - r^2 \delta g_{xy} \partial_{x} \delta g_{ty} - r^4f(r) \delta g_{xy} \partial_{x} \delta g_{yr} \nonumber \\
&+& \delta g_{ty} (r^4 \partial_{t} \delta g_{yr} + (12 r^5 - 6 r^2 r_0^3)  \delta g_{yr}) \bigg).
\label{transverseboundaryterm}
\end{eqnarray}
There is a contribution at each asymptotic boundary consisting of this term, plus $S_{GH}$, which we label $S_{\mathrm{UV}}^\sigma$. Explicitly
\begin{equation}
    \begin{aligned}
    \label{eq:StransverseUV}
        S_{\mathrm{UV}}^{\sigma}=r_0^3\int d^3x\Biggl(&\delta g_{ty}^{\sigma (s)}\left(\frac{3}{2}B_{ty}^\sigma+\frac{3i}{4\pi r_0}\partial_xB_{xy}^- - \delta g_{ty}^{\sigma (s)}\right)\\
        &\,+\delta g_{xy}^{\sigma (s)}\left(\frac{3i}{4\pi}\left(B_{xy}^- -\frac{1}{r_0}\partial_xB_{ty}^-\right) -\frac{1}{2r_0}\partial_t B_{xy}^{\bar{\sigma}}-\frac{1}{2}\delta g_{xy}^{\sigma (s)}\right)\Biggr).
    \end{aligned}
\end{equation}
There is also, in principle, a contribution from \eqref{transverseboundaryterm} from the horizon end of each of our two integrals. Each horizon contribution has a logarithmically divergent piece plus a piece $\sim(r-r_0)^0$. For the integral along the upper branch, we simply take the $r\rightarrow r_0$ limit of $S_{\mathrm{boundary}}$ with $\delta g_{\mu\nu}\rightarrow \delta g_{\mu\nu}^{2}$. For the second integral, we first take the $r\rightarrow r_0$ limit of $S_{\mathrm{boundary}}$ with $\delta g_{\mu\nu}\rightarrow\delta g_{\mu\nu}^1$, and then subsequently circle the horizon by analytically continuing $\log(r-r_0)\rightarrow \log(r-r_0)-2\pi i$ in this expression.\footnote{We can think of this replacement as being the effect on $\delta g_{\mu\nu}^1$ of circling the horizon, which is consistent with the boundary conditions we imposed on $\delta g_{xy}$ previously.} These horizon contributions from the two endpoints cancel out. In other words, the boundary conditions we have imposed near the horizon are such that there is no contribution from  $S_{\mathrm{boundary}}$ at the horizon.

\paragraph{}Finally we return to the contributions from $S_{\mathrm{bulk}}$. Recalling that the constraint equation $\text{momentum}_y$ is radially conserved when partially on-shell, these contributions will naturally involve the relative diffeomorphisms between the asymptotic boundaries and the horizon endpoints. The analytic continuation of the diffeomorphism field around the horizon is trivial and so combining the expressions for both branches gives
\begin{equation}
\label{transversebulkaction}
S_{\mathrm{bulk}} =   -\frac{1}{2}\int d^3x  \left( \xi^1_y \mathrm{momentum}^1_y -   \xi^2_y \mathrm{momentum}^2_y \right),
\end{equation}
where
\begin{equation}
\label{eq:transconstraintsexplicit}
    \text{momentum}_y^\sigma=-r_0^3\left(3\partial_t B_{ty}^\sigma+\frac{3i}{2\pi}\partial_xB_{xy}^{-}-\frac{3i}{2\pi r_0}\partial_x\left(\partial_xB_{ty}^- -\partial_tB_{xy}^-\right)-\frac{1}{r_0}\partial_x\partial_t B_{xy}^{\bar{\sigma}}\right).
\end{equation}

\paragraph{}Combining this with the UV contribution to the action~\eqref{eq:StransverseUV} we obtain the hydrodynamical action as $S_{\mathrm{bulk}} + S_{\mathrm{UV}}^{1} - S_{\mathrm{UV}}^{2}$. Up to first order in derivatives we obtain the action
\begin{equation}
\begin{aligned}
S_\mathrm{hydro}[\xi^{\pm}_y, \delta g_{ty}^{\pm (s)}, \delta g_{xy}^{\pm (s)} ]   =r_0^3 \int &d^3{x} \bigg( 3 B_{ty}^{+} B_{ty}^{-}   + \frac{3 i}{4 \pi} B_{xy}^{-} B_{xy}^{-}  + \frac{3 i}{2 \pi r_0}  B_{ty}^{-} \partial_{x} B_{xy}^{-} \\
&+ \frac{1}{r_0} B_{xy}^{+} \partial_{t} B_{xy}^{-}- \delta g_{xy}^{+ (s)} \delta g_{xy}^{- (s)} - 2 \delta g_{ty}^{+ (s)} \delta g_{ty}^{- (s)} \bigg).
\label{transverseactionads4}
\end{aligned}
\end{equation}
Note that \eqref{transverseactionads4} includes terms that cannot be expressed solely in terms of the fields $B^{\pm}_{xy}$ and $B^{\pm}_{ty}$. As we will discuss in Section~\ref{sec:tests}, this is related to the non-linear nature of diffeomorphisms.  

\section{Longitudinal perturbations}
\label{sec:longitudinal}

\paragraph{} We now consider the following class of metric perturbations describing the longitudinal channel 
\begin{eqnarray}
\delta ds^2 &=& \delta g^{\sigma}_{tt}(t, r, x) dt^2 + 2 \delta g^{\sigma}_{tr}(t,r, x) dt dr + 2 \delta g^{\sigma}_{tx}(t,r,x)
 dt dx + 2 \delta g_{xr}^\sigma(t,r,x) dx dr \nonumber \\
 &+& \delta g^{\sigma}_{rr}(t,r,x) dr^2 + \delta g^{\sigma}_{xx}(t,r,x) dx^2 + \delta g^{\sigma}_{yy}(t,r,x) dy^2,  
 \end{eqnarray}
where $\sigma = 1,2$ is an index labelling the perturbation on each contour. To obtain the hydrodynamic action we will essentially follow the same strategy as above. While the algebra is more complicated due to the larger number of fields, the only real differences are in what equations we impose when going partially on-shell, and the horizon boundary conditions.

\paragraph{}We first decompose the perturbations as 
\begin{eqnarray}
\label{eq:longdecomp}
\delta g^{\sigma}_{tt} &=& -D(r) h^{\sigma}_{tt}- D(r) D'(r) \zeta^{\sigma}_{r} - D'(r) \zeta^{\sigma}_{t} + 2 \partial_{t} \zeta^{\sigma}_t, \nonumber \\
\delta g^{\sigma}_{tr} &=& D'(r) \zeta^{\sigma}_r + \partial_{r} \zeta^{\sigma}_t + \partial_{t} \zeta^{\sigma}_r, \nonumber \\
\delta g^{\sigma}_{tx} &=& r^2 h^{\sigma}_{tx} + \partial_{x} \zeta^{\sigma}_t + \partial_{t} \zeta^{\sigma}_x, \nonumber \\
\delta g^{\sigma}_{xr} &=& - \frac{2}{r} \zeta^{\sigma}_x + \partial_{x} \zeta^{\sigma}_r + \partial_{r} \zeta^{\sigma}_x, \nonumber \\
\delta g^{\sigma}_{xx} &=&  \frac{r^2}{2}(h^{\sigma}_p + h^{\sigma}_m) + 2 r D(r) \zeta^{\sigma}_r + 2 r \zeta^{\sigma}_t + 2 \partial_{x} \zeta^{\sigma}_x, \nonumber \\
\delta g^{\sigma}_{yy} &=& \frac{r^2}{2}(h^{\sigma}_{p} - h^{\sigma}_{m}) + 2 r D(r) \zeta^{\sigma}_{r} + 2 r \zeta^{\sigma}_{t}, \nonumber \\
\delta g^{\sigma}_{rr} &=& 2 \partial_{r} \zeta^{\sigma}_r, 
\end{eqnarray} 
where all fields are functions of $t,r,x$. As is now familiar this can be interpreted as a decomposition into a solution in radial gauge plus a diffeomorphism generated by $\zeta_t^{\sigma}$, $\zeta_x^{\sigma}$, and $\zeta_r^{\sigma}$. 

\paragraph{}In the UV we will constrain the diffeomorphism at large $r$ such that the spacetime is asymptotically AdS$_4$ (see Appendix~\ref{app:uvbcs} for more details). Near the horizon, we will impose radial gauge in ingoing coordinates, which implies the diffeomorphism fields take the form 
\begin{eqnarray}
\zeta^{\sigma}_r(t, r\rightarrow r_0, x)&=& T^{\sigma}(t, x)+O((r_0-r)^2), \nonumber \\
\zeta^{\sigma}_x(t, r\rightarrow r_0, x) &=& X^{\sigma}(t, x) r^2 + \partial_{x} T^{\sigma}(t, x) r+O((r_0-r)^2), \nonumber \\
\zeta^{\sigma}_t(t,r\rightarrow r_0,x) &=& - D(r) T^{\sigma}(t, x) - r \partial_{t} T^{\sigma}(t, x)  + R^{\sigma} (t, x)+O((r_0-r)^2).
\label{IRdiffeoslongitudinal}
\end{eqnarray}
The fields $R^{\sigma}(t,x), X^{\sigma}(t,x), T^{\sigma}(t,x)$ generate residual gauge transformations that preserve radial gauge on the horizon.
\subsubsection*{Hydrodynamic variables}

\paragraph{} We will seek to construct a hydrodynamic action for longitudinal perturbations
\begin{equation}
S_{\mathrm{hydro}}[\xi_{t}^{\sigma}, \xi_{x}^{\sigma}, \delta g_{tt}^{\sigma (s)}, \delta g_{tx}^{\sigma (s)} , \delta g_{xx}^{\sigma (s)}, \delta g_{yy}^{\sigma (s)}].
\end{equation}
The hydrodynamical variables that will naturally arise in our analysis are
\begin{eqnarray} 
\xi^{\sigma}_t(t,x) &=& [-\zeta^{t, \sigma}]^{\infty}_{r_0} = -\zeta^{t, \sigma}(t,r=\infty,x) + T^{\sigma}(t,x),  \nonumber \\
\xi^{\sigma}_x(t,x) &=& [\zeta^{x, \sigma} - \partial_{x}\zeta^{t, \sigma}/r]^{\infty}_{r_0}  = \zeta^{x, \sigma}(t,r=\infty,x) - X^{\sigma}(t,x).
\label{hydrodeflong}
\end{eqnarray}
These again can be interpreted as relative diffeomorphisms between the asymptotic boundaries and the black hole horizon, that preserve the appropriate gauge on each surface. After inverting \eqref{eq:longdecomp}, they can equivalently be expressed as the Wilson line-like fields
\begin{eqnarray}
\xi_t^\sigma(t,x)&=&-\int^\infty_{r_0}\frac{\delta g_{rr}^\sigma(t,r,x)}{2} dr,\nonumber \\
\xi_x^\sigma(t,x)&=&\int^\infty_{r_0}\left(\frac{\delta g_{xr}^\sigma(t,r,x)}{r^2}-\frac{\partial_x\delta g_{rr}^\sigma(t,r,x)}{2r}\right)dr,
\end{eqnarray}
which vanish identically if one works in radial gauge throughout the bulk.
It is again convenient to arrange the fields into symmetric and antisymmetric combinations as in~\eqref{hydrora} and~\eqref{metricra}. It is particularly convenient to define the following quantities
\begin{equation}
\begin{aligned}
B_{tt}^{\sigma} &\,= \delta g_{tt}^{ \sigma (s)} - 2 \partial_{t} \xi_{t}^{\sigma},\quad\quad\quad\quad\quad\quad
B_{tx}^{\sigma} = \delta g_{tx}^{\sigma (s)} -  \partial_{x} \xi_{t}^{\sigma} - \partial_{t} \xi_{x}^{\sigma}, \\
B_{p}^{\sigma} &\,= \delta g_{p}^{\sigma (s)} - 2 \partial_{x} \xi_{x}^{\sigma},\quad\quad\quad\quad\quad\quad
B_{m}^{\sigma} = \delta g_{m}^{\sigma (s)} - 2 \partial_{x} \xi_{x}^{\sigma}, 
\end{aligned}
\end{equation}
where $\delta g_{p}^{\sigma (s)} = \delta g_{xx}^{\sigma (s)} + \delta g_{yy}^{\sigma (s)}$ and $\delta g_{m}^{\sigma (s)} = \delta g_{xx}^{\sigma (s)} - \delta g_{yy}^{\sigma (s)}$.\footnote{We similarly define $B_{tt}^\pm$, $B_{tx}^\pm$, $B_{p}^\pm$ and $B_{m}^\pm$ as the symmetric and antisymmetric linear combinations of these, following the conventions in \eqref{hydrora} and \eqref{metricra}.}

\subsubsection*{Equations of motion}

\paragraph{}Our next step in computing the hydrodynamic action is to evaluate the gravitational action for metric perturbations when partially on-shell i.e.~after imposing a subset of the Einstein equations.

\paragraph{}On each branch there are 7 Einstein equations governing the longitudinal metric perturbations. These can be arranged into 4 dynamical equations that are second order in radial derivatives and 3 constraint equations that are first order. The second order equations can be taken to be the equations $\delta E^{\sigma}_{rr},  \delta E^{\sigma}_{rx},  \delta E^{\sigma}_{xx},  \delta E^{\sigma}_{yy} $. Explicitly these are
\begin{eqnarray}
\delta E^{\sigma}_{rr} &=& - \frac{2 \partial_{r} h^{\sigma}_p + r \partial_{r}^2 h^{\sigma}_p}{2 r},   \nonumber  \\ 
\delta E^{\sigma}_{rx} &=& 2 r \partial_{r} h^{\sigma}_{tx} + \frac{1}{4} \partial_{r} \partial_{x} (h^{\sigma}_{m} - h^{\sigma}_{p}) + \frac{r^2}{2} \partial_{r}^2 h^{\sigma}_{tx}, \nonumber \\
\delta E^{\sigma}_{xx} - \delta E^{\sigma}_{yy} &=& \frac{- 4 r^3 + r_0^3}{2} \partial_{r} h^{\sigma}_m + \partial_{r} \left(r^2\partial_{x} h^{\sigma}_{tx}\right) - \frac{r^4f(r)}{2} \partial_{r}^2 h^{\sigma}_m - r \partial_{t} h^{\sigma}_m - r^2 \partial_{t} \partial_{r} h^{\sigma}_m, \nonumber \\
\delta E^{\sigma}_{xx} + \delta E^{\sigma}_{yy} &=& 6 r^2 h^{\sigma}_{tt} - 2 r \partial_{x} h^{\sigma}_{tx} + 2 r^3 \partial_{r} h^{\sigma}_p  - \frac{r_0^3}{2} \partial_{r} h^{\sigma}_p + 6 r^3 \partial_{r} h^{\sigma}_{tt} - r^2 \partial_{r} \partial_{x} h^{\sigma}_{tx}   \nonumber \\
&+& \frac{r^4f(r)}{2} \partial_r^2 h^{\sigma}_p+  r^4f(r) \partial_{r}^2 h^{\sigma}_{tt} + r \partial_{t} h^{\sigma}_p + r^2 \partial_{t} \partial_{r} h^{\sigma}_p.
\label{dynamicalads4longitudinal}
\end{eqnarray}
It is convenient to consider the following three combinations of Einstein equations as the constraint equations
\begin{eqnarray}
\mathrm{trace}^{\sigma} &=& 4 r^2 D(r) \delta E^{\sigma}_{rr} + 4 r^2 \delta E^{\sigma}_{tr},\\
\mathrm{momentum}_x^{\sigma} &=&  2 r^2 D(r) \delta E^{\sigma}_{rx} + 2 r^2 \delta E_{tx}^\sigma,  \nonumber \\
\mathrm{energy}^{\sigma} &=&  - 2 r^2 D(r) \delta E^{\sigma}_{tr} - 2 r^2 \delta E^{\sigma}_{tt} - 2 r^3 D(r) \partial_{t} \delta E^{\sigma}_{rr} - 2 r^3 \partial_{t} \delta E^{\sigma}_{tr} \nonumber\\
&+& 2 r D(r) \partial_{x} \delta E^{\sigma}_{rx} + 2 r \partial_{x} \delta E_{tx}^{\sigma}, \nonumber 
\label{constraintsads4longitudinal}
\end{eqnarray}
whose names we will explain shortly. These combinations have been chosen because they are radially constant on solutions to the dynamical equations, due to the following identities
\begin{eqnarray}
&\partial_r&\left(\text{trace}^\sigma\right)+2(2r^3+r_0^3)\delta E_{rr}^\sigma-\frac{4}{r}\left(\delta E_{xx}^\sigma +\delta E_{yy}^\sigma\right)+4\partial_x\delta E_{rx}^\sigma+4r^2\partial_t\delta E_{rr}^\sigma=0,\nonumber\\
&\partial_r&\left(\text{momentum}_x^\sigma\right)+2\partial_x\delta E_{xx}^\sigma+2r^2\partial_t\delta E_{rx}^\sigma=0,\\
&\partial_r&\left(\text{energy}^\sigma\right)+2r^2f\partial_x\delta E_{rx}^\sigma-3r r_0^3\partial_t\delta E_{rr}^\sigma+2\partial_t\left(\delta E_{xx}^\sigma+\delta E_{yy}^\sigma\right)+\frac{2}{r}\partial_x^2\delta E_{xx}^\sigma-2r^3\partial_t^2\delta E_{rr}^\sigma=0.\nonumber
\end{eqnarray}

\paragraph{}The first constraint should be interpreted as requiring the CFT energy-momentum tensor to be traceless, while the latter two require covariant conservation of the energy and the $x$-component of momentum. Explicitly
\begin{eqnarray}    \text{trace}^\sigma&=&4\partial_r(r^3fh_{tt}^\sigma)+(4r^3-r_0^3)\partial_rh_p^\sigma+2\partial_r\partial_t(r^2h_p^\sigma)-\frac{2}{r^2}\partial_r\partial_x(r^4h_{tx}^\sigma)+\partial_x^2(h_p^\sigma-h_m^\sigma),\nonumber\\
\text{momentum}_x^\sigma&=&-\partial_t\left(r^4\partial_r h_{tx}^\sigma+\frac{r^2}{2}\partial_x(h_p^\sigma-h_m^\sigma)\right)-\partial_x\left(r^2\partial_r(r^2fh_{tt}^\sigma)+\frac{r^4f}{2}\partial_r(h_p^\sigma-h_m^\sigma)\right),\nonumber\\
\text{energy}^\sigma&=&-\partial_t\biggl(2r^2\partial_r(r^2fh_{tt}^\sigma)+\frac{(2r^3+r_0^3)^2}{2r^2}\partial_r\left(\frac{r^3h_p^\sigma}{2r^3+r_0^3}\right)+r^2\partial_r\partial_t(rh_p^\sigma)+r\partial^2_x(h_p^\sigma-h_m^\sigma)\biggr)\nonumber\\
&-&\partial_x\biggl(r^4f^2\partial_r\left(\frac{h_{tx}^\sigma}{f}\right)+\partial_r\partial_x(r^3fh_{tt}^\sigma)+\frac{r^3f}{2}\partial_r\partial_x(h_p^\sigma-h_m^\sigma)-2r^2\partial_t h_{tx}^\sigma\biggr).\nonumber
\end{eqnarray}

\subsubsection*{Partially on-shell solution}

\paragraph{} To obtain the hydrodynamic action we will go partially on-shell. This means that we impose the second order equations of motion (as we did for the transverse perturbations), as well as the constraint equation $\mathrm{trace}^{\sigma} = 0$. The reason for imposing this constraint is intuitively simple: it ensures that the stress tensor of our hydrodynamic action is traceless even off-shell. We do not impose the constraints corresponding to energy and momentum conservation: these will arise as the equations of motion of the hydrodynamic action that we ultimately derive.

\paragraph{}The simplest second order equation is $\delta E_{rr} = 0$, which can be solved exactly for $h_p^\sigma(t,r,x)$. Since this field couples to others which cannot be obtained exactly, it will be convenient to expand it in an expansion in boundary derivatives $\partial_{t}, \partial_{x}$:
\begin{equation}
h_{p}^{\sigma}(t,r,x) = \frac{\alpha^{\sigma}_1(t,x)+\alpha^\sigma_{11}(t,x)}{r} + \alpha^{\sigma}_2(t,x)+\alpha^{\sigma}_{21}(t,x).
\end{equation}
As for transverse perturbations, the additional subscript $1$ on the $\alpha^\sigma$s denotes a term first order in derivatives.\footnote{Again, these first order quantities could be absorbed by appropriate redefinitions of the zeroeth order $\alpha_i^\sigma$ but we will keep them explicit.} The remaining dynamical equations in~\eqref{dynamicalads4longitudinal} can be solved for $h^{\sigma}_{tx}, h^{\sigma}_{m}, h^{\sigma}_{tt}$ in a derivative expansion. Expanding~\eqref{dynamicalads4longitudinal} to zeroth order in boundary derivatives and solving for the zeroth order solutions $h^{\sigma}_{tx, 0}, h^{\sigma}_{m, 0}, h^{\sigma}_{tt, 0}$ gives 
\begin{eqnarray}
h_{tx,0}^{\sigma}(t ,r, x) &=&  \alpha^{\sigma}_3(t,x) + \frac{\alpha^{\sigma}_4(t,x)}{r^3}, \nonumber \\
h_{m,0}^{\sigma}(t,r,x)&=& \alpha^{\sigma}_5(t,x) + \alpha^{\sigma}_6(t,x) \mathrm{log}f(r), \nonumber \\
h_{tt,0}^{\sigma}(t,r,x) &=& \frac{(2 r^3 + r_0^3)}{4r^4f(r)} \alpha^{\sigma}_1(t,x) + \frac{\alpha^{\sigma}_7(t,x)+r \alpha^{\sigma}_8(t,x)}{r^3f(r)}.
\label{longitudinalsolzero} 
\end{eqnarray}
The corrections at first order are firstly
\begin{equation}
h^{\sigma}_{tx,1} = \alpha^{\sigma}_{31} + \frac{\alpha^{\sigma}_{41}}{r^3} + \frac{ \partial_{x} \alpha^{\sigma}_1 }{4 r^2}- \frac{ \partial_{x} \alpha^{\sigma}_6 }{2 r} - \frac{1}{2}f(r) H(r) \partial_{x} \alpha^{\sigma}_6,
\end{equation}
secondly
\begin{eqnarray}
h^{\sigma}_{m,1} &=& \alpha^{\sigma}_{51} + \alpha^{\sigma}_{61} \mathrm{log}f(r) + \frac{2 \partial_x \alpha_4^\sigma}{r r_0^3}  + 2 H(r) \bigg( \partial_{x} \alpha^{\sigma}_3 + \frac{\partial_{x} \alpha^{\sigma}_4}{r_0^3} - \frac{\partial_{t} \alpha^{\sigma}_5}{2} - \mathrm{log}f(r) \frac{\partial_{t} \alpha^{\sigma}_6}{2} \bigg), \nonumber  
\end{eqnarray}
and finally
\begin{equation}
h^{\sigma}_{tt,1} = \frac{(2 r^3 + r_0^3)}{4 r^4f(r)} \alpha^{\sigma}_{11}  + \frac{\alpha^{\sigma}_{71}+r \alpha^{\sigma}_{81}}{r^3f(r)}  + \frac{2 r^3 \partial_{x} \alpha^{\sigma}_3 -  \partial_{x} \alpha^{\sigma}_4 -  r^3 \partial_{t} \alpha^{\sigma}_2  }{2 r^4f(r)}.
\label{hvvfirstorder}
\end{equation}
Using these partially on-shell solutions we can compute the constraint equations 
\begin{eqnarray}
\label{eq:constrainteqslongalpha}
\mathrm{trace}^{\sigma} &=& 4 (\alpha^{\sigma}_8 + \alpha^{\sigma}_{81}) + 2 \partial_{t} \alpha^{\sigma}_1, \\
\mathrm{momentum}_x^{\sigma} &=& \frac{3}{2} r_0^3 \partial_{x} (\alpha^{\sigma}_6+\alpha_{61}^\sigma) + \partial_{x} (\alpha^{\sigma}_7+\alpha_{71}^\sigma) + 3 \partial_{t} (\alpha^{\sigma}_4+\alpha^\sigma_{41}), \nonumber \\
\mathrm{energy}^{\sigma} &=& 3 r_0^3 \partial_{x} (\alpha^{\sigma}_3+\alpha_{31}^\sigma) + 3 \partial_{x} (\alpha^{\sigma}_4+\alpha^\sigma_{41}) - \frac{3}{2} r_0^3 \partial_{t} (\alpha^{\sigma}_2+\alpha_{21}^\sigma) + 2 \partial_{t} (\alpha^{\sigma}_7+\alpha^\sigma_{71}) - \partial_x^2 \alpha^{\sigma}_8. \nonumber 
\end{eqnarray}
As mentioned above, when going partially on-shell we will also impose the equations $\mathrm{trace}^{\sigma}  =0$. This corresponds to setting $\alpha^{\sigma}_8 = 0$ in the zeroth order solution~\eqref{longitudinalsolzero} and setting $\alpha^{\sigma}_{81} = - \partial_{t} \alpha^{\sigma}_{1}/2$ in the first order solution~\eqref{hvvfirstorder}. Our partially on-shell solution on each branch therefore depends on 7 integration constants (arbitrary functions of $(t,x)$) at each order in derivatives. We will now fix these functions by imposing boundary conditions at the asymptotic boundary and horizon of each branch.

\subsubsection*{Asymptotic and horizon boundary conditions}
\paragraph{} We will first impose boundary conditions in the UV, in exactly the same way as we did in the transverse case. We expand the bulk diffeomorphism fields for large $r$ as\footnote{With these asymptotics the hydrodynamic fields are $\xi_t^{\sigma} = \zeta_t^{\sigma (2)} + T^{\sigma}$ and $\xi_x^\sigma = \zeta_x^{\sigma (2)} - X^{\sigma}$.}  
\begin{equation}
\label{eq:asymptoticzetalong}
\zeta_{M}^{\sigma}(t, r, x) = \zeta_{M}^{\sigma (2)}(t,x) r^{2} +  \zeta_{M}^{\sigma (1)}(t,x) r + \dots.  
\end{equation}
We take $t$ and $x$ derivatives of $\zeta^{t,\sigma(2)}$ and $\zeta^{x,\sigma(2)}$ to be zeroeth order in the derivative expansion. We restrict the terms in this expansion such that, upon switching to Schwarzschild coordinates, the partially on-shell solutions are in Fefferman-Graham gauge. This allows us to fix functions appearing in the asymptotic bulk metric in terms of the field theory metric. More details of this calculation can be found in Appendix \ref{app:uvbcs}.

\paragraph{} At zeroeth order in derivatives this fixes 3 integration constants on each branch in terms of the field theory metric and the asymptotic diffeomorphisms
\begin{equation}
    \begin{aligned}
    \label{eq:longUVBCs1}
\alpha_2^\sigma&\,=\delta g_p^{\sigma (s)} +2\delta g_{tt}^{\sigma (s)} -2\partial_x\zeta_x^{\sigma (2)}-4\partial_t\zeta_t^{\sigma (2)},\\
\alpha_3^\sigma&\,=\delta g_{tx}^{\sigma (s)}-\partial_x\zeta_t^{\sigma (2)}-\partial_t\zeta_x^{\sigma (2)},\\
\alpha_5^\sigma&\,=\delta g_m^{\sigma (s)}-2\partial_x\zeta_x^{\sigma (2)}.
    \end{aligned}
\end{equation}
At first order in derivatives
\begin{equation}
    \begin{aligned}
        \label{eq:longUVBCs2}
        \alpha_{21}^{\sigma}=0,\quad\quad\quad
        \alpha_{31}^\sigma=\frac{\sqrt{3}\pi}{12r_0}\partial_x\alpha_6^\sigma,\quad\quad\quad
        \alpha_{51}^\sigma=-\frac{\pi}{\sqrt{3}r_0}\left(\partial_x\alpha_3^\sigma+\frac{\partial_x\alpha_4^\sigma}{r_0^3}-\frac{\partial_t\alpha_5^\sigma}{2}\right).
    \end{aligned}
\end{equation}
Note that these asymptotic boundary conditions do not fix any of the parameters in the solution for $h^{\sigma}_{tt}$. 

\paragraph{}There remain 4 unfixed integration constants on each branch of the contour, at each order in derivatives. We will fix these by imposing boundary conditions at the horizon. As for transverse perturbations, we will impose that some metric perturbations vanish near the horizon, while others should be continuous after an analytic continuation around the horizon.

\paragraph{}The first horizon boundary conditions are that
\begin{equation}
\delta g^{\sigma}_{tx}(r_0) = 0,  \hspace{2.0cm} \delta g^{\sigma}_{tt}(r_0) = 0, \hspace{2.0cm} 2 r_0^2 \partial_{r} \delta g^{\sigma}_{tt}(r_0) = \partial_{t} \delta g^{\sigma}_{p}(r_0),
\label{irbcslongitudinal} 
\end{equation}
on each branch. The first two of these are natural analogues of the boundary condition $\delta g_{ty}^\sigma(r_0)=0$ we imposed on transverse perturbations, which was inspired by the gauge field case in \cite{CGL}. The third is more unusual. This condition ensures that the variational principle of the quadratic Einstein-Hilbert action on the horizon is satisfied (in radial gauge): see Appendix \ref{app:varprinciple}.\footnote{These are mixed, rather than Dirichlet, boundary conditions as we are not including a Gibbons-Hawking term at the horizon.} Indeed, we will see shortly that after imposing these non-trivial boundary conditions we will end up with a hydrodynamic action whose equations of motion are exactly the Einstein equations that we have not yet imposed.\footnote{See Section~\ref{sec:altBCs} for a discussion of a different set of horizon boundary conditions.}

\paragraph{}The conditions \eqref{irbcslongitudinal} are used to fix three of the remaining integration constants on each branch, at each order in the derivative expansion. In imposing these boundary conditions we take $R^{\sigma}$ to be zeroth order in derivatives, and $T^{\sigma}$ and $X^{\sigma}$ to be order minus one (recall these residual diffeomorphisms were introduced in~\eqref{IRdiffeoslongitudinal}). At zeroeth order we then obtain 
\begin{equation}
 \alpha_{1}^{\sigma} = - 4 R^{\sigma},  \quad\quad\quad\quad
 \alpha_4^{\sigma} = - r_0^3\left( B_{tx}^\sigma + \partial_{x} T^{\sigma}\right), \quad\quad\quad\quad
\alpha_{7}^{\sigma} = 3 r_0^3 \partial_{t}T^{\sigma},
\end{equation}
where $T^\sigma$ and $R^\sigma$ are the residual horizon diffeomorphisms. And at first order
\begin{equation}
\begin{aligned}
\alpha^{\sigma}_{11}&\,= - 2 \partial_{x} B_{tx}^{\sigma} + \frac{4}{3} \partial_{t} B_{tt}^{\sigma} + \frac{2}{3} \partial_{t} B_{p}^{\sigma} - 2 \partial_{x}^2 T^{\sigma},\quad\quad \alpha^\sigma_{41}=-\frac{r_0^2}{12}(\sqrt{3}\pi-6)\partial_x\alpha_6^\sigma,\\ \alpha^{\sigma}_{71}&\,=0.
\end{aligned}
\end{equation}

\paragraph{}The second set of horizon boundary conditions that we impose concern the field $\delta g_m^\sigma=\delta g_{xx}^\sigma-\delta g_{yy}^\sigma$. Like $\delta g_{xy}$ in the transverse sector, on each branch this diverges near the horizon as $\sim a(t,x)\log^2(r-r_0)+b(t,x)\log(r-r_0)+c(t,x)+\ldots$.\footnote{As in the transverse case, this can be understood as the derivative expansion of the outgoing solution on the future horizon.} As for $\delta g_{xy}$ we impose that this field should be continuous after analytically continuing it around the horizon. In practice this means we take $\log(r-r_0)\rightarrow\log(r-r_0)-2\pi i$ in the near-horizon expansion of $\delta g_m^1$ and then match the constant and log terms to the near-horizon expansion of $\delta g_m^2$. This automatically fixes the coefficients of the $\log^2(r-r_0)$ terms on each branch to be equal. At the first two orders in the derivative expansion this gives the following two explicit conditions
\begin{equation}
\alpha_{6}^{\sigma} =-\frac{i}{2 \pi} B_{m}^{-},\quad\quad\quad\quad\quad\quad \alpha^{\sigma}_{61} = -\frac{\sqrt{3} i}{12 r_0} \partial_{t} B_{m}^{-} +\frac{i}{\pi r_0} \partial_{x} B_{tx}^{-} + \frac{1}{3 r_0} \partial_{t} B_{m}^{\bar{\sigma}},
\end{equation}
where $\bar{\sigma} = 1$ if $\sigma=2$ and $\bar{\sigma} = 2$ is $\sigma = 1$. 

\paragraph{}We have now fixed all integration constants in the partially on-shell solution in terms of sources and residual gauge transformations. These results are summarised in Appendix \ref{app:ads4nearhorizon}. The final step is to evaluate the gravitational action on these solutions, which will again reduce to a (2+1)-dimensional action dependent only on the hydrodynamic fields $\xi_t^\sigma$, $\xi_x^\sigma$ and the boundary metric $\delta g_{\mu\nu}^{\sigma (s)}$.

\subsection*{Linear hydrodynamical action}
\paragraph{} For the longitudinal perturbations, the partially on-shell action has a piece linear in the perturbation amplitude. The contribution from $S_{\mathrm{EH}}$ can be reduced to a boundary term. Combining this with $S_{\mathrm{GH}}$ and evaluating at the asymptotic boundaries yields
\begin{equation}
S_{\mathrm{hydro}}^{(1)} =  \int d^3 x  \left( r_0^3 \delta g_{tt}^{- (s)} + \frac{r_0^3}{2} \delta g_{p}^{-(s)} \right).
\label{linearactionads4}
\end{equation}
As we discussed for the transverse action, we must be careful as it is not a priori obvious if the partially on-shell action is continuous around the horizon on the CGL contour. However, an explicit calculation shows that the boundary term arising from $S_{\mathrm{EH}}$ vanishes as $r\rightarrow r_0$ on both branches (up to total derivative terms in the $t$ and $x$ coordinates). In other words, after a trivial analytic continuation there is no contribution from the horizon and so \eqref{linearactionads4} is the full answer at linear order in the perturbation amplitude. From the definition of the stress tensor~\eqref{tmunudef}, we can read off the energy density $\langle T^{tt,\sigma} \rangle = 2 r_0^3$ and pressure $\langle T^{xx,\sigma} \rangle = \langle T^{yy,\sigma} \rangle = r_0^3$ of the equilibrium state. 

\subsection*{Quadratic hydrodynamical action}
\paragraph{} To obtain the action at quadratic order in amplitude, it is helpful to integrate the Einstein-Hilbert action by parts to obtain (on each branch)
\begin{equation}
S_{\mathrm{EH}} = S_{\mathrm{bulk}} + S_{\mathrm{boundary}},
\end{equation}
where the bulk action is
\begin{eqnarray}
\label{eq:Sbulklongit}
S_{\mathrm{bulk}} &=& \int d^4x \frac{r^2}{2}  \delta E_{M N} \delta g^{M N}  \nonumber \\
 &=&\frac{1}{2}  \int d^{4}x \bigg( \partial_{r} \left(( -\zeta^{x} + \partial_{x} \zeta^t/r) \mathrm{momentum}_x + \zeta^{t} \mathrm{energy} - \frac{1}{2}(\zeta^r + r \partial_{t} \zeta^t)\mathrm{trace} \right) \nonumber \\
&+& r^2 D(r) h_{tt} \delta E_{rr} - 2r^2 h_{tx} \delta E_{rx} - \frac{h_{m}}{2} (\delta E_{xx} - \delta E_{yy})  - \frac{h_p}{2}(\delta E_{xx} + \delta E_{yy}) \bigg),
\end{eqnarray}
and we will return to the boundary term soon. Since our partially on-shell solution satisfies all of the dynamical equations, this piece of the action manifestly also reduces to a boundary term. 

\paragraph{}As for the transverse perturbations, we will split the integral over the entire CGL contour into two integrals -- one over the upper branch, and one over the lower branch plus the circle around the horizon -- allowing us to carefully evaluate any contributions from discontinuities in the action around the horizon.

\paragraph{}We begin with the contributions from the explicit boundary term $S_{\mathrm{boundary}}$ which in radial gauge is
\begin{eqnarray}
\label{longitudinalboundarytermradialgauge}
S_{\mathrm{boundary}} &=& \int d^{3}x \frac{1}{8 r^2} \bigg(- 4rf(r) \delta g_{m}^2 - 2rf(r) \delta g_{p}^2 - 8 r \delta g_{tx}^2 + 6 r^2 \delta g_{tt} \partial_{r} \delta g_{p} \nonumber \\
&-& 2 \delta g_{tx}(\partial_{x} \delta g_{m} - \partial_{x} \delta g_{p} + 6 r^2 \partial_{r} \delta g_{tx}) + \delta g_{m}(3r^2f(r) \partial_{r} \delta g_{m} + \partial_{t} \delta g_{m}) \nonumber \\
&+& \delta g_{p}(- 4 r \delta g_{tt} + r^2f(r) \partial_{r} \delta g_{p} + 2 r^2 \partial_{r} \delta g_{tt} - \partial_{t} \delta g_{p}) \bigg).
\end{eqnarray}
The boundary terms at asymptotic infinity on each branch consist of this boundary term, together with $S_{\mathrm{GH}}$. Combining these gives
\begin{eqnarray}
S_{\mathrm{UV}}^{\sigma} &=& \int d^{3} x \Biggl(\frac{3 i r_0^3}{16 \pi}  \delta g^{\sigma (s)}_{m}  B_{m}^{-} - \frac{3 r_0^3}{2} \delta g^{\sigma (s)}_{tx} \partial_t \xi^{\sigma}_x - \frac{3 r_0^3}{4}  (\delta g^{\sigma (s)}_{p} + 2 \delta g^{\sigma (s)}_{tt}) \partial_t \xi^{\sigma}_t  \nonumber \\
&-&\frac{r_0^3}{8} (\delta g^{\sigma (s)}_{m})^2 + \frac{r_0^3}{2} \delta g^{\sigma (s)}_{p} \delta g^{\sigma (s)}_{tt} + r_0^3 (\delta g^{\sigma (s)}_{tt})^2 + \frac{r_0^3}{2} (\delta g^{\sigma (s)}_{tx})^2\\
&+&\frac{3 i r_0^2}{8 \pi} \delta g_{tx}^{\sigma (s)} \partial_{x} B_{m}^{-} - \frac{3 i r_0^2}{8 \pi}\delta g_{m}^{\sigma (s)}  \partial_{x} B_{tx}^{-} - \frac{r_0^2}{8}\delta g_{m}^{\sigma (s)} \partial_{t} B_{m}^{\bar{\sigma}}  \Biggr)\nonumber,
\label{sboundary0}
\end{eqnarray}
at each asymptotic boundary.

\paragraph{}At the horizon end of each integral, there is no Gibbons-Hawking term and so we simply have the contribution from \eqref{longitudinalboundarytermradialgauge}. Evaluating this for the partially on-shell solution on each branch as $r\rightarrow r_0$ gives a constant plus a logarithmically divergent term. As when computing the transverse action, we analytically continue the contribution from the first branch around the horizon by taking $\log(r-r_0)\rightarrow \log(r-r_0)-2\pi i$. Upon doing this, the sum of the two horizon contributions from $S_{\mathrm{boundary}}$ cancels. This is a consequence of the horizon boundary conditions that we have imposed.

\paragraph{}The final piece of the action is $S_{\mathrm{bulk}}$ in \eqref{eq:Sbulklongit}. Since the energy and momentum$_x$ constraints are radially conserved for our partially on-shell solutions (and the trace constraint is identically satisfied), it naturally involves the relative diffeomorphisms $\xi_{t}^\sigma$ and $\xi^\sigma_{x}$ between the asymptotic boundaries and the horizon. Since the analytic continuation of the diffeomorphism field around the horizon is trivial, the total contribution from both branches is
\begin{equation}
S_{\mathrm{bulk}} = -\frac{1}{2} \int d^{3}x\left(\xi_x^{1} \mathrm{momentum}^{1}_x + \xi_{t}^{1} \mathrm{energy}^{1} - \xi_x^{2} \mathrm{momentum}^{2}_x - \xi_{t}^{2} \mathrm{energy}^{2}  \right),
\label{sbulk0}
\end{equation}
where the remaining constraints are
\begin{eqnarray}
\label{eq:longconstraintsexplicit}
\mathrm{momentum}^\sigma_{x} &=& -\frac{3 i r_0^3}{4 \pi} \partial_{x}B_{m}^{-} - 3 r_0^3 \partial_{t} B_{tx}^{\sigma}+\frac{3 i r_0^2}{2 \pi} \partial_{x}^2 B_{tx}^{-} + \frac{r_0^2}{2} \partial_{t} \partial_{x} B_{m}^{\bar{\sigma}} - \frac{3 i r_0^2}{4 \pi} \partial_{t} \partial_{x} B_{m}^{-}, \nonumber \\
\mathrm{energy}^{\sigma} &=& - \frac{3 r_0^3}{2} \partial_{t} B_{p}^{\sigma} - 3 r_0^3 \partial_{t} B_{tt}^{\sigma}- \frac{3 i r_0^2}{4 \pi} \partial^2_{x} B_{m}^{-}.
\end{eqnarray}

\paragraph{}The total quadratic action is $S_{\mathrm{bulk}} + S_{\mathrm{UV}}^{1}  - S_{\mathrm{UV}}^{2}$. At zeroeth order in derivatives it will be convenient to organise this as
\begin{equation}
S_{\mathrm{hydro, 0}}[\xi_{\mu}^{\sigma}, \delta g_{\mu}^{\sigma (s)}]= S_{\mathrm{diss}}[ \xi_{\mu}^-, \delta g_{\mu \nu}^{- (s)}]+ S_{\mathrm{non-diss}}[\xi_{\mu}^{1}, \delta g_{\mu \nu}^{1 (s)}] - S_{\mathrm{non-diss}}[\xi_{\mu}^{2}, \delta g_{\mu \nu}^{2 (s)}],
\label{longitudinalactionads4}
\end{equation}
where $S_{\mathrm{diss}}$ is a dissipative piece of the action that couples fields on the two contours and $S_{\mathrm{non-diss}}$ are non-dissipative terms that involve only fields on a given contour. We have explicitly
\begin{eqnarray}
S_{\mathrm{diss}}[ \xi_{\mu}^{-}, \delta g_{\mu \nu}^{- (s)}]&=r_0^3& \int d^3x \frac{3 i}{16 \pi} (B_m^{-})^2,\nonumber \\
S_{\mathrm{non-diss}}[\xi_{\mu}^{\sigma}, \delta g_{\mu \nu}^{\sigma (s)} ] &=r_0^3&  \int d^{3}x \bigg(  3 (\partial_{t} \xi_t^{\sigma} )^2 + \frac{3}{2} (\partial_{t} \xi_x^{\sigma})^2 + 3  (\partial_{t} \xi_t^{\sigma}) (\partial_{x} \xi_x^{\sigma}) \nonumber \\  
&-& 3  \partial_{t} \xi_x^{\sigma}  \delta g_{tx}^{\sigma (s)} - 3 \partial_{t} \xi_t^{\sigma} (\delta g^{\sigma (s)}_{tt} + \delta g^{\sigma (s)}_{p}/2) \nonumber \\
&-&  \frac{1}{8} (\delta g_{m}^{\sigma (s)})^2 + \frac{1}{2} \delta g^{\sigma (s)}_{p} \delta g^{\sigma (s)}_{tt} + (\delta g^{\sigma (s)}_{tt})^2 + \frac{1}{2} (\delta g^{\sigma (s)}_{tx})^2 \bigg).
\end{eqnarray}
At first order in derivatives
\begin{equation}
S_{\mathrm{hydro, 1}}[\xi_{\mu}^{\sigma}, \delta g_{\mu}^{\sigma (s)}] =r_0^2 \int d^{3} x \left(- \frac{1}{4} B_m^- \partial_{t} B_m^{+} + \frac{3 i}{4 \pi} B_{tx}^{-} \partial_{x} B_{m}^{-} \right).
\label{longitudinalactionfirstorder}
\end{equation}
This completes our derivation of the Schwinger-Keldysh action. In the next Section we will analyse its properties, and perform several consistency checks.

\section{The Schwinger-Keldysh action} 
\label{sec:tests}

\paragraph{} We have now constructed the full Schwinger-Keldysh action, to quadratic order in perturbations and to first order in the hydrodynamic derivative expansion.\footnote{The hydrodynamic fields $\xi_\mu^\sigma$ are order $-1$ in the derivative expansion.} The action is given by the sum of the linear term~\eqref{linearactionads4}, the quadratic action for transverse perturbations~\eqref{transverseactionads4} and the quadratic action for longitudinal perturbations~\eqref{longitudinalactionads4}. Recall that we have taken the spatial dependence of all fields to lie in the $x$ direction (without loss of generality).

\paragraph{} The structure of the hydrodynamic action appears, at this stage, to be rather complicated. In particular we note that the quadratic action for both transverse~\eqref{transverseactionads4} and longitudinal perturbations~\eqref{longitudinalactionads4} cannot be expressed solely in the terms of the fields $B_{\mu \nu}^{\sigma}$. This is related to the fact that diffeomorphisms act non-linearly, and non-trivially mix terms in our linear and quadratic actions. By taking into account the non-linear nature of diffeomorphisms, we will now express our hydrodynamical action in a manifestly diffeomorphism invariant manner.

\paragraph{} In particular, we define the fields ${\mathcal B}^{\sigma}_{\mu \nu}$ to be the perturbations of the induced metric under a diffeomorphism. Suppressing, momentarily, the branch index we define 
\begin{equation}\label{eq:def_Bmn}
{\mathcal B}_{\mu \nu}(x) = \frac{\partial y^{\alpha}}{\partial x^{\mu} } \frac{\partial y^{\beta}}{\partial x^{\nu}}  g^{(s)}_{\alpha \beta}(y(x)) - \eta_{\mu \nu}, 
\end{equation}
where $y^{\mu}= x^{\mu} - \xi^{\mu}$, and $g_{\mu \nu}^{(s)} = \eta_{\mu \nu} + \delta g_{\mu \nu}^{(s)}$. To quadratic order in the fields $\delta g_{\mu \nu}^{(s)}, \xi^{\mu}$ one has explicitly
%
%
%
%
%
\begin{equation}
{\mathcal B}_{\mu \nu}(x) = \delta g_{\mu \nu}^{(s)} - \partial_{\mu} \xi_{\nu} - \partial_{\nu} \xi_{\mu} - \partial_{\rho} \delta g^{(s)}_{\mu \nu} \xi^{\rho} - \delta g^{(s)}_{\mu \rho} \partial_{\nu} \xi^{\rho} - \delta g^{(s)}_{\rho \nu} \partial_{\mu} \xi^{\rho} + \partial_{\mu} \xi^{\alpha} \partial_{\nu} \xi_{\alpha},
\label{bmunu}
\end{equation}
where all indices on $\xi^\mu$ are raised and lowered using the Minkowski metric $\eta_{\mu\nu}$. We note that the objects ${\mathcal B}_{\mu \nu}(x)$ in \eqref{eq:def_Bmn} are, by construction,  invariant under the diffeomorphisms
\begin{equation}
\delta \xi^{\mu}(x) = \epsilon^{\mu}(y(x)), \qquad \delta g^{(s)}_{\mu \nu}(y(x)) \to \delta g^{(s)}_{\mu \nu}(y(x)) + \nabla_{\mu}( g^{(s)}_{\nu \rho} \epsilon^{\rho}) + \nabla_{\nu} (g^{(s)}_{\mu \rho}\epsilon^{\rho}),
\label{diffeosymmetry}
\end{equation} 
where the covariant derivatives are defined with respect to $y^{\mu}$. 
Note that, for our Schwinger-Keldysh action, we define two fields ${\mathcal B}^{\sigma}_{\mu \nu}(x)$ which are given by~\eqref{bmunu} with the appropriate sources $\delta g_{\mu \nu}^{\sigma (s)}$ and diffeomorphism fields $\xi_{\mu}^{\sigma}$ for each branch.\footnote{We also define $\mathcal{B}_p^\sigma=\mathcal{B}_{xx}^\sigma+\mathcal{B}^\sigma_{yy}$ and $\mathcal{B}_m^\sigma=\mathcal{B}_{xx}^\sigma-\mathcal{B}_{yy}^\sigma$.} As previously mentioned these fields coincide with the objects $B^{\sigma}_{\mu \nu}$ previously introduced in Sections~\ref{sec:transverse} and~\ref{sec:longitudinal} to linear order in the amplitude of perturbations.

\paragraph{} With these definitions, the combined linear and quadratic action can be expressed in a manifestly diffeomorphism invariant way. In particular, the full hydrodynamic action (linear plus quadratic terms) can be written at zeroth order in derivatives as
\begin{equation}
\begin{aligned}
S_{\mathrm{hydro},0}[{\mathcal B}_{\mu \nu}^{\sigma}] =r_0^3 \int d^{3} x \bigg(& {\mathcal B}_{tt}^{-} + \frac{1}{2} {\mathcal B}_{p}^{-} + {\mathcal B}_{tx}^{+} {\mathcal B}_{tx}^{-} - \frac{1}{4} {\mathcal B}_{m}^{+} {\mathcal B}_{m}^{-}    + 2  {\mathcal B}_{tt}^{+} {\mathcal B}_{tt}^{-} + \frac{1}{2} {\mathcal B}_{tt}^{+} {\mathcal B}_{p}^{-}  \\
&+ \frac{1}{2} {\mathcal B}_{p}^{+} {\mathcal B}_{tt}^{-}-{\mathcal B}_{xy}^{+} {\mathcal B}_{xy}^{-}  +   {\mathcal B}_{ty}^{+} {\mathcal B}_{ty}^{-} + \frac{3 i }{16 \pi} ({\mathcal B}^{-}_{m})^2  + \frac{3 i }{4\pi} ({\mathcal B}_{xy}^{-})^2 \bigg) .
\label{zeroorderbmunu}
\end{aligned}
\end{equation}
The first order correction to the action is 
\begin{eqnarray}
S_{\mathrm{hydro}, 1}[{\mathcal B}_{\mu \nu}^{\sigma}] =r_0^2 \int d^{3} x \left( {\mathcal B}_{xy}^{+} \partial_{t} {\mathcal B}_{xy}^{-} +  \frac{3 i}{2 \pi} {\mathcal B}_{ty}^{-} \partial_{x} {\mathcal B}_{xy}^{-} - \frac{1}{4} {\mathcal B}_{m}^{-} \partial_{t} {\mathcal B}_{m}^{+} + \frac{3 i}{4 \pi} {\mathcal B}_{tx}^{-} \partial_{x} {\mathcal B}_{m}^{-} \right). \nonumber 
\label{firstorderbmunu}
\end{eqnarray}
We emphasise that the linear in ${\mathcal B}_{tt}^{-}$ and $ {\mathcal B}_{p}^{-}$ terms in~\eqref{zeroorderbmunu} contribute to the quadratic (in terms of $\delta g_{\mu \nu}^{(s)}$, $\xi_{\mu}^{\sigma}$) hydrodynamical action due to the non-linear terms in \eqref{bmunu}. 

\paragraph{}The rotational symmetry can be made manifest by writing the action as 
\begin{equation}\label{actnn}
\begin{aligned}
    S_{\mathrm{hydro}}=r_0^3\int d^3x\Biggl(&\,\mathcal{B}_{tt}^{-}+\frac{1}{2}\mathcal{B}_{ii}^{-}+\mathcal{B}_{ti}^{+}\mathcal{B}_{ti}^{-}+2\mathcal{B}_{tt}^{+}\mathcal{B}_{tt}^{-}+\frac{1}{2}\mathcal{B}_{tt}^{+}\mathcal{B}_{ii}^{-}+\frac{1}{2}\mathcal{B}_{ii}^{+}\mathcal{B}_{tt}^{-}-\frac{1}{2}\mathcal{B}_{ij}^{+}\mathcal{B}_{ij}^{-}\\
    &\,+\frac{1}{4}\mathcal{B}_{ii}^{+}\mathcal{B}_{jj}^{-}+\frac{3i}{8\pi}(\mathcal{B}_{ij}^{-})^2-\frac{3i}{16\pi}\mathcal{B}_{ii}^{-}\mathcal{B}_{jj}^{-}+\frac{1}{2r_0}\mathcal{B}_{ij}^{+}\partial_t\mathcal{B}_{ij}^{-}\\
    &\,-\frac{1}{4r_0}\mathcal{B}_{ii}^{+}\partial_t\mathcal{B}_{jj}^{-}+\frac{3i}{2\pi r_0}\mathcal{B}_{ti}^{-}\partial_j \mathcal{B}_{ij}^{-}-\frac{3i}{4\pi r_0}\mathcal{B}_{ti}^{-}\partial_i\mathcal{B}_{jj}^{-}\Biggr),
\end{aligned}
\end{equation}
after an integration by parts, where $i,j$ label the field theory spatial coordinates. 

\paragraph{}To make the Weyl symmetry of the action~\eqref{actnn} manifest, we first define, on each Schwinger--Keldysh branch,
\begin{equation}
 \mathcal G_{\mu\nu}(x)
 \equiv
 \eta_{\mu\nu}+\mathcal B_{\mu\nu}(x)
 =
 \frac{\partial y^\alpha}{\partial x^\mu}
 \frac{\partial y^\beta}{\partial x^\nu}
 g^{(s)}_{\alpha\beta}(y(x)),
 \label{eq:ads4-weyl-pullback}
\end{equation}
where we have again suppressed the branch label for readability. Let us now define the fields 
\begin{equation}
 \Psi\equiv\log(-\mathcal G_{tt}),
 \qquad
 \widetilde{\mathcal G}_{\mu\nu}
 \equiv e^{-\Psi}\mathcal G_{\mu\nu}
 =\frac{\mathcal G_{\mu\nu}}{-\mathcal G_{tt}},
 \label{eq:ads4-weyl-normalised}
\end{equation}
which, by construction, are invariant under the boundary diffeomorphisms \eqref{diffeosymmetry}. Under independent Weyl transformations (on each branch) $g^{(s)}_{\alpha\beta}(y)\to
e^{-2\Omega(y)}g^{(s)}_{\alpha\beta}(y)$, with the maps
$y$ held fixed, the above fields transform as
\begin{equation}
\mathcal{G}_{\mu\nu}(x)\to e^{-2\Omega(y(x))}\mathcal{G}_{\mu\nu}(x), \quad
 \Psi(x)\to
 \Psi(x)-2\Omega(y(x)),
 \quad
 \widetilde{\mathcal G}_{\mu\nu}(x)
 \to\widetilde{\mathcal G}_{\mu\nu}(x).
 \label{eq:ads4-weyl-invariance}
\end{equation}
The action~\eqref{actnn} can be expressed compactly in terms of the Weyl-invariant fields $\widetilde{\mathcal G}_{\mu \nu}$. It is convenient to introduce 
\begin{equation}
\widetilde{\mathcal G}_{\langle ij \rangle } 
 \equiv\widetilde{\mathcal G}_{ij}
 -\frac12\delta_{ij}
             \widetilde{\mathcal G}_{kk}.
 \label{eq:ads4-weyl-QV}
\end{equation}
The action~\eqref{actnn} is then equivalent (up to total derivatives) to
\begin{align}
 S_{\mathrm{hydro}}
 ={}&
 r_0^3\int d^3x\,
 \left(
 \sqrt{-\widetilde{\mathcal G}^{1}}
 -\sqrt{-\widetilde{\mathcal G}^{2}}
  +\frac{3i}{8\pi}\widetilde{\mathcal G}_{\langle ij \rangle }^-\widetilde{\mathcal G}_{\langle ij \rangle }^- 
+ \frac{1}{2 r_0}\widetilde{\mathcal G}_{\langle ij \rangle }^+\partial_t\widetilde{\mathcal G}_{\langle ij \rangle }^-
 +\frac{3i }{2\pi r_0}\widetilde{\mathcal G}_{ti}^-\partial_j\widetilde{\mathcal G}_{\langle ij \rangle }^-
 \right),
 \label{eq:ads4-weyl-final}
\end{align}
expanded to quadratic order in the fields $\mathcal B^{\sigma}_{\mu \nu}$. This form of the hydrodynamic action is manifestly invariant under both boundary diffeomorphisms and Weyl transformations. The first two terms in~\eqref{eq:ads4-weyl-final} are the ideal part of the hydrodynamic action. 
\subsection*{Constitutive relations and equations of motion} 

\paragraph{}From the off-shell action $S_{\mathrm{hydro}}[\xi^\sigma_\mu,\delta g_{\mu\nu}^{\sigma (s)}]$, we can now obtain the hydrodynamics of the theory. First we will derive expressions for the off-shell stress tensor in terms of the hydrodynamical fields $\xi^\sigma_\mu$. We will then show explicitly that the equations of motion of the hydrodynamical fields are simply the covariant conservation of this stress tensor and, furthermore, that they are equivalent to the Einstein equations that we have not yet imposed. In other words, integrating out the hydrodynamic fields corresponds to studying classical solutions of the gravitational action. In the following we will reduce notational complexity by keeping to our convention of the spatial dependence of all fields lying in the $x$ direction.\footnote{Our notation explicitly incorporates the symmetry of $\delta g^{(s)}_{\mu\nu}$, and therefore of $B_{\mu\nu}$ and $\mathcal{B}_{\mu\nu}$. To undo this (which is necessary, for example, to compute the correctly normalised off-diagonal stress tensor), one should interpret $\mathcal{B}_{\mu\nu}$ as $(\mathcal{B}_{\mu\nu}+\mathcal{B}_{\nu\mu}$)/2 and so on.}

\paragraph{}Recalling that the energy-momentum tensor on each branch is defined by~\eqref{tmunudef}, varying the hydrodynamic action gives the following contributions at linear order in amplitudes around the equilibrium state
\begin{eqnarray}
\delta T^{tt ,\sigma} &=&  5 r_0^3 \delta g^{\sigma (s)}_{tt} - 6 r_0^3 \partial_{t} \xi^{\sigma}_t,  \nonumber \\
\delta T^{tx, \sigma} &=& r_0^3 \delta g^{\sigma(s)} _{tx}  - 3 r_0^3 \partial_{t} \xi_x^{\sigma} + \frac{3 i r_0^2}{4 \pi} \partial_{x} B^{-}_{m}, \nonumber \\
\delta T^{xx, \sigma} + \delta T^{yy , \sigma} &=&  3 r_0^3 \delta g_{tt}^{\sigma (s)} -  r_0^3 \delta g_{xx}^{\sigma (s)} - r_0^3 \delta g_{yy}^{\sigma (s)} - 6 r_0^3 \partial_{t} \xi_{t}^{\sigma}, \nonumber \\
\delta T^{xx, \sigma} - \delta T^{yy, \sigma} &=& - r_0^3 \delta g_{m}^{\sigma (s)}  + \frac{3 i r_0^3}{2 \pi} B_{m}^{-} - r_0^2 \partial_{t} B_{m}^{\bar{\sigma}}  - \frac{3 i r_0^2}{\pi} \partial_{x} B_{tx}^{-},  \nonumber \\
\delta T^{ty, \sigma}  &=& r_0^3 \delta g_{ty}^{\sigma (s)} - 3 r_0^3 \partial_{t} \xi_{y}^{\sigma} + \frac{3 i r_0^2}{2 \pi}  \partial_{x} B_{xy}^{-}, \nonumber \\
\delta T^{xy, \sigma} &=& - r_0^3 \delta g_{xy}^{\sigma (s)} + \frac{3 i r_0^3}{2 \pi} B_{xy}^{-}  - r_0^2 \partial_{t} B^{\bar{\sigma}}_{xy} - \frac{3 i r_0^2}{2 \pi} \partial_{x} B_{ty}^{-}.
\label{longconstitrelations}
\end{eqnarray}
At linear order in amplitudes, the condition that the stress tensor is traceless on each branch $T_{\mu}^{\;\;\mu,\sigma}=0$ requires
\begin{equation}
 \delta T^{tt, \sigma} - \delta T^{xx,\sigma} - \delta T^{yy, \sigma} -2 r_0^3 \delta g^{\sigma (s)} _{t t} - r_0^3 \delta g^{\sigma (s)}_{xx} - r_0^3 \delta g^{\sigma (s)}_{yy} = 0.    
\end{equation}
It is straightforward to check from \eqref{longconstitrelations} that our energy-momentum tensor satisfies this even off-shell.\footnote{Recall that this was our motivation for imposing the trace$^\sigma$ constraint when going partially on-shell.} This is due to the Weyl invariance of the action.

\paragraph{}In contrast to this, the off-shell energy-momentum tensor is not covariantly conserved: linearising $\nabla_\mu T^{\mu\nu,\sigma}=0$ around equilibrium implies
\begin{eqnarray}
\partial_{t} \delta T^{tt, \sigma} + \partial_{x} \delta T^{xt, \sigma} - 2 r_0^3 \partial_{t} \delta g^{\sigma (s)}_{tt} + \frac{3 r_0^3}{2} \partial_{t} \delta g_{p}^{\sigma (s)} - r_0^3 \partial_{x} \delta g^{\sigma (s)}_{tx} = 0, \nonumber \\
\partial_{t} \delta T^{tx, \sigma} + \partial_{x} \delta T^{xx, \sigma} -\frac{3 r_0^3}{2} \partial_{x} \delta g^{\sigma (s)}_{tt} + r_0^3 \partial_{x} \delta g_{xx}^{\sigma (s)} + 2 r_0^3 \partial_{t} \delta g^{\sigma (s)}_{tx} = 0, \nonumber \\
\partial_{t} \delta T^{ty, \sigma} + \partial_{x} \delta T^{xy, \sigma} + 2 r_0^3 \partial_{t} \delta g_{ty}^{\sigma (s)} + r_0^3 \partial_{x} \delta g^{\sigma (s)}_{xy} = 0, 
\label{conservationequations}
\end{eqnarray}
which are not identically satisfied by the off-shell stress tensor \eqref{longconstitrelations}. In fact, for the stress tensor \eqref{longconstitrelations}, these three conditions are respectively the Einstein equations $-$energy$^\sigma=0$, $-$momentum$_x^\sigma=0$ and $-$momentum$_y^\sigma=0$ given explicitly in \eqref{eq:transconstraintsexplicit} and \eqref{eq:longconstraintsexplicit} before. The partially on-shell solutions we have constructed do not yet satisfy these. However, varying the hydrodynamic action with respect to the fields $\xi_\mu^\sigma$ yields
\begin{equation}
\begin{aligned}
\label{eq:hydroequationsexplicit}
    &\frac{\delta S_{\mathrm{hydro}}}{\delta\xi_t^\sigma}=\pm r_0^3\left(- \frac{3}{2} \partial_{t} B_{p}^{\sigma} - 3 \partial_{t} B_{tt}^{\sigma}- \frac{3 i}{4 \pi r_0} \partial^2_{x} B_{m}^{-}\right),\\
    &\frac{\delta S_{\mathrm{hydro}}}{\delta\xi_x^\sigma}=\pm r_0^3\left(-\frac{3 i}{4 \pi} \partial_{x}B_{m}^{-} - 3 \partial_{t} B_{tx}^{\sigma}+\frac{3 i }{2 \pi r_0} \partial_{x}^2 B_{tx}^{-} + \frac{1}{2r_0} \partial_{t} \partial_{x} B_{m}^{\bar{\sigma}} - \frac{3 i}{4 \pi r_0} \partial_{t} \partial_{x} B_{m}^{-}\right),\\
    &\frac{\delta S_{\mathrm{hydro}}}{\delta\xi_y^\sigma}=\pm r_0^3\left(-3\partial_t B_{ty}^\sigma-\frac{3i}{2\pi}\partial_xB_{xy}^{-}+\frac{3i}{2\pi r_0}\partial_x\left(\partial_xB_{ty}^- -\partial_tB_{xy}^-\right)+\frac{1}{r_0}\partial_x\partial_t B_{xy}^{\bar{\sigma}}\right),
\end{aligned}
\end{equation}
where the lower sign is for $\sigma=1$ and the upper sign for $\sigma=2$. Comparing these to equations \eqref{eq:transconstraintsexplicit} and \eqref{eq:longconstraintsexplicit}, we see that the hydrodynamic equations of motion are simply
\begin{equation}
    \frac{\delta S_{\mathrm{hydro}}}{\delta\xi_t^\sigma}=\pm\mathrm{energy}^\sigma,\quad\quad\quad\quad\quad\quad \frac{\delta S_{\mathrm{hydro}}}{\delta\xi_i^\sigma}=\pm\mathrm{momentum}_i^\sigma.
\end{equation}
In other words the hydrodynamic equations of motion are the local conservation of the stress tensor \eqref{conservationequations}, due to the symmetry discussed in~\eqref{diffeointro}, as well as being precisely the remaining Einstein constraint equations that were not yet imposed when deriving the hydrodynamic action. So putting the hydrodynamic fields on-shell corresponds to studying solutions that satisfy all of the gravitational equations of motion.

\subsection*{Response functions}

\paragraph{}We will now finally integrate out the hydrodynamic fields and compute two-point functions of the stress tensor by varying with respect to the sources (the asymptotic metric on the two branches). We check that our action indeed gives the correct retarded and symmetric correlators for this field theory, in the hydrodynamic limit.

\paragraph{}For external sources $\phi_{i}^{\sigma}(x)$ coupling to operators $O_{i}^{\sigma}(x)$ the generating functional on the Schwinger-Keldysh contour is in general given by (here $i$ labels distinct operators and $\sigma$ the branch of the contour)\footnote{See e.g. Section II of~\cite{Liu:2018kfw} for a review} 
\begin{equation}
W[\phi_{i}^{\sigma}] =  \int d^{3} x_1 d^{3} x_2 \left(  \phi_{i}^{-}(x_1) G^{R}_{ij}(x_1, x_2) \phi_{j}^{+} (x_2) + \frac{i}{2} \phi_{i}^{-}(x_1) G_{ij}^{S}(x_1, x_2) \phi_{j}^{-} (x_2) \right), \nonumber 
\end{equation}
where $G_{ij}^{R}, G_{ij}^{S}$ are the retarded and symmetric Green's functions for the operators $O_{i}$.\footnote{Our conventions for Green's function definitions are taken from~\cite{Liu:2018kfw}. In particular the retarded Green's function is defined by $G^{R}_{ij}(x_1, x_2) = i \theta(t_1-t_2) \Delta_{ij}(x_1,x_2)$. Note this differs by a minus sign from the conventions of~\cite{Kovtun:2012rj}.} In Fourier space, for a translationally invariant system
\begin{equation}
W[\phi^{\sigma}_i] = \int \frac{d\omega dk}{(2\pi)^2} \left(  \phi_{i}^{-}(-\omega, -k) G_{i j}^{R}(\omega, k) \phi_{j}^{+} (\omega, k) + \frac{i}{2} \phi_{i}^{-}(-\omega, -k) G_{i j}^{S}(\omega, k) \phi_{j}^{-} (\omega, k) \right). \nonumber 
\end{equation}
In the thermal state of temperature $T$, the Green's functions satisfy the fluctuation-dissipation relation 
\begin{equation}
G^{S}_{ij}(\omega, k) = \frac{1}{2} \mathrm{coth}\bigg(\frac{\omega}{2 T}\bigg) \Delta_{ij}(\omega, k), \hspace{2.0cm} i \Delta_{ij}(\omega, k) = G_{ij}^R(\omega, k) - G^A_{ij}(\omega, k),
\label{fdt}
\end{equation}
where $G^A_{ij}(\omega, k) = G^{R}_{j i}(-\omega, -k)$ is the advanced Green's function.

\paragraph{}To obtain the generating functional from our hydrodynamic action, we integrate out the fields $\xi_\mu^\sigma$ by imposing their equations of motion. In practice it is easiest to do this in Fourier space, yielding a generating functional of the form
\begin{equation}
W[\delta g_{\mu \nu}^{\sigma (s)}] = \int \frac{d \omega d k}{(2\pi)^2} \left( \delta g_{\mu \nu}^{- (s)}(-\omega, -k) P^{\mu \nu \alpha \beta} \delta g_{\alpha \beta}^{+ (s)} (\omega, k) + \frac{i}{2} \delta g_{\mu \nu}^{- (s)}(-\omega, -k) Q^{\mu \nu \alpha \beta} \delta g_{\alpha \beta}^{- (s)}(\omega,k) \right) \nonumber .
\end{equation}
We can then read off the Green's functions of the energy momentum tensor as 
\begin{equation}
G^R_{T^{\mu \nu} T^{\alpha \beta}}(\omega, k)= 4 P^{\mu \nu \alpha \beta}(\omega, k), \hspace{2.0cm} G^S_{T^{\mu \nu} T^{\alpha \beta}}(\omega, k) = 4 Q^{\mu \nu \alpha \beta}(\omega, k). \nonumber 
\end{equation}
We note that there are multiple definitions of the stress tensor two point function that are commonly used, varying in how they treat factors of $\sqrt{-g}$, and so differing from one another by contact terms. Our treatment is equivalent to that in the variational approach to classical hydrodynamics as described in \cite{Kovtun:2012rj} (up to an overall minus sign in the definition of each Green's function). In this treatment, the two-point functions are symmetric under the exchange of the two operators. 

\paragraph{}We obtain the following expressions for the retarded two-point functions of the momentum density
\begin{equation}
\label{eq:ourretarded2pt}
G^R_{T^{ty} T^{ty}}(\omega, k) =- \frac{r_0^2k^2}{i\omega - \frac{k^2}{3r_0}}-2r_0^3, \quad\quad\quad G^R_{T^{tx} T^{tx}}(\omega, k) =- \frac{3r_0^3\omega^2}{\omega^2-\frac{k^2}{2} + i\frac{\omega k^2}{3r_0}}+r_0^3. 
\end{equation}
These are the correct retarded two-point functions. Following the variational method of \cite{Kovtun:2012rj}, one obtains the following expressions for the hydrodynamic two-point functions of a conformal fluid, at first order in the derivative expansion\footnote{To obtain these expressions we aligned the wavevector with the $x$ direction, accounted for the minus sign difference in the definition, and set the bulk viscosity $\zeta=0$ as is required by symmetry in a CFT. The first of these equations is written explicitly as equation (2.31b) in \cite{Kovtun:2012rj}.}
\begin{equation}
    G^R_{T^{ty}T^{ty}}(\omega,k)=-\frac{\eta k^2}{i\omega-\frac{\eta k^2}{\varepsilon+p}}-\varepsilon,\quad\quad\quad G^R_{T^{tx}T^{tx}}(\omega,k)=-\frac{(\varepsilon+p)\omega^2}{\omega^2-k^2\frac{dp}{d\varepsilon}+i\frac{\eta\omega k^2}{\varepsilon+p}}+p,
\end{equation}
where $\varepsilon$ and $p$ are the energy density and pressure of the thermal state, and $\eta$ is the fluid's shear viscosity. These coincide exactly with our expressions \eqref{eq:ourretarded2pt} after inserting the energy density and pressure given in Section \ref{sec:background} and the previously known result for the shear viscosity of this state $\eta=s/(4\pi)=r_0^2$ \cite{Herzog:2002fn,Herzog:2003ke}.

\paragraph{}We can similarly compute the retarded two-point functions of the other components of the stress tensor. These should be related to those presented in equation \eqref{eq:ourretarded2pt} by Ward identities. In Appendix \ref{app:explicit2ptfunctions} we show the expressions obtained by computing them explicitly from our hydrodynamic action. In all cases they coincide exactly with the answer obtained by the variational method of classical hydrodynamics.

\paragraph{}From our hydrodynamic action we obtain the following expressions for the symmetric two-point functions of the momentum density
\begin{equation}
\label{eq:oursymmetric2pt}
    G^S_{T^{ty}T^{ty}}(\omega,k)=\frac{3r_0^3}{2\pi}\frac{k^2\left(1-\frac{2k^2}{3r_0^2}\right)}{\omega^2+\frac{k^4}{9r_0^2}},\quad\quad\quad G^S_{T^{tx}T^{tx}}(\omega,k)=\frac{3r_0^3}{2\pi}\frac{\omega^2k^2\left(1-\frac{2k^2}{3r_0^2}\right)}{\left(\omega^2-\frac{k^2}{2}\right)^2+\frac{\omega^2k^4}{9r_0^2}}.
\end{equation}
Notice that in both the numerator and denominator, the correction to the leading term in the derivative expansion is suppressed by two powers of derivatives compared to the leading term. Since our calculation of the action was only valid to first order in derivatives, we cannot trust these corrections: we would need to extend our gravitational calculation to the next order to reliably compute them. They are simply shown for completeness. The symmetric correlators for other components of the energy-momentum tensor are presented explicitly in Appendix \ref{app:explicit2ptfunctions}.

\paragraph{}It is straightforward to check that the symmetric and retarded correlators that we obtain from our action are indeed related by the fluctuation-dissipation relation \eqref{fdt}, within the accuracy of our calculation. Specifically, inserting our results \eqref{eq:ourretarded2pt} in the right hand side of \eqref{fdt} and then expanding the hyperbolic function to leading order in $\omega$, gives our symmetric correlators \eqref{eq:oursymmetric2pt} if one neglects the untrustworthy terms suppressed by two powers of derivatives in the numerator and denominator. In fact, it is only the corrections in the numerator that differ between these two expressions.

\section{Symmetries of the hydrodynamic action and horizon symmetries} 
\label{sec:horizonsyms}
\paragraph{}As we highlighted in the introduction, one of our main motivations for deriving a hydrodynamic action is to understand whether such actions realise conjectured symmetries related to many-body quantum chaos~\cite{Blake:2017ris, Blake:2021wqj}. Their gravitational manifestation is believed to be symmetries of the black hole horizon, with a specific proposal given in~\cite{Knysh:2024asf}. In a companion paper~\cite{Blake:2026axf} we perform a detailed analysis of such symmetries in lower dimensional theories of gravity (JT gravity and AdS$_3$ gravity) where hydrodynamic actions can be computed exactly (rather than to a fixed order in the derivative expansion). In those examples we explicitly demonstrate how non-trivial symmetries of the hydrodynamic action, including those conjectured in~\cite{Knysh:2024asf}, arise from infrared diffeomorphisms that preserve certain horizon boundary conditions. To repeat this for AdS$_4$-Schwarzschild would require an action for the stress tensor that is non-perturbative in derivatives, which we do not have. However, we can identify the infrared diffeomorphisms that preserve our horizon boundary conditions. In this Section we do this, and perform a non-trivial consistency check required for these to be symmetries of the hydrodynamic action presented in Section~\ref{sec:tests}. 

\paragraph{} In particular, consider the effect of diffeomorphisms $\chi_{M}^{\sigma}(t,r,x,y)$ on each branch of the CGL contour which act trivially near the asymptotic boundary but preserve radial gauge near the horizon. Such diffeomorphisms can be parameterised as
\begin{equation}\label{eq:para}
\begin{split}
\chi^{\sigma}_r(t,r \to r_0, x, y) &= \delta T^{\sigma}(t,x, y) + \mathcal{O}((r-r_0)^2), \\
\chi^{\sigma}_x(t, r \to r_0, x, y)  &=r^2 \delta X^{\sigma}(t,x, y)+r\partial_x \delta T^{\sigma}(t,x, y) + \mathcal{O}((r-r_0)^2),   \\
\chi^{\sigma}_y(t, r \to r_0, x, y)  &=r^2 \delta Y^{\sigma}(t,x, y)+r\partial_y \delta T^{\sigma}(t,x, y) + \mathcal{O}((r-r_0)^2),   \\
\chi^{\sigma}_t(t, r \to r_0, x, y) &=-D(r)\delta T^{\sigma}(t,x, y)-r\partial_t \delta T^{\sigma}(t,x, y) + \delta R^{\sigma}(t, x, y) + \mathcal{O}((r-r_0)^2).
\end{split}
\end{equation}
For generality we now allow $y$ dependence for diffeomorphisms,\footnote{The boundary conditions~\eqref{bcsintro} are rotationally invariant and continue to respect the variational principle (in the sense discussed in Appendix~\ref{app:varprinciple}) in the presence of $y$-dependence.} and the hydrodynamic fields should be identified as
\begin{eqnarray} 
\xi^{\sigma}_t(t,x,y) &=& [-\zeta^{t, \sigma}]^{\infty}_{r_0} = -\zeta^{t, \sigma}(t,r=\infty,x, y) + T^{\sigma}(t,x, y),  \nonumber \\
\xi^{\sigma}_x(t,x, y) &=& [\zeta^{x, \sigma} - \partial_{x}\zeta^{t, \sigma}/r]^{\infty}_{r_0}  = \zeta^{x, \sigma}(t,r=\infty,x, y) - X^{\sigma}(t,x, y), \nonumber \\
\xi^{\sigma}_y(t,x, y) &=& [\zeta^{y, \sigma} - \partial_{y}\zeta^{t, \sigma}/r]^{\infty}_{r_0}  = \zeta^{y, \sigma}(t,r=\infty,x, y) - Y^{\sigma}(t,x, y).
\label{hydrodefsec6}
\end{eqnarray}
$\delta R^{\sigma}$, $\delta X^{\sigma}$, $\delta Y^{\sigma}$ and $\delta T^{\sigma}$ can be thought of as shifts of the corresponding residual horizon diffeomorphisms, and hence give rise to non-trivial transformations of the hydrodynamic fields. 

\paragraph{} We anticipate that such diffeomorphisms that also preserve our horizon boundary conditions will correspond to symmetries of the hydrodynamic action. First, let us identify diffeomorphisms that preserve the non-trivial horizon conditions
\begin{equation}
\delta g^{\sigma}_{tt}(r_0) = \delta g^{\sigma}_{tx}(r_0) = \delta g^{\sigma}_{ty}(r_0) = 0, \hspace{2.0cm} 2 r_0^2 \partial_{r} \delta g^{\sigma}_{tt}(r_0) = \partial_{t} \delta g^{\sigma}_{p}(r_0).
\label{discussionbc1}
\end{equation}
These conditions are preserved under a diffeomorphism $\delta g^{\sigma}_{MN} \to \delta g^{\sigma}_{MN} + \nabla_{M} \chi^{\sigma}_{N} + \nabla_{N} \chi^{\sigma}_{M}$ where 
\begin{equation}
\begin{split}
\delta T^{\sigma}(t,x,y)
&=
b^{\sigma}_1(x,y)+b^{\sigma}_2(x,y)e^{-\kappa_0 t}
+d^{\sigma}_1(x,y)e^{\kappa_0 t},
\\
\delta R^{\sigma}(t,x,y)
&=
b^{\sigma}_3(x,y)e^{\kappa_0 t}
-r_0\kappa_0 b^{\sigma}_2(x,y)e^{-\kappa_0 t},
\\
\delta X^{\sigma}(t,x,y)
&=
b^{\sigma}_x(x,y)
-\frac{1}{r_0}\partial_x b^{\sigma}_2(x,y)e^{-\kappa_0 t}
-\frac{1}{\kappa_0r_0^2}\partial_x b^{\sigma}_3(x,y)e^{\kappa_0 t}, \\
\delta Y^{\sigma}(t,x,y)
&=
b^{\sigma}_y(x,y)
-\frac{1}{r_0}\partial_y b^{\sigma}_2(x,y)e^{-\kappa_0 t}
-\frac{1}{\kappa_0r_0^2}\partial_y b^{\sigma}_3(x,y)e^{\kappa_0 t},
\end{split} \label{dd1}
\end{equation}
provided that $b^{\sigma}_3(x,y)$ and $d^{\sigma}_1(x,y)$ are related through 
\begin{equation}
2\left(-\nabla^2+3r_0^2\right)b^{\sigma}_3
+
3r_0^2\left(3r_0^2+\nabla^2\right)d^{\sigma}_1
=0, \label{dd3}
\end{equation}
with $\nabla^2=\partial_x^2+\partial_y^2$, and where we have defined the equilibrium surface gravity $\kappa_0 = D'(r_0)/2 = 2 \pi T$ in terms of the temperature. From~\eqref{hydrodefsec6}, the diffeomorphisms \eqref{dd1} give rise to the following transformations of the hydrodynamic fields
\begin{equation}
\begin{split}
\delta_H\xi^{\sigma}_t(t, x, y)
&=
b^{\sigma}_1(x,y)+b^{\sigma}_2(x,y)e^{-\kappa_0 t} + d_1^{\sigma}(x,y) e^{\kappa_0 t},
\\
\delta_H\xi^{\sigma}_x(t, x, y)
&=
-b^{\sigma}_x(x,y)
+\frac{1}{r_0}\partial_x b^{\sigma}_2(x,y)\,e^{-\kappa_0 t}
+\frac{1}{\kappa_0r_0^2}\partial_x b^{\sigma}_3(x,y)\,e^{\kappa_0 t}, 
\\
\delta_H\xi^{\sigma}_y(t,x,y)
&=
-b^{\sigma}_y(x,y)
+\frac{1}{r_0}\partial_y b^{\sigma}_2(x,y) \,e^{-\kappa_0 t}
+\frac{1}{\kappa_0r_0^2}\partial_y b^{\sigma}_3(x,y)\,e^{\kappa_0 t}. 
\label{ads4sym}
\end{split}
\end{equation}
For comparison with~\cite{Knysh:2024asf} it is instructive to express the horizon diffeomorphism $\chi^{\sigma}$ as 
\begin{equation}\label{eq:hor_vec}
\chi^{\sigma}(t, r_0, x, y) = f^{\sigma}(t, x, y) \partial_{t} + Z^{\sigma}(t, x, y) \partial_{r}  + \tilde{X}^{\sigma}(t, x, y) \partial_{x} + \tilde{Y}^{\sigma}(t, x, y) \partial_{y}. 
\end{equation}
In this notation, the diffeomorphisms~\eqref{dd1} are equivalent to 
\begin{eqnarray}
f^{\sigma}(t,x, y) &=& b^{\sigma}_1(x,y) + b^{\sigma}_2(x,y) e^{-\kappa_0 t} + d^{\sigma}_1(x, y) e^{\kappa_0 t}, \nonumber \\
Z^{\sigma}(t,x, y) &=& \bar{b}_3^{\sigma}(x, y) e^{\kappa_0 t}, \nonumber \\
\tilde{X}^{\sigma}(t, x, y) &=& \bar{b}_{x}^{\sigma}(x,y) - \frac{1}{\kappa_0  r_0^2} \partial_{x} \bar{b}^{\sigma}_3(x, y) e^{\kappa_0 t}, \nonumber \\
\tilde{Y}^{\sigma}(t, x, y) &=& \bar{b}_{y}^{\sigma}(x,y) - \frac{1}{\kappa_0  r_0^2} \partial_{y} \bar{b}^{\sigma}_3(x, y) e^{\kappa_0 t}, 
\label{horizonchiA}
\end{eqnarray}
where $\bar{b}_{i}^{\sigma} = b_{i}^{\sigma} + \partial_{i} b^{\sigma}_1/r_0$ and $\bar{b}^{\sigma}_3 = b^{\sigma}_3 - \kappa_0 r_0 d^{\sigma}_1$.

\paragraph{}On each branch of the CGL contour, the exponentially decaying and time-independent transformations in~\eqref{horizonchiA} are precisely the corresponding symmetries proposed in~\cite{Knysh:2024asf} and that have previously been referred to as supertranslations and superrotations \cite{Donnay:2015abr,Donnay:2016ejv,Chandrasekaran:2018aop, Marjieh:2021gln}. However the exponentially growing mode has a different profile than in~\cite{Knysh:2024asf}. For the case of AdS$_4$-Schwarzschild, the symmetries proposed in~\cite{Knysh:2024asf} have $d^{\sigma}_1(x,y) = 0$ and $b_3^{\sigma}(x,y)$ unconstrained. This distinction is related to our choice of boundary conditions~\eqref{discussionbc1}. In the next section we will discuss alternative horizon boundary conditions which are preserved precisely by the horizon diffeomorphisms of~\cite{Knysh:2024asf}. 

\paragraph{} If the only boundary conditions we had imposed were~\eqref{discussionbc1} then we would anticipate the hydrodynamic action having two independent copies of the symmetries~\eqref{ads4sym}: one for each branch of the contour. However our boundary conditions also required that $\delta g_{xy}^{\sigma}$ and $\delta g_{m}^{\sigma}$ are continuous on the CGL contour. In order to make SO(2) rotational invariance manifest, we can package these metric components into the symmetric traceless tensor
\begin{equation}
     \delta g^{\sigma}_{\langle ij \rangle} \equiv \delta g^{\sigma}_{ij} - \frac{\delta_{ij}}{2}\delta g^{\sigma}_{kk},
\end{equation}
where $i,j\in\{x,y\}$. Under the diffeomorphisms~\eqref{dd1} the difference between the horizon values of these tensors on the two branches is given by
\begin{equation}
    \delta_{\chi} (\delta g_{\langle ij\rangle}^1 - \delta g_{\langle ij\rangle}^2 ) = r_0^2 \partial_i\bar{b}_j^- + r_0^2 \partial_j\bar{b}_i^- - r_0^2 \delta_{ij}\partial_l \bar{b}^-_l + 2e^{\kappa_0 t}\Delta_{ij}\mathfrak{d}^-,
    \label{gtensorchange}
\end{equation}
where we have defined
\begin{equation}
\begin{split}
    &\mathfrak{d}^\sigma\equiv r_0d_1^\sigma - \frac{2}{3 r_0}b_3^\sigma, \quad\quad\quad\quad\quad\quad\Delta_{ij} \equiv \partial_i\partial_j - \frac{1}{2}\delta_{ij}\nabla^2. 
\end{split}    
\end{equation}
Demanding that the diffeomorphisms preserve continuity at the horizon therefore imposes\footnote{Note the first of these conditions is the conformal Killing equation in $\mathbb{R}^2$.}
\begin{eqnarray}\label{eq:conf_kill}
   \partial_i \bar{b}_j^- + \partial_j\bar{b}_i^- - \delta_{ij}\partial_l\bar{b}^-_l = 0, \qquad\qquad\qquad \Delta_{ij}\mathfrak{d}^- = 0.
\end{eqnarray}
For non-zero spatial wavevector $k_i$ the first equation in~\eqref{eq:conf_kill} implies $\bar{b}^{-}_i = 0$ or equivalently
\begin{equation}
\label{eq:biminus}
    b_i^- = -\frac{1}{r_0}\partial_i b_1^-.
\end{equation}
This condition therefore fixes the shifts $b_i^{-}$ in~\eqref{ads4sym} in terms of $b^{-}_1$, and hence for time-independent shifts the off-diagonal symmetries are reduced to a single function's worth of freedom. Likewise, for $k_i\ne0$ the second equation in~\eqref{eq:conf_kill} implies that for the exponentially growing mode
\begin{equation}
    d_1^- = \frac{2}{3r_0^2}b_3^-.
\end{equation}
Upon recalling that $b_3^{-}$ and $d_1^{-}$ are also related through \eqref{dd3}, this implies $b_3^- = d_1^- = 0$. Thus at non-zero spatial momentum the exponentially growing symmetry is purely diagonal (i.e.~the exponentially growing mode must be the same on both branches). Note that the exponentially decaying modes do not appear in~\eqref{gtensorchange} and hence are not restricted by continuity of $ \delta g^{\sigma}_{\langle ij \rangle}$. As such the boundary conditions still allow for independent symmetries of this kind on each branch. 

\paragraph{}We expect that the transformations \eqref{ads4sym} are symmetries of the hydrodynamic action in that they leave the Lagrangian invariant (up to total derivatives). We expect that they should be thought of as gauge symmetries in that the off-shell stress tensor is invariant under them. For the exponentially growing and decaying modes we cannot test this explicitly as the transformations couple terms at all orders in derivatives, while we only have the action to first order. However for the time-independent shift symmetries in~\eqref{ads4sym} we can test this. Using the SO(2)-covariant form of the action \eqref{actnn}, the time-independent transformations in \eqref{ads4sym} leave the Lagrangian invariant up to total derivatives provided \eqref{eq:biminus} is satisfied. It is straightforward to obtain the SO(2)-covariant form of the constitutive relations to first order in derivatives (i.e.~to restore $y$-dependence to~\eqref{longconstitrelations}) starting from the action \eqref{actnn} and using~\eqref{tmunudef}. Under the shifts \(\delta_H\xi_t^\sigma=b_1^\sigma\) and
\(\delta_H\xi_i^\sigma=-b_i^\sigma\) the only components of the stress tensor that change are

\begin{equation}
\delta_H (\delta {T^{\langle ij\rangle}}^{\sigma}) =
\frac{3ir^3_0}{2\pi} (\partial_i \bar{b}_j^- + \partial_j\bar{b}_i^- - \delta_{ij}\partial_l\bar{b}^-_l),
\label{constitchange1}
\end{equation}
and
\begin{equation}
\delta_H(\delta T^{ti, \sigma})
=
\frac{3ir_0^2}{2\pi}\nabla^2b_i^{-}.
\label{constitchange2}
\end{equation}
The first equation~\eqref{constitchange1} is manifestly invariant under shifts corresponding to horizon symmetries satisfying~\eqref{eq:conf_kill}. While the second equation~\eqref{constitchange2} contains a non-trivial shift, it arises at higher order in derivatives than our calculation is valid for.\footnote{It can, for example, be cancelled by a second order term $\sim
-\frac{3ir_0}{2\pi} \nabla^2B_{ti}^{-}$ in $\delta T^{ti, \sigma}$.}
Thus the off-shell stress tensor is invariant under the time-independent horizon symmetry to the order controlled by our derivative expansion. The hydrodynamic equations of motion are also invariant under this transformation, to the order that we are working.

\section{Alternative horizon boundary conditions}
\label{sec:altBCs}

\paragraph{}In this Section we will go back and examine the hydrodynamic action one obtains upon imposing a slightly different set of horizon boundary conditions. Specifically, we will replace the third condition in \eqref{irbcslongitudinal} by
\begin{equation}
\label{eq:alternateIRBC}
    \partial_r\delta g_{tt}^\sigma(r_0)=0.
\end{equation}
As we explain in detail in Appendix~\ref{app:knysh}, and will shortly see explicitly, these modified boundary conditions are related in a precise way to the horizon symmetries defined in~\cite{Knysh:2024asf}. In lower-dimensional (non-dissipative) holographic theories, these conditions are also consistent with the variational principle~\cite{Blake:2026axf}, but for Schwarzschild-AdS$_4$ this is not the case (Appendix \ref{app:varprinciple}). It is nevertheless instructive to understand what changes in our calculation if we instead impose \eqref{eq:alternateIRBC}, keeping all other horizon boundary conditions the same.  

\paragraph{}Changing the boundary conditions does not affect the structure of the partially on-shell solutions that we use to construct the action, only the relation between the integration constants $\alpha_i^\sigma$ and the hydrodynamic fields and sources. In particular, imposing the horizon boundary condition \eqref{eq:alternateIRBC} changes two of the expressions for the integration constants at first order in derivatives to\footnote{The expressions for the other integration constants are the same as in equation \eqref{eq:alphaslongitudinalappendix1} and \eqref{eq:alphaslongitudinalappendix2}.}
\begin{equation}
\alpha^{\sigma}_{11}= - 2 \partial_{x} B_{tx}^{\sigma} + \frac{8}{3} \partial_{t} B_{tt}^{\sigma} + \frac{4}{3} \partial_{t} B_{p}^{\sigma} - 2 \partial_{x}^2 T^{\sigma},\quad\quad\quad \alpha^{\sigma}_{71}=-\frac{r_0^2}{2}\partial_t(2B_{tt}^\sigma+B_p^\sigma).
    \end{equation}

\paragraph{}This change in integration constants means that the energy$^\sigma$ and momentum$^\sigma_x$ constraint equations \eqref{eq:constrainteqslongalpha} (i.e.~the unimposed Einstein equations for longitudinal perturbations), expressed in terms of the hydrodynamic fields and sources, are also altered. One now finds the constraint equations become, to first order in derivatives, 
\begin{equation}
    \begin{aligned}
\mathrm{momentum}^\sigma_{x}&\,= -\frac{3 i r_0^3}{4 \pi} \partial_{x}B_{m}^{-} - 3 r_0^3 \partial_{t} B_{tx}^{\sigma} + \frac{3ir_0^2}{2\pi}\partial_x^2B_{tx}^{-}-\frac{3ir_0^2}{4\pi}\partial_x\partial_t B_m^{-}-\frac{r_0^2}{2}\partial_x\partial_t\left(2B_{tt}^\sigma +B_p^\sigma -B_m^{\bar{\sigma}}\right) ,\\
\mathrm{energy}^{\sigma}&\,=- \frac{3 r_0^3}{2} \partial_{t} B_{p}^{\sigma} - 3 r_0^3 \partial_{t} B_{tt}^{\sigma}
-\frac{3ir_0^2}{4\pi}\partial_x^2B_m^{-}-r_0^2\partial_t^2\left(2B_{tt}^\sigma+B_p^\sigma\right).
    \end{aligned}
    \label{newconstraintsexplicit}
\end{equation}
These agree to leading order derivatives with the expressions~\eqref{eq:longconstraintsexplicit} obtained for our original boundary conditions, but have additional terms at first order. 

\paragraph{}Following the same steps as in Section \ref{sec:longitudinal}, we can compute the hydrodynamic action $S_{\mathrm{hydro}}$ with these boundary conditions. It turns out, to first order in derivatives, to be identical to the action in equation \eqref{longitudinalactionads4}.\footnote{The additional terms in~\eqref{newconstraintsexplicit} contribute only a total derivative to the Lagrangian. } As a consequence, the correlation functions obtained with this horizon boundary condition will also be identical to those in Section \ref{sec:tests}. 

\paragraph{} Whilst the first order action is identical to that previously obtained, we expect at higher orders to obtain a distinct action. We can again ask about the symmetries of such an action. This time, the horizon boundary conditions are preserved by diffeomorphisms of the form~\eqref{dd1} with $d^{\sigma}_1(x,y) = 0$ (note \eqref{dd3} no longer applies, and $b_3^{\sigma}(x,y)$ is unfixed). Evaluating the diffeomorphism on the horizon one then obtains
\begin{eqnarray}
f^{\sigma}(t,x, y) &=& b^{\sigma}_1(x,y) + b^{\sigma}_2(x,y) e^{-\kappa_0 t},  \nonumber \\
Z^{\sigma}(t,x, y) &=& \bar{b}_3^{\sigma}(x, y) e^{\kappa_0 t}, \nonumber \\
\tilde{X}^{\sigma}(t, x, y) &=& \bar{b}_{x}^{\sigma}(x,y) - \frac{1}{\kappa_0  r_0^2} \partial_{x} b^{\sigma}_3(x, y) e^{\kappa_0 t}, \nonumber \\
\tilde{Y}^{\sigma}(t, x, y) &=& \bar{b}_{y}^{\sigma}(x,y) - \frac{1}{\kappa_0  r_0^2} \partial_{y} b^{\sigma}_3(x, y) e^{\kappa_0 t}, 
\label{horizonchiB}
\end{eqnarray}
where $\bar{b}_{i}^{\sigma} = b_{i}^{\sigma} + \partial_{i} b^{\sigma}_1/r_0$. These are precisely the horizon symmetries identified in equations (3.25) to (3.28) of~\cite{Knysh:2024asf}. We review the analysis of~\cite{Knysh:2024asf} from our perspective in Appendix~\ref{app:knysh} and demonstrate that their conditions for horizon symmetries reduce, for an AdS black brane in radial gauge, to identifying diffeomorphisms that preserve the boundary conditions~\eqref{bcsintro2}. These transformations correspond to shifts of the hydrodynamic fields 
\begin{equation}
\begin{split}
\delta_H\xi^{\sigma}_t(t, x, y)
&=
b^{\sigma}_1(x,y)+b^{\sigma}_2(x,y)e^{-\kappa_0 t}, 
\\
\delta_H\xi^{\sigma}_x(t, x, y)
&=
-b^{\sigma}_x(x,y)
+\frac{1}{r_0}\partial_x b^{\sigma}_2(x,y)\,e^{-\kappa_0 t}
+\frac{1}{\kappa_0r_0^2}\partial_x b^{\sigma}_3(x,y)\,e^{\kappa_0 t}, 
\\
\delta_H\xi^{\sigma}_y(t,x,y)
&=
-b^{\sigma}_y(x,y)
+\frac{1}{r_0}\partial_y b^{\sigma}_2(x,y) \,e^{-\kappa_0 t}
+\frac{1}{\kappa_0r_0^2}\partial_y b^{\sigma}_3(x,y)\,e^{\kappa_0 t}. 
\label{ads4sym2}
\end{split}
\end{equation}
\paragraph{}After also imposing continuity of $\delta g^{\sigma}_{xy}$ and $\delta g^{\sigma}_{m}$ these symmetries are no longer fully independent on the two branches: the off-diagonal symmetries are constrained by the conditions~\eqref{eq:conf_kill} with $\mathfrak{d}^-=-2 b_3^{-}/3 r_0$. For non-zero spatial wavevector, these conditions again break a subset of the off-diagonal symmetries by imposing the conditions $\bar{b}^{-}_{i} = 0$ and $b_3^{-} = 0$. The exponentially decaying solutions are again independent on the two branches. Since the form of the time-independent shifts -- and the hydrodynamic action at first order in derivatives -- are unchanged from the previous boundary conditions, the results of Section~\ref{sec:horizonsyms} regarding the invariance of the action and off-shell stress tensor continue to hold with these boundary conditions.

\paragraph{}However, with these horizon boundary conditions there is an important conceptual subtlety in whether the transformations \eqref{ads4sym2} are indeed realised as symmetries of the hydrodynamic action. Specifically, this subtlety affects the exponentially growing mode of \eqref{ads4sym2} that is conjectured to be related to quantum chaos \cite{Knysh:2024asf}. The origin of this subtlety is that with the alternative boundary conditions studied in this Section it is no longer obvious that the hydrodynamic equations of motion are equivalent to the unimposed Einstein constraint equations.\footnote{For the previous boundary conditions, this followed from the variational principle argument in Appendix \ref{app:varprinciple}.} 

\paragraph{}This issue is already manifest in our calculation: since the hydrodynamic action is unchanged (to first order in derivatives) the equations of motion of the hydrodynamic action $S_{\mathrm{hydro}}$ are still given by \eqref{eq:hydroequationsexplicit}. For the previous boundary conditions, these equations were exactly the constraint equations that we had not yet imposed. However, with the new boundary conditions this is no longer the case. The equations of motion of the hydrodynamic action~\eqref{eq:hydroequationsexplicit} differ from the unimposed Einstein equations (the constraint equations)~\eqref{newconstraintsexplicit} at first order in derivatives.

\paragraph{}If we are only interested in viewing the action to a fixed order in a derivative expansion, then the additional terms that appear in~\eqref{newconstraintsexplicit} are not as troubling as they may initially appear. Indeed, one can see they involve a derivative acting on $\partial_{t} (2 B_{tt}^{\sigma} + B^{\sigma}_{p})$. This combination of fields vanishes upon imposing the equation $\mathrm{energy}^{\sigma} = 0$ (or equivalently $\delta S_{\mathrm{hydro}}/\delta \xi_t^{\sigma}=0$) at zeroeth order in derivatives. As a result, solving the pair of bulk constraint equations~\eqref{newconstraintsexplicit} to first order in derivatives remains equivalent to imposing the pair of equations~\eqref{eq:hydroequationsexplicit} to first order in derivatives.

\paragraph{}This on-shell, order-by-order equivalence is not a coincidence. From the form of the two different boundary conditions, the corresponding actions can differ by terms proportional to $\partial_t\delta g^\sigma_p(t,r_0,x)$. But for fully on-shell solutions to the Einstein equations, we expect this quantity to vanish for generic $\omega$. To see this, first note that $\delta E_{tt}^\sigma$ on the horizon on each branch is\footnote{The importance of this equation for quantum chaos was noted in \cite{Blake:2018leo}. The form presented here relies on sufficient regularity of the metric near the horizon, which we have explicitly checked is valid for the solutions at first order in derivatives.}
\begin{equation}
(2 r_0 \partial_{t} - \partial_{x}^2) \delta g^{\sigma}_{tt}(t, r_0, x) + (2 \pi T - \partial_{t})(\partial_{t} \delta g^{\sigma}_{p}(t, r_0, x) - 2 \partial_{x} \delta g_{tx}^{\sigma}(t, r_0,x)) = 0.
\label{Evvequation}
\end{equation}
Using the other horizon boundary conditions $\delta g_{tt}^{\sigma}(t, r_0, x) = \delta g_{tx}^{\sigma}(t, r_0, x) = 0 $ we see that imposing this equation for generic frequency $\omega \neq 2 \pi T i$ then forces $\partial_{t} \delta g^{\sigma}_{p}(t, r_0,x) = 0$. At zeroeth order the hydrodynamic actions are identical, and integrating out the fields corresponds to putting the Einstein equations fully on-shell, and so we know that when either zeroeth order action is put fully on-shell $\partial_t\delta g^\sigma_{p,0}(t,r_0,x)=0$ where $\delta g^\sigma_{p,n}$ denotes the solution at the $n$th order in the derivative expansion. Since the first order actions can differ by terms proportional to $\partial_t\delta g^\sigma_{p,0}(t,r_0,x)$, if we put the lower order equations on-shell then the first order actions will become the same. 

\paragraph{}We expect this argument can be iterated at higher orders, relying on the variational principle argument that the equations of motion of our original hydrodynamic action are equivalent to the Einstein constraint equations, to ensure that imposing the hydrodynamic equations of either action order-by-order means that $\partial_t\delta g^\sigma_p(t,r_0,x)=0$ order-by-order. However, we do not have a full proof: the variational principle argument, and the regularity conditions we assumed in writing \eqref{Evvequation}, rely on the asymptotics of the on-shell solutions, which we have only checked to first order in derivatives.

\paragraph{}Even if this argument does hold to all orders, the subtlety for the exponentially growing mode is that when $\omega=2\pi Ti$, equation \eqref{Evvequation} no longer imposes a condition on $\delta g_p^\sigma(t,r_0,x)$ and so there is no guarantee that the equations of motion of the hydrodynamic action with boundary condition \eqref{eq:alternateIRBC} are equivalent to the unimposed Einstein equations for these modes. This is problematic: we have postulated symmetries that are gauge transformations and therefore solutions to the Einstein equations. But if the hydrodynamic equations are different from the Einstein equations, there is no guarantee these will actually be solutions to the hydrodynamic equations and therefore symmetries of the hydrodynamic action. Ultimately, understanding if this symmetry of \cite{Knysh:2024asf} is realised in the action with these boundary conditions may require an explicit non-perturbative (in derivatives) action for the stress tensor dynamics.\footnote{The fact that the time-independent and exponentially decaying horizon symmetries are the same for either boundary condition is consistent with our observation that it is only modes with $\omega=2\pi Ti$ for which this subtlety arises.}

 \section{Discussion}
 \label{sec:discussion} 
 
 \paragraph{} We have constructed an explicit hydrodynamical action describing the dissipative stress tensor dynamics of the thermal state of a holographic quantum field theory. Our calculation is valid to quadratic order in perturbations about equilibrium, and to first order in derivatives. This work represents a technical breakthrough, and as a result there are many immediate generalisations to explore. In the context of AdS$_4$-Schwarzschild it would clearly be instructive to push our methodology further. It is important, in particular, to verify that our approach continues to work at higher orders in the derivative expansion.

 \paragraph{}It is likely that by working very explicitly throughout, we have masked some simpler underlying structures.\footnote{For example, there are ingoing and outgoing solutions related by time-reversal symmetry, but we instead explicitly computed all partially on-shell solutions to first order in the derivative expansion by brute force.} By working more abstractly we expect it should be possible to prove that our approach (or one similar to it) produces a consistent hydrodynamic action to all orders, without having to explicitly solve the equations of motion order-by-order in derivatives.\footnote{See \cite{Bu:2020jfo} for related work in the case of a U(1) field.} This would also allow a more direct verification of KMS symmetry rather than (as we have done here) simply checking consistency with the fluctuation-dissipation relation to a given order. Such a formulation may also be helpful in understanding how to derive a non-perturbative (in derivatives) action for the stress tensor dynamics.\footnote{For this state such an action would be non-local and so we mean it in an appropriately formal sense.} It would also likely make it easier to precisely relate our action with others in the literature: in particular the formulation of \cite{Crossley:2015evo} in terms of a physical and fluid spacetime (that, at least heuristically, are our asymptotic boundary and horizon), and the approach of \cite{Jana:2020vyx,Ghosh:2020lel,He:2021jna,He:2022jnc,He:2022deg,Loganayagam:2022zmq,Loganayagam:2022teq}. We expect that there is a more natural parameterisation of our calculation (an off-shell analogue of the fluid/gravity correspondence) that will directly produce the formulation of \cite{Crossley:2015evo}, and it would be very helpful to find this.

 \paragraph{} Whilst we have focused on metric perturbations of the AdS$_4$-Schwarzschild black brane, it should also be possible to use our methodology to obtain dissipative hydrodynamic actions for other holographic theories. As well as the obvious generalisations to higher dimensional AdS$_d$-Schwarzschild geometries, it would be of interest to consider states with different symmetries and therefore structurally different hydrodynamic actions. For example, including additional U(1) charges or breaking translational symmetry.\footnote{The fluid-gravity correspondence has been applied to such examples in \cite{Erdmenger:2008rm,Banerjee:2008th,Blake:2015epa, Blake:2015hxa, Burikham:2016roo, Gouteraux:2023uff}.}

\paragraph{} We have emphasised throughout that an important ingredient in constructing our hydrodynamic action was the horizon boundary conditions. We motivated our first choice of boundary conditions using the variational principle for quadratic gravity, ensuring that the hydrodynamic equations of motion were equivalent to the unimposed Einstein constraint equations. However, we do not believe this choice is unique and it would be interesting to study this further. In particular, one could consider relaxing radial gauge near the horizon, or adding a Gibbons-Hawking term on a cutoff surface near the horizon and then imposing Dirichlet conditions there.\footnote{Here we specifically mean $\delta g^{\sigma}_{tt}(r_c) = \delta g^{\sigma}_{ti}(r_c) = \delta g^{\sigma}_{p}(r_c)$ in radial gauge where $r_c = r_0 - \epsilon$.} With our second choice of boundary conditions, we saw that the hydrodynamic equations coincided with the Einstein equations if both were strictly imposed only order-by-order in a derivative expansion and there may also be a more general family of horizon boundary conditions that satisfy this weaker condition. Such choices are reminiscent of the frame transformation freedom of classical hydrodynamics and it would be worthwhile to explore this further.

\paragraph{} More conceptually, an important aspect of the fluid-gravity correspondence is that it can be applied to obtain non-linear hydrodynamics from non-linear gravity \cite{Bhattacharyya:2007vjd}. It is important to understand how our approach should be generalised to include non-linearities. In particular, the boundary conditions we apply in the interior are imposed at the horizon of the equilibrium black hole, and it is not immediately clear how to generalise this to non-linear gravity. 

\paragraph{} One of our central motivations for constructing a hydrodynamic action was to analyse whether it possesses non-trivial symmetries that have been conjectured in the context of quantum chaos. We identified the horizon symmetries that preserve the boundary conditions we imposed. These included time-independent shifts of the hydrodynamic fields as well exponentially growing and decaying modes in time that are reminiscent of those conjectured from quantum chaos considerations.  We were able to provide a non-trivial test of these symmetries by verifying that the constitutive relations arising from our action were invariant under the specified time-independent shifts. We were not at this stage able to test explicitly if the exponential modes are symmetries of the action, as this requires knowledge of the action to all orders in derivatives. We did however highlight that, in contrast to lower dimensional examples~\cite{Blake:2026axf}, the boundary conditions we motivated from the variational principle in radial gauge are not invariant under the horizon symmetries of~\cite{Knysh:2024asf}. This suggests that the exponentially growing symmetry of the hydrodynamic action may not necessarily be that conjectured in~\cite{Knysh:2024asf}. Or, at least, that if there are multiple ways of formulating the action they likely have different exponentially growing symmetries.

\paragraph{}To close, it is useful to present our identification of horizon symmetries in a broader context. For generic holographic systems the hydrodynamic modes for U(1) current or stress tensor dynamics will continue to correspond to relative U(1) gauge transformations or diffeomorphisms between the horizon and the asymptotic boundaries of the CGL contour. When going partially on-shell, the horizon boundary condition on a field may be independent on each branch of the contour or may couple the fields on each branch. In the U(1) case, the condition $A^{\sigma}_{t}(r_0) = 0$ is an example of the first type while the continuity of $A_{i}$ after analytic continuation around the horizon is an example of the second.\footnote{In this paper, the horizon conditions~\eqref{discussionbc1} are an example of the first type while the continuity of $\delta g^{\sigma}_{xy}, \delta g^{\sigma}_{m}$ on the CGL contour are an example of the second type.} The first type of boundary conditions will be preserved by some gauge transformations acting separately on each branch, which we denote $G$. Acting with these will induce shifts of the hydrodynamic fields that are symmetries of the action. If one only had the first type of boundary conditions then the symmetries would be two independent actions of $G$ (one on each branch of the SK contour). However, the second type of boundary conditions couple the two branches and partially break the symmetry.  In the U(1) case, $A^{\sigma}_t(r_0) = 0$ are preserved on each branch by independent near-horizon gauge transformations $A^{\sigma}_{M} \to A^{\sigma}_{M} + \partial_{M} \lambda^{\sigma}(\vec{x})$, but $A_i$ only remains continuous if $\partial_{i} \lambda^{-}(\vec{x}) = 0$.\footnote{In Section~\ref{sec:horizonsyms} we studied the corresponding off-diagonal symmetry breaking for the stress tensor case.} There has recently been significant activity in re-interpreting the symmetries of SK actions in the language of spontaneous breaking of strong to weak symmetries~\cite{Gu:2024wgc, Ogunnaike:2023qyh, Huang:2024rml, Firat:2025upx}. It would be very interesting to formulate horizon symmetries, and more generally hydrodynamic symmetries related to quantum chaos, more directly in this language.

\acknowledgments

\paragraph{}We are grateful to Yanyan Bu, Luca Delacr\'{e}taz, Akash Jain, Jong Yeon Lee, Hong Liu, Andrew Lucas, M\'{a}rk Mezei, Natalia Pinzani-Fokeeva and Xiyang Sun for helpful discussions. MB acknowledges support from UK Research and Innovation (UKRI) under the UK government's Horizon Europe guarantee (EP/Y00468X/1). The work of AD is supported by the STFC Consolidated Grant ST/T000600/1 -- ``Particle Theory at the Higgs Centre''. The work of RD was partially supported by the STFC Ernest Rutherford Grant ST/R004455/1. There is no underlying data associated with this manuscript. For the purpose of open access, the authors have applied a Creative Commons Attribution (CC BY) licence to any Author Accepted Manuscript version arising from this submission.

\appendix

\section{Asymptotic boundary conditions}
\label{app:uvbcs} 

\paragraph{} In this Appendix we give more details on the boundary conditions we impose on our partially on-shell solutions near the asymptotic boundary on each branch. These are motivated by the usual holographic dictionary which is simplest to implement in the Schwarzschild coordinates $(\tau,r,x,y)$. The principle is the same for both transverse and longitudinal perturbations, but the former are easier to implement in practice so we explain them first.

\paragraph{}The Fourier modes of the transverse metric perturbations in Schwarzschild coordinates on each branch (denoted $\delta G$) are related to those in our ingoing coordinates (denoted $\delta g$) by
\begin{equation}
    \begin{aligned}
        \delta G_{\tau y}=e^{-i\omega r_*}\delta g_{ty},\quad\quad\quad \delta G_{xy}=e^{-i\omega r_*}\delta g_{xy},\quad\quad\quad \delta G_{ry}=e^{-i\omega r_*}\left(\delta g_{ry}+\frac{\delta g_{ty}}{D}\right),
    \end{aligned}
\end{equation}
where an explicit expression for the tortoise coordinate $r_*$ is given in equation \eqref{eq:Hdefn}. We firstly want to restrict the form of the asymptotic gauge transformation $\zeta_y$ such that $\delta G_{ry}$ is in Fefferman-Graham gauge (i.e.~vanishes asymptotically). An expansion of the form in equation \eqref{zetayexpansion} ensures that $\delta G_{ry}\sim O(r^0)$ which is enough for our purposes.\footnote{The gauge can be imposed systematically at higher orders by relating the coefficients $\zeta_y^{\sigma (n<2)}$ to $\zeta_y^{\sigma (2)}$, but this will not be necessary for our purposes.}
Substituting this expansion into the expressions for $\delta G_{\tau y}$ and $\delta G_{xy}$ gives
\begin{eqnarray}
    \delta G_{\tau y}^\sigma&=&\left(\beta_2^\sigma+\partial_t\zeta_y^{\sigma (2)}+\beta_{21}^\sigma-\frac{\pi}{2\sqrt{3}r_0}\partial_x\beta_4^\sigma\right)r^2+O(r^0),\\
    \delta G_{xy}^\sigma&=&\left(\beta_3^\sigma+\partial_x\zeta_y^{\sigma (2)}+\beta_{31}^{\sigma}+\frac{1}{3r_0^4}\left(1-\frac{\pi}{2\sqrt{3}}\right)\partial_x\beta_1^\sigma-\frac{\pi}{2\sqrt{3}r_0}\left(\partial_t\beta_3^\sigma-\partial_x\beta_2^\sigma\right)\right)r^2+O(r^0).\nonumber
\end{eqnarray}
Finally, using the usual holographic dictionary, we identify the $r^2$ in each expansion with the field theory metric perturbation $\delta g_{ty}^{\sigma (s)}$ and $\delta g_{xy}^{\sigma (s)}$. Doing this order-by-order in a derivative expansion yields the conditions \eqref{eq:transverseUVBC1} and \eqref{eq:transverseUVBC2} in the main text.

\paragraph{}For longitudinal perturbations, the Fourier modes of the metric in the two coordinate systems are related by
\begin{equation}
    \begin{aligned}
        &\delta G_{\tau\tau}=e^{-i\omega r_*}\delta g_{tt},\quad\quad\delta G_{\tau x}=e^{-i\omega r_*}\delta g_{tx}, \quad\quad\delta G_{\tau r}=e^{-i\omega r_*}\left(\delta g_{tr}+\frac{\delta g_{tt}}{D}\right),\\
        &\delta G_{xx}=e^{-i\omega r_*}\delta g_{xx},\quad\quad\delta G_{yy}=e^{-i\omega r_*}\delta g_{yy},\quad\quad\delta G_{xr}=e^{-i\omega r_*}\left(\delta g_{xr}+\frac{\delta g_{tx}}{D}\right),\\
        &\delta G_{rr}=e^{-i\omega r_*}\left(\delta g_{rr}+\frac{2\delta g_{tr}}{D}+\frac{\delta g_{tt}}{D^2}\right).        
    \end{aligned}
\end{equation}
We restrict the form of the gauge transformations in \eqref{eq:asymptoticzetalong} by
\begin{equation}
    \begin{aligned}
        \zeta_r^{\sigma (2)}=\zeta_r^{\sigma (1)}=0,\quad\quad
        \zeta_r^{\sigma (0)}=-\zeta_t^{\sigma (2)},\quad\quad
        \zeta_t^{\sigma (1)}=\partial_t\zeta_t^{\sigma (2)}.
    \end{aligned}
\end{equation}
This ensures that $\delta G_{rr}\sim O(r^{-3})$, $\delta G_{\tau r}\sim O(r^{-1})$ and $\delta G_{xr}\sim O(r^0)$, which is sufficient for our purposes.\footnote{Again, this can be extended systematically to higher orders by solving for higher order coefficients in \eqref{eq:asymptoticzetalong} in terms of $\zeta_r^{\sigma (-1)}$, $\zeta_t^{\sigma (2)}$ and $\zeta_x^{\sigma (2)}$.} Substituting these into the expressions for $\delta G^\sigma_{\mu\nu}$ gives
\begin{equation}
\begin{aligned}
    \delta G^\sigma_{\tau\tau}=&\,-2\zeta_r^{\sigma (-1)}r^2+\ldots,\\
    \delta G^\sigma_{xx}+\delta G^\sigma_{yy}=&\,\left(\alpha^\sigma_2+4\zeta_r^{\sigma (-1)}+2\partial_x\zeta_x^{\sigma (2)}+4\partial_t\zeta_t^{\sigma (2)}+\alpha_{21}^\sigma\right)r^2+\ldots,\\
    \delta G^\sigma_{xx}-\delta G^\sigma_{yy}=&\,\left(\alpha^\sigma_5+2\partial_x\zeta_x^{\sigma (2)}+\frac{\pi}{\sqrt{3}r_0}\left(\partial_x\alpha_3^{\sigma}+\frac{\partial_x\alpha_4^\sigma}{r_0^3}-\frac{\partial_t\alpha_5^\sigma}{2}\right)+\alpha_{51}^\sigma\right)r^2+\ldots,\\
    \delta G^\sigma_{\tau x}=&\,\left(\alpha^\sigma_3+\partial_x\zeta_t^{\sigma (2)}+\partial_t\zeta_x^{\sigma (2)}+\alpha^\sigma_{31}-\frac{\pi}{4\sqrt{3}r_0}\partial_x\alpha_6^\sigma\right)r^2+\ldots,\\
\end{aligned}    
\end{equation}
and we identify the leading term in each as the corresponding field theory metric perturbation $\delta g_{\mu\nu}^{\sigma (s)}$. This leads to the expression $\zeta_r^{\sigma (-1)}=-\delta g_{tt}^{\sigma (s)}/2$, in addition to the boundary conditions \eqref{eq:longUVBCs1} and \eqref{eq:longUVBCs2} in the main text.

\paragraph{}Our UV boundary conditions are simplest to impose in Schwarzschild coordinates, while our horizon boundary conditions are simplest to impose in ingoing coordinates. The complication of switching between coordinate systems that we have implemented in this Appendix could be avoided by working throughout in the `regular' coordinates of \cite{Arean:2023ejh}.

 \section{Summary of partially on-shell solutions}
 \label{app:ads4nearhorizon}

 \paragraph{}In this Appendix, we gather together for reference the partially on-shell solutions for the fields $h_{\mu\nu}$ after imposing all boundary conditions. 

\paragraph{}For transverse perturbations, the boundary conditions fix the integration constants to
 \begin{equation}
     \begin{aligned}
&\,\beta_1^{\sigma} =  3 r_0^3 B_{ty}^{\sigma},\quad\quad\quad\quad\quad\quad \beta^{\sigma}_{11}= -i\frac{(\sqrt{3} \pi - 6)}{4\pi} r_0^2 \partial_{x} B_{xy}^{-},\\
&\,\beta_2^{\sigma} = \delta g_{ty}^{\sigma (s)} - \partial_t \zeta_y^{\sigma (2)},  \quad\quad \;\beta_{21}^\sigma=-\frac{\sqrt{3}i}{12r_0}\partial_x B_{xy}^{-},\\
&\,\beta_3^{\sigma} = \delta g_{xy}^{\sigma (s)} - \partial_x \zeta_y^{\sigma (2)},\quad\quad\;
\beta_{31}^{\sigma}=-\frac{1}{r_0}\partial_xB_{ty}^\sigma+\frac{\sqrt{3}\pi}{6r_0}\partial_t B_{xy}^{\sigma},\\
&\,\beta_4^{\sigma} =  -\frac{i}{2 \pi} B_{xy}^{-},\quad\quad\quad\quad\quad\; \beta^{\sigma}_{41} = -\frac{\sqrt{3} i}{12 r_0} \partial_{t} B_{xy}^{-} +\frac{i}{2 \pi r_0} \partial_{x} B_{ty}^{-} + \frac{1}{3 r_0} \partial_{t} B_{xy}^{\bar{\sigma}}.
     \end{aligned}
 \end{equation}
Inserting these into the solutions \eqref{hxyzeroorder} and \eqref{hxyfirstorder} yields the explicit expressions
\begin{equation}
    \begin{aligned}
        h_{ty}^\sigma=&\,\delta g_{ty}^{\sigma (s)} - \partial_t \zeta_y^{\sigma (2)}-\frac{r_0^3}{r^3}B_{ty}^{\sigma} +\frac{i}{2\pi}\left(\frac{1}{r}\left(1-\frac{r_0^2}{r^2}\right)+r_*(r)f(r)\right) \partial_{x} B_{xy}^{-},\\       h_{xy}^\sigma=&\,\delta g_{xy}^{\sigma (s)} - \partial_x \zeta_y^{\sigma (2)}-r_*(r)\partial_t B_{xy}^{\sigma}-\frac{1}{r}\partial_{x}B_{ty}^\sigma\\
        &\,-\frac{i}{2\pi} \left(\left(1-r_*(r) \partial_t\right)B_{xy}^{-}-\frac{1}{r_0} \partial_{x} B_{ty}^{-} + \frac{2\pi i}{3 r_0} \partial_{t} B_{xy}^{\bar{\sigma}}\right) \mathrm{log}f(r).
    \end{aligned}
\end{equation}

\paragraph{}For longitudinal perturbations we considered two different horizon boundary conditions. For the original ones in equation \eqref{irbcslongitudinal}, the integration constants are
\begin{equation}
    \begin{aligned}
\label{eq:alphaslongitudinalappendix1}
    \alpha_{1}^{\sigma} &\,= - 4 R^{\sigma},\quad\quad\quad\quad\quad\quad\quad\quad\quad\quad 
    \alpha_2^\sigma=\delta g_p^{\sigma (s)} +2\delta g_{tt}^{\sigma (s)} -2\partial_x\zeta_x^{\sigma (2)}-4\partial_t\zeta_t^{\sigma (2)},\\
\alpha_3^\sigma&\,=\delta g_{tx}^{\sigma (s)}-\partial_x\zeta_t^{\sigma (2)}-\partial_t\zeta_x^{\sigma (2)},\quad\;\; 
\alpha_4^{\sigma} = - r_0^3\left( B_{tx}^\sigma + \partial_{x} T^{\sigma}\right),\\ 
\alpha_5^\sigma&\,=\delta g_m^{\sigma (s)}-2\partial_x\zeta_x^{\sigma (2)},\quad\quad\quad\quad\quad\;
        \alpha_{6}^{\sigma} =-\frac{i}{2 \pi} B_{m}^{-},\\
\alpha_{7}^{\sigma}&\,= 3 r_0^3 \partial_{t}T^{\sigma},\quad\quad\quad\quad\quad\quad\quad\quad\;\;\; \alpha^{\sigma}_8 = 0,
    \end{aligned}
    \end{equation}
    at zeroeth order in derivatives and 
\begin{equation}
    \begin{aligned}
\label{eq:alphaslongitudinalappendix2}
\alpha^{\sigma}_{11}&\,= - 2 \partial_{x} B_{tx}^{\sigma} + \frac{4}{3} \partial_{t} B_{tt}^{\sigma} + \frac{2}{3} \partial_{t} B_{p}^{\sigma} - 2 \partial_{x}^2 T^{\sigma},\quad\quad\;\;\;\alpha_{21}^{\sigma}=0,\\
\alpha_{31}^\sigma&\,=-\frac{\sqrt{3} i}{24 r_0}\partial_x B_m^{-},\quad\quad\alpha^\sigma_{41}=\frac{i r_0^2}{24\pi}(\sqrt{3}\pi-6)\partial_x B_m^{-},\quad\quad\alpha_{51}^\sigma=\frac{\pi}{2\sqrt{3}r_0}\partial_t B_m^\sigma,\\ 
\alpha^{\sigma}_{61} &\,= -\frac{\sqrt{3} i}{12 r_0} \partial_{t} B_{m}^{-} +\frac{i}{\pi r_0} \partial_{x} B_{tx}^{-} + \frac{1}{3 r_0} \partial_{t} B_{m}^{\bar{\sigma}},\quad\quad\quad\alpha^{\sigma}_{71}=0,\quad\quad\quad \alpha^{\sigma}_{81} = 2 \partial_{t} R^\sigma,
    \end{aligned}
    \end{equation}
at first order in derivatives, where $\bar{\sigma} = 1$ if $\sigma=2$ and $\bar{\sigma} = 2$ is $\sigma = 1$. Inserting these into the solutions in Section \ref{sec:longitudinal} gives
\begin{equation}
    \begin{aligned}
    h_p=&\,\delta g_p^{\sigma (s)} +2\delta g_{tt}^{\sigma (s)} -2\partial_x\zeta_x^{\sigma (2)}-4\partial_t\zeta_t^{\sigma (2)}+\frac{1}{r}\left(-4R^\sigma- 2 \partial_{x} B_{tx}^{\sigma} + \frac{4}{3} \partial_{t} B_{tt}^{\sigma} + \frac{2}{3} \partial_{t} B_{p}^{\sigma} - 2 \partial_{x}^2 T^{\sigma}\right),\\
    h_{tt}=&\,- \frac{1}{2 rf(r)}\left(4R^\sigma+  \frac{1}{3}\partial_{t} \left(B_p^\sigma +2B_{tt}^\sigma+12\partial_t T^\sigma\right)\right)+ \frac{2 \partial_tR^\sigma}{r^2f(r)}+ \frac{3 r_0^3 \partial_{t}T^{\sigma}}{r^3f(r)}  \\
    &\,+ \frac{r_0^3}{4 r^4f(r)}\left(-4R^\sigma + \frac{2}{3} \partial_{t} \left(B_p^\sigma+2B_{tt}^{\sigma}\right)\right),\\
    h_{tx}=&\,  \delta g_{tx}^{\sigma (s)}-\partial_x\zeta_t^{\sigma (2)}-\partial_t\zeta_x^{\sigma (2)} - \frac{ r_0^3\left( B_{tx}^\sigma + \partial_{x} T^{\sigma}\right)}{r^3} - \frac{ \partial_{x} R^\sigma}{ r^2}+ \frac{i}{4\pi}\left(\frac{1}{r}\left(1-\frac{r_0^2}{r^2}\right)+r_*(r)f(r)\right)\partial_{x} B_m^{-},\\
    h_m=&\, \delta g_m^{\sigma (s)}-2\partial_x\zeta_x^{\sigma (2)}- r_*(r)\partial_{t}B_m^\sigma- \frac{2}{r} \partial_x \left( B_{tx}^\sigma + \partial_{x} T^{\sigma}\right)  \\
    &\,-\frac{i}{2 \pi}\left( \left(1-r_*(r)\partial_t\right)B_{m}^{-}-\frac{2}{r_0} \partial_{x} B_{tx}^{-} + \frac{2\pi i}{3 r_0} \partial_{t} B_{m}^{\bar{\sigma}}\right) \mathrm{log}f(r).
    \end{aligned}
\end{equation}
We note that the only metric perturbation that we did not explicitly impose a near-horizon boundary condition on is $\delta g_{p}$. Near the horizon, this has the form
\begin{equation}
    \begin{aligned}
    \label{eq:appgplusBC}
        \delta g_p(r\rightarrow r_0)\rightarrow r_0^2\left(1+\frac{2}{3r_0}\partial_t\right)(B_p^\sigma+2B_{tt}^\sigma)-2r_0\partial_x B_{tx}^\sigma.
    \end{aligned}
\end{equation}
It is not continuous around the horizon, although due to the first equation in \eqref{eq:hydroequationsexplicit} when fully on-shell its $t$-derivative is (to the order we are working). It would be interesting to better understand this detail in the context of the usual prescription for obtaining the generating functional of holographic theories \cite{Skenderis:2008dh,Skenderis:2008dg}.

\paragraph{}For the alternative horizon boundary condition described in Section \ref{sec:altBCs}, the only different integration constants are $\alpha_{11}$ and $\alpha_{71}$. This results in changes to only $h_p$ and $h_{tt}$, which are now 
\begin{equation}
    \begin{aligned}
    h_p=&\,\delta g_p^{\sigma (s)} +2\delta g_{tt}^{\sigma (s)} -2\partial_x\zeta_x^{\sigma (2)}-4\partial_t\zeta_t^{\sigma (2)}+\frac{1}{r}\left(-4R^\sigma- 2 \partial_{x} B_{tx}^{\sigma} + \frac{8}{3} \partial_{t} B_{tt}^{\sigma} + \frac{4}{3} \partial_{t} B_{p}^{\sigma} - 2 \partial_{x}^2 T^{\sigma}\right),\\
    h_{tt}=&\,- \frac{1}{2 rf(r)}\left(4R^\sigma-  \frac{1}{3}\partial_{t} \left(B_p^\sigma +2B_{tt}^\sigma-12\partial_t T^\sigma\right)\right)+ \frac{2 \partial_tR^\sigma}{r^2f(r)}- \frac{r_0^2 \partial_{t}\left(B_{p}^\sigma+2B_{tt}^\sigma-6r_0T^{\sigma}\right)}{2r^3f(r)}  \\
    &\,+ \frac{r_0^3}{4 r^4f(r)}\left(-4R^\sigma + \frac{4}{3} \partial_{t} \left(B_p^\sigma+2B_{tt}^{\sigma}\right)\right).
    \end{aligned}
\end{equation}
With these boundary conditions, near the horizon
\begin{equation}
    \begin{aligned}
        \delta g_p(r\rightarrow r_0)\rightarrow r_0^2\left(1+\frac{4}{3r_0}\partial_t\right)(B_p^\sigma+2B_{tt}^\sigma)-2r_0\partial_x B_{tx}^\sigma.
    \end{aligned}
\end{equation}
The discussion below equation \eqref{eq:appgplusBC} continues to apply with these alternative boundary conditions.

\section{Variational principle near the horizon}
\label{app:varprinciple} 

\paragraph{}In this Appendix we consider the variational principle for the action for small amplitude metric perturbations, and use this to motivate the boundary conditions we apply in the main text. 

\paragraph{}After a linear variation $\delta g_{MN}\rightarrow\delta g_{MN}+\delta \chi_{MN}$ in the quadratic (in $\delta g$) bulk action $S_{\mathrm{EH}}$, integrating by parts yields a bulk term that is proportional to a linear combination of $\delta \chi_{MN}$s multiplied by the linearized Einstein equations, plus a boundary term. Explicitly
\begin{equation}
\delta S_{\mathrm{EH}} = \int d^{4}x \bigg[ r^2 \delta E_{MN} \, \delta \chi^{MN} \bigg] + S^{\mathrm{lin, bound}},
\end{equation}
where $\delta E_{MN}$ are the linearised Einstein equations for $\delta g_{MN}$ and $S^{\mathrm{lin, bound}}$ is a boundary term receiving contributions from the asymptotic boundaries and the horizon. 

\paragraph{} It is the boundary term that interests us. At the asymptotic boundaries we must add the variation of the boundary term $S_{\mathrm{GH}}$. After doing this, the action is stationary for Dirichlet boundary conditions (i.e.~fixed boundary metric) provided the linearized Einstein equations are satisfied. This is the usual one-sided story. We have two additional complications. The first is that on the CGL contour we also have the boundary terms near the horizon, and the second is that in deriving the hydrodynamic action we do not impose all of the linearised Einstein equations for $\delta g_{MN}$, but only a subset of them.

\paragraph{}By imposing horizon boundary conditions such that the variation at the horizon vanishes, we ensure that the equations of motion of the hydrodynamic action are precisely equivalent to the unimposed Einstein constraint equations.\footnote{This is desirable as putting the hydrodynamic action on-shell yields the generating function for correlators, which in the usual formulation of AdS/CFT is obtained by solving all of Einstein's equations.} To see this, we can consider parameterising the variation $\delta \chi_{MN}$ as in~\eqref{transverseansatz} and \eqref{eq:longdecomp}. For clarity we will suppress the branch index in the following discussion. That is, for transverse perturbations we take
\begin{eqnarray}
\delta \chi_{ty}(t,r,x)&=& r^2 \delta h_{ty}+ \partial_{t} \delta \zeta_{y}, \nonumber \\
\delta \chi_{yr}(t, r, x)&=& - \frac{2}{r} \delta \zeta_{y}+ \partial_{r} \delta \zeta_{y}, \nonumber \\
\delta \chi_{xy}(t,r,x)  &=& r^2 \delta h_{xy} + \partial_{x} \delta \zeta_{y},
\end{eqnarray}
and likewise for longitudinal perturbations
\begin{eqnarray}
\delta \chi_{tt}(t, r, x)&=& -D(r) \delta h_{tt}- D(r) D'(r) \delta \zeta_{r} - D'(r) \delta \zeta_{t} + 2 \partial_{t} \delta \zeta_t, \nonumber \\
\delta \chi_{tr}(t,r,x) &=& D'(r) \delta \zeta_r + \partial_{r} \delta \zeta_t + \partial_{t} \delta \zeta_r, \nonumber \\
\delta \chi_{tx}(t,r,x) &=& r^2 \delta h_{tx} + \partial_{x} \delta \zeta_t + \partial_{t} \delta \zeta_x, \nonumber \\
\delta \chi_{xr}(t,r,x) &=& - \frac{2}{r} \delta\zeta_x + \partial_{x} \delta\zeta_r + \partial_{r} \delta\zeta_x, \nonumber \\
\delta \chi_{xx}(t,r,x) &=&  \frac{r^2}{2} (\delta h_p + \delta h_m) + 2 r D(r) \delta \zeta_r + 2 r \delta \zeta_t + 2 \partial_{x} \delta \zeta_x, \nonumber \\
\delta \chi_{yy}(t,r,x) &=& \frac{r^2}{2}(\delta h_{p} - \delta h_{m}) + 2 r D(r) \delta \zeta_{r} + 2 r \delta \zeta_{t}, \nonumber \\
\delta \chi_{rr}(t,r,x) &=& 2 \partial_{r} \delta \zeta_r.
\end{eqnarray} 
Following similar manipulations as in the main text one finds that after imposing the dynamical equations and trace identity for $\delta g_{MN}$ the bulk contribution to the action reduces to
\begin{eqnarray}
\label{eq:posvar}
\int d^{4} x \, r^2 \delta E_{MN} \delta \chi^{MN} &=& - \int d^{4}x \partial_{r}(\delta \zeta^{y} \mathrm{momentum}_{y} + (\delta \zeta^{x} - \partial_{x} \delta \zeta^{t}/r) \mathrm{momentum}_{x} - \delta \zeta^{t} \mathrm{energy})  \nonumber \\
&=& -\int d^{3} x \bigg[\delta \xi_y \mathrm{momentum}_{y} + \delta \xi_x \mathrm{momentum}_{x} + \delta \xi_t \mathrm{energy}\bigg],
\end{eqnarray}
where $\mathrm{momentum}_{y}$, $\mathrm{momentum}_{x}$, and $\mathrm{energy}$ are the constraint equations for $\delta g_{MN}$, and $\delta \xi_{\mu}$ are identified as variations of the hydrodynamic fields. In other words, varying the bulk term in the partially on-shell action with respect to the hydrodynamic fields yields \eqref{eq:posvar}. Thus provided the variation in the boundary terms vanishes, varying the partially on-shell with respect to the hydrodynamic fields gives precisely the unimposed Einstein constraint equations. 

\paragraph{}We now want to identify boundary conditions near the horizon of the CGL contour such that the boundary term vanishes. To deal with this carefully, we split the contour into two by inserting a surface near the horizon on the upper branch, as described before equation \eqref{transverseboundaryterm} in the main text. This produces two boundary terms, one associated to each branch, arising from the integration by parts of the variation of $S_{\mathrm{EH}}$ (since we do not include a Gibbons-Hawking term here). On each branch, the terms involving transverse perturbations of the fields are
\begin{equation}
    \begin{aligned}
        \label{eq:variedactiontrans}
S^{\mathrm{lin, bound}}_{\mathrm{hor,trans}}&\,=\int d^3x\Biggl(-(\partial_r\delta g_{ty})\delta \chi_{ty}+\delta g_{ty}\left(-2\partial_r\delta \chi_{ty}-\frac{1}{r^2}\partial_x\delta \chi_{xy}-\frac{2}{r}\delta \chi_{ty}\right)\\
&\,+f(r)(\partial_r\delta g_{xy})\delta \chi_{xy}+\delta g_{xy}\left(2f(r)\partial_r\delta \chi_{xy}+\frac{1}{r^2}\partial_t\delta \chi_{xy}-\frac{4f(r)}{r}\delta \chi_{xy}\right)\Biggr),      
    \end{aligned}
\end{equation}
where we have fixed radial gauge near the horizon: $\delta g_{yr}=\delta \chi_{yr}=0$. Imposing Dirichlet boundary conditions $\delta g_{ty}^\sigma(r_0)=0$, as we did in the main text, causes the terms on the first line to vanish.\footnote{Technically we assume the off-shell variations have similar asymptotics to the partially on-shell solutions at first order in derivatives.\label{eq:IRBCfootnote1}}  Furthermore, remembering that $f(r)\rightarrow0$ on the horizon, imposing that $\delta g_{xy}$ be continuous after an analytic continuation around the horizon causes the terms on the second line to cancel between the two branches.

\paragraph{}Similarly, on each branch the terms involving longitudinal perturbations of the fields are
\begin{equation}
    \begin{aligned}
    \label{eq:variedactionlong}
S^{\mathrm{lin, bound}}_{\mathrm{hor,long}}=\int &\,d^3x\Biggl((\partial_r\delta g_p)\left(\frac{\delta \chi_{tt}}{2}+\frac{f(r)}{4}\delta \chi_p\right)+\delta g_p\left(\frac{\partial_r\delta \chi_{tt}}{2}-\frac{\partial_t\delta \chi_p}{4r^2}-\frac{f(r)}{2r}\delta \chi_p\right)\\
&\,-(\partial_r\delta g_{tx})\delta \chi_{tx}+\delta g_{tx}\left(-2\partial_r\delta \chi_{tx}+\frac{1}{r^2}\partial_x\delta \chi_{yy}-\frac{2}{r}\delta \chi_{tx}\right)\\
&\,+\frac{f(r)}{4}(\partial_r\delta g_m)\delta \chi_m+\delta g_m\left(\frac{f(r)}{2}\partial_r\delta \chi_m+\frac{1}{4r^2}\partial_t\delta \chi_m-\frac{f(r)}{r}\delta \chi_m\right)\\
&\,+\delta g_{tt}\left(\partial_r\delta \chi_p-\frac{1}{r}\delta \chi_p\right)\Biggr),
    \end{aligned}
\end{equation}
after imposing radial gauge $\delta g_{rM}=\delta \chi_{rM}=0$ near the horizon. The terms on the second line (with $\delta g_{tx}$ replacing $\delta g_{ty}$) and third line (with $\delta g_{m}$ replacing $\delta g_{xy}$) are analogous to those in the transverse action: they will similarly vanish for the Dirichlet boundary conditions $\delta g_{tx}^\sigma(r_0)=0$ and the continuity condition we impose on $\delta g_m$ after analytic continuation around the horizon.\footnote{Again, we assume that off-shell variations of fields have similar asymptotics to the partially on-shell solutions at first order in derivatives. \label{eq:IRBCfootnote2}} In the main text we also imposed the Dirichlet condition $\delta g_{tt}^\sigma(r_0)=0$. After further imposing this, and recalling that $f(r)\rightarrow0$ on the horizon, we are left with the following boundary variation on each branch\footnote{The boundary conditions imposed so far do not require $\delta g_p$ to be continuous across the two branches, and so the contributions from each branch to $S^{\mathrm{lin, bound}}_{\mathrm{hor,long}}$ do not necessarily cancel. Indeed, as explained in Appendix \ref{app:ads4nearhorizon}, even after imposing all boundary conditions $\delta g_p$ is not continuous.}
\begin{equation}
    S^{\mathrm{lin, bound}}_{\mathrm{hor,long}}=\frac{1}{2}\int d^3x\delta g_p\left(\partial_r\delta \chi_{tt}-\frac{1}{2r_0^2}\partial_t\delta \chi_p\right).
    \label{longvariation}
\end{equation}
This motivates the remaining horizon condition
\begin{equation}
\label{eq:IRBCappendix}
    2 r_0^2 \partial_{r} \delta g^{\sigma}_{tt}(r_0) = \partial_{t} \delta g^{\sigma}_{p}(r_0),
\end{equation}
that we imposed on the fields at the horizon in the main text (equation \eqref{irbcslongitudinal}): restricting our action to metric perturbations satisfying this condition causes the boundary variation to vanish. 

\paragraph{}The discussion in this Appendix is consistent with what we presented in the main text. For the boundary conditions~\eqref{irbcslongitudinal}, the equations of motion of the hydrodynamic action are precisely the unimposed Einstein constraint equations (the minus signs in~\eqref{eq:hydroequationsexplicit} arise from the orientation of the radial integral along the CGL contour). In contrast, for the second set of horizon boundary conditions~\eqref{eq:alternateIRBC} that we considered in the main text, the variation in~\eqref{longvariation} is non-zero, and precisely accounts for the discrepancy between the equations of motion of the hydrodynamic action and the constraint equations that we discussed in Section~\ref{sec:altBCs}. 

\paragraph{}Finally, let us comment on the fact that~\eqref{longvariation} naively suggests that there is a simpler choice of third boundary condition -- $\delta g^{\sigma}_{p}(r_0) = 0$ -- for which the equations of motion of the hydrodynamic action should simply be the unimposed Einstein constraint equations. The calculation outlined in the main text can be repeated for such boundary conditions. However, at zeroth order in derivatives, one finds that imposing these boundary conditions automatically imposes the energy constraint equation. It may be possible to circumvent this, and leave the energy equation off-shell as expected for conventional hydrodynamic actions, by imposing Dirichlet boundary conditions on a cut-off surface slightly away from the horizon. In JT and AdS$_3$ gravity this is indeed possible \cite{Blake:2026axf}, and it would be interesting to investigate this possibility for the dissipative hydrodynamics of AdS$_4$-Schwarzschild.

\paragraph{}The equations \eqref{eq:variedactiontrans} and \eqref{eq:variedactionlong} that we used to motivate the horizon boundary condition \eqref{eq:IRBCappendix} are exact in derivatives, and so we expect \eqref{eq:IRBCappendix} to be the boundary condition for which the hydrodynamic equations of motion are equal to the Einstein constraint equations, even at higher order in derivatives. However, we caution that we do not have a complete proof of this: as noted in footnotes \ref{eq:IRBCfootnote1} and \ref{eq:IRBCfootnote2}, our derivation made assumptions on the asymptotics of the partially on-shell solutions and have not proven that higher order corrections are consistent with these.

\section{Explicit expressions for two-point functions}
\label{app:explicit2ptfunctions}

\paragraph{}In this Appendix we present explicit expressions for the retarded and symmetric two-point correlators obtained from our hydrodynamic action. In the transverse channel the retarded Green's functions are given by
\begin{equation}
    \begin{aligned}
G^R_{T^{ty} T^{ty}}(\omega, k)  &=  -  \frac{6 r^4_0 \omega - i r_0^3 k^2}{3 r_0 \omega +  i k^2}, \quad\quad\quad\quad\quad\quad
G^R_{T^{ty} T^{xy}}(\omega, k) = \frac{3 i r_0^3 k \omega}{3 r_0 \omega +  i k^2}, \\
G^R_{T^{xy} T^{xy}}(\omega, k) &=  - \frac{3 r^4_0 \omega + i r_0^3 k^2 - 3 i r_0^3 \omega^2}{3 r_0 \omega +  i k^2}.
\end{aligned}
\end{equation}
The symmetric Green's functions are
\begin{equation}
\begin{aligned}
G^S_{T^{ty} T^{ty}}(\omega, k)  &=  \frac{9 r_0^3}{2 \pi} \frac{k^2(3 r_0^2 - 2k^2)}{9 r_0^2 \omega^2 + k^4},\quad\quad
G^S_{T^{ty} T^{xy}}(\omega, k) =\frac{9 r_0^3}{2 \pi} \frac{\omega k (3 r_0^2 - 2k^2)}{9 r_0^2 \omega^2 + k^4},\\
G^S_{T^{xy} T^{xy}}(\omega, k) &=  \frac{9 r_0^3}{2 \pi} \frac{\omega^2(3 r_0^2 - 2k^2)}{9 r_0^2 \omega^2 + k^4}.
\end{aligned}
\end{equation}
In the longitudinal channel one has the retarded Green's functions
\begin{eqnarray}
G^R_{T^{tt} T^{tt}}(\omega, k) &=& 4 r_0^3 \frac{6 r_0 k^2 -3 r_0 \omega^2 - i \omega k^2}{3 r_0 k^2  - 6 r_0 \omega^2  - 2 i \omega k^2}, \hspace{1.0cm} 
G^R_{T^{tt} T^{tx}}(\omega, k) = \frac{18 k \omega r_0^4 }{3 r_0 k^2 - 6 r_0 \omega^2 - 2 i \omega k^2}, \nonumber \\
G^R_{T^{tt} T^{xx}}(\omega, k) &=& 2 r_0^3 \frac{ 3 r_0 k^2 + 3 r_0 \omega^2  - 2 i \omega k^2}{3 r_0 k^2 - 6 r_0 \omega^2 - 2 i \omega k^2}, \hspace{1.0cm}
G^R_{T^{tt} T^{yy} }(\omega, k) = 2 r_0^3 \frac{3 r_0 k^2 + 3 r_0 \omega^2 + 4 i \omega k^2 }{3 r_0 k^2 - 6 r_0 \omega^2 - 2 i \omega k^2}, \nonumber \\
G^R_{T^{tx} T^{tx} }(\omega, k) &=& r_0^3 \frac{3 r_0 k^2 + 12 r_0 \omega^2 - 2 i \omega k^2 }{3 r_0 k^2 - 6 r_0 \omega^2 - 2 i \omega k^2}, \hspace{1.0cm} 
G^R_{T^{tx} T^{xx} }(\omega, k) = 3 r_0^3 \frac{3 r_0 \omega k  - 2 i \omega^2 k}{3 r_0 k^2 - 6 r_0 \omega^2 - 2 i \omega k^2},\nonumber \\
G^R_{T^{tx} T^{yy} }(\omega, k) &=& 3 r_0^3 \frac{3 r_0 \omega k + 2 i \omega^2 k }{3 r_0 k^2 - 6 r_0 \omega^2 - 2 i \omega k^2}, \hspace{1.0cm}
G^R_{T^{xx} T^{xx} }(\omega, k) = r_0^3 \frac{15 r_0 \omega^2 - 3 r_0 k^2 + 2 i \omega k^2 - 6 i \omega^3}{3 r_0 k^2 - 6 r_0 \omega^2 - 2 i \omega k^2}, \nonumber \\
G^R_{T^{xx} T^{yy} }(\omega, k) &=&  r_0^3 \frac{3 r_0 \omega^2 + 3 r_0 k^2 - 2 i \omega k^2 + 6 i \omega^3 }{3 r_0 k^2 - 6 r_0 \omega^2 - 2 i \omega k^2},  
G^R_{T^{yy} T^{yy} }(\omega, k) = r_0^3 \frac{15 r_0 \omega^2 - 3 r_0 k^2 + 14 i \omega k^2 -6 i \omega^3 }{3 r_0 k^2 - 6 r_0 \omega^2 - 2 i \omega k^2}.  \nonumber \\
\end{eqnarray}
The symmetric Green's functions are
\begin{eqnarray}
G^S_{T^{tt} T^{tt}}(\omega, k) &=& \frac{18 r_0^3}{\pi} \frac{3 r_0^2 k^4 - 2 k^6}{9 r_0^2(k^2 - 2 \omega^2)^2 +  4 
\omega^2 k^4}, \hspace{1.0cm} 
G^S_{T^{tt} T^{tx}}(\omega, k) = \frac{18 r_0^3}{\pi} \frac{3 r_0^2 \omega k^3 - 2 k^5 \omega}{9 r_0^2(k^2 - 2 \omega^2)^2 +  4 
\omega^2 k^4}, \nonumber \\
G^S_{T^{tt} T^{xx}}(\omega, k) &=& \frac{18 r_0^3}{\pi} \frac{3 r_0^2 \omega^2 k^2 - 2 k^4 \omega^2}{9 r_0^2(k^2 - 2 \omega^2)^2 +  4 
\omega^2 k^4}, \hspace{1.0cm}
G^S_{T^{tt} T^{yy} }(\omega, k) = \frac{18 r_0^3}{\pi} \frac{k^2( 3 r_0^2 - 2 k^2)(k^2 - \omega^2)}{9 r_0^2(k^2 - 2 \omega^2)^2 +  4 
\omega^2 k^4} , \nonumber \\
G^S_{T^{tx} T^{tx} }(\omega, k) &=& \frac{18 r_0^3}{\pi} \frac{3 r_0^2 \omega^2 k^2 - 2 k^4 \omega^2}{9 r_0^2(k^2 - 2 \omega^2)^2 +  4 
\omega^2 k^4}, \hspace{1.0cm} 
G^S_{T^{tx} T^{xx} }(\omega, k) =\frac{18 r_0^3}{\pi} \frac{3 r_0^2 \omega^3 k - 2 k^3 \omega^3}{9 r_0^2(k^2 - 2 \omega^2)^2 +  4 
\omega^2 k^4}, \nonumber \\
G^S_{T^{tx} T^{yy} }(\omega, k) &=& \frac{18 r_0^3}{\pi} \frac{\omega k( 3 r_0^2 - 2 k^2)(k^2 - \omega^2)}{9 r_0^2(k^2 - 2 \omega^2)^2 +  4 
\omega^2 k^4} , \hspace{1.0cm}
G^S_{T^{xx} T^{xx} }(\omega, k) = \frac{18 r_0^3}{\pi} \frac{\omega^4( 3 r_0^2 - 2 k^2)}{9 r_0^2(k^2 - 2 \omega^2)^2 +  4 
\omega^2 k^4}, \nonumber \\
G^S_{T^{xx} T^{yy} }(\omega, k) &=& \frac{18 r_0^3}{\pi} \frac{\omega^2( 3 r_0^2 - 2 k^2)(k^2 - \omega^2)}{9 r_0^2(k^2 - 2 \omega^2)^2 +  4 
\omega^2 k^4} , \hspace{1.0cm} 
G^S_{T^{yy} T^{yy} }(\omega, k) = \frac{18 r_0^3}{\pi} \frac{( 3 r_0^2 - 2 k^2)(k^2 - \omega^2)^2}{9 r_0^2(k^2 - 2 \omega^2)^2 +  4 
\omega^2 k^4}. \nonumber
\end{eqnarray}
As explained in the main text, we have explicitly checked the retarded Green's functions agree with those computed using variational hydrodynamics\footnote{We do not know of a reference where these are all explicitly presented, but the reader can easily verify that our results are consistent with equation (2.34) of \cite{Kovtun:2012rj}.} and that the fluctuation-dissipation relation~\eqref{fdt} is satisfied to the order we are working at (see the discussion at the end of Section \ref{sec:tests} for what this means in practice).

\section{Review of horizon symmetries of \cite{Knysh:2024asf}}
\label{app:knysh}

 \paragraph{} It has previously been suggested that hydrodynamic actions of holographic theories should exhibit specific symmetries arising from gravitational symmetries of black hole horizons~\cite{Knysh:2024asf}. In this Appendix we review the definition of horizon symmetries in~\cite{Knysh:2024asf}, and explain how they are related to the symmetries we studied in Sections~\ref{sec:horizonsyms} and~\ref{sec:altBCs}.

\paragraph{} We first review the definition of horizon symmetries in~\cite{Knysh:2024asf}, presenting them more explicitly for the case of a black brane that is relevant to us.\footnote{Note that~\cite{Knysh:2024asf} also considers horizon symmetries for more general geometries.} We start from a background metric describing a black brane
\begin{equation}
ds^2 = -D(r) dt^2 + 2 dt dr + r^2 dx^{i} dx^{i}, \nonumber 
\end{equation}
with $i=1,\ldots,d-2$ spatial dimensions and where $D(r)$ has a horizon at $r = r_0$.  The horizon $r=r_0$ can be parameterised by coordinates $(t, x^{i})$ which can be identified with their bulk extensions. The horizon metric in terms of $(t,x^{i})$ is
\begin{equation}
ds^2 = r_0^2 dx^{i} dx^{i}, \nonumber 
\end{equation}
which is degenerate and has a null direction in the horizon
\begin{equation}
\hat{l}^{\mu} = \bigg(\frac{\partial}{\partial t}\bigg)^{\mu}. \nonumber 
\end{equation}
The extension of this vector into the bulk is  
\begin{equation}
{l}^{M} = \bigg(\frac{\partial}{\partial t}\bigg)^{M}, \nonumber 
\end{equation}
where we note that $l^{M}$ satisfies (at the horizon)
\begin{equation}
l^{N} \nabla_{N} l^{M} = \kappa_0 l^{M}, \nonumber 
\end{equation}
where $2 \kappa_0 = 4 \pi T = D'(r_0)$. 

\paragraph{} To define horizon symmetries as in~\cite{Knysh:2024asf} one considers performing a diffeomorphism of the bulk metric 
\begin{equation}
g_{M N} \to g_{M N} +  \delta g_{M N}, \hspace{2.0cm} \delta g_{M N} = \nabla_{M} \chi_{N} + \nabla_{N} \chi_{M}, \nonumber
\label{diff1}
\end{equation}
which acts near the horizon as\footnote{We use $\tilde{X}^{i}$ to distinguish these fields from the residual diffeomorphism $X$ discussed in the main text. They are denoted by $Y^{i}$ in~\cite{Knysh:2024asf}.}
\begin{equation}
\chi = f(t, x^{i}) \bigg(\frac{\partial}{\partial t}\bigg) + Z(t, x^{i})\bigg(\frac{\partial}{\partial r}\bigg) + \tilde{X}^{i}(t,x^{i}) \bigg(\frac{\partial}{\partial x^{i}}\bigg). 
\label{diff2}
\end{equation}
Such a diffeomorphism is defined in~\cite{Knysh:2024asf} to be a horizon symmetry if the following hold:
\begin{enumerate} 
\item The vector $\partial/\partial t$ remains a null vector of the induced metric on the horizon. In the above setting this is equivalent to the conditions 
\begin{equation}
\delta g_{t \mu}(r_0) = 0,
\label{symcond1}
\end{equation}
 where $\mu = (t, x^{i})$. 
\item We further require that, evaluated at the horizon, 
\begin{equation}
\delta(l^{B} {\nabla}_{B} l^{A}) = 0, \nonumber 
\end{equation}
where $\delta$ denotes the change under the diffeomorphism~\eqref{diff2} to linear order in $\chi$. The $t$-component of this vector is given by 
\begin{eqnarray}
\delta(l^{B} {\nabla}_{B} l^{t }) = - \kappa_0 \delta g_{tr}(r_0) - \partial_{r} \delta g_{tt}(r_0)/2 + \partial_{t} \delta g_{tr}(r_0) &=& 0, 
\label{symcond2}
\end{eqnarray}
whilst the $(r, x^{i})$ components evaluate to
\begin{eqnarray}
\delta(l^{B} {\nabla}_{B} l^{r}) &=& - \kappa_0 \delta g_{tt}(r_0) + \partial_{t} \delta g_{tt}(r_0)/2 = 0, \nonumber \\
\delta(l^{B} {\nabla}_{B} l^{i})  &=&\frac{1}{2 r_0^2}(-2 \kappa_0 \delta g_{tx}(r_0) - \partial_{i} \delta g_{tt}(r_0) + 2 \partial_{t} \delta g_{ti}(r_0))= 0. 
\label{kappaconstraints}
\end{eqnarray} 
\end{enumerate} 
Note that after imposing $\delta g_{t\mu}(r_0) = 0$ for $\mu = (t, x^{i})$ then the conditions in~\eqref{kappaconstraints} are already satisfied. The independent conditions defining a horizon symmetry are then simply~\eqref{symcond1} and~\eqref{symcond2}. These conditions can be solved for horizon symmetries of the metric in any dimension by noting that 
\begin{eqnarray}
\delta g_{tt}(r_0) &=& - 2 \kappa_0 Z + 2 \partial_{t} Z  = 0, \nonumber \\
\delta g_{t i}(r_0) &=& \partial_{i} Z + r_0^2 \partial_{t} \tilde{X}^{i} = 0, \nonumber \
\end{eqnarray}
which can be used to solve for $Z(t, x^{i}), \tilde{X}^{i}(t, x^{i})$. The additional requirement of~\eqref{symcond2} is then 
\begin{equation}
\partial_{t}^2 f + \kappa_0 \partial_{t} f + \frac{1}{2} D''(r_0) Z= 0, \nonumber 
\end{equation}
which can be solved for $f(t, x^{i})$. This leads to the horizon symmetries
\begin{eqnarray}
f(t,x^{i}) &=& a_1(x^{i}) + a_2(x^{i}) e^{-\kappa_0 t} - \frac{D''(r_0)}{4 \kappa_0^2} a_3(x^{i}) e^{\kappa_0 t}, \nonumber \\
Z(t, x^{i}) &=& a_3(x^{i}) e^{\kappa_0 t}, \nonumber \\
\tilde{X}^{i}(t,x^{i}) &=& a_{i}(x^{i}) - \frac{1}{\kappa_0 r_0^2} \partial_{{i}} a_3(x^{i}) e^{\kappa_0 t}, 
\label{hongsymmetry} 
\end{eqnarray} 
which were originally identified in~\cite{Knysh:2024asf}. In summary, the horizon symmetries of~\cite{Knysh:2024asf} are precisely IR diffeomorphisms of the form~\eqref{diff2} that satisfy~\eqref{symcond1} and~\eqref{symcond2}. Note that for the case of the AdS$_4$-Schwarzschild black hole discussed in the main text one has $D''(r_0) =0 $ and there is no exponentially growing mode in $f(t, x^i)$. 

\paragraph{} In Sections~\ref{sec:horizonsyms} and~\ref{sec:altBCs} we identified horizon symmetries in a slightly different way. We first imposed radial gauge near the horizon, and then identified residual diffeomorphisms that preserve both radial gauge and our boundary conditions at the horizon. We found that for the boundary conditions~\eqref{bcsintro2} studied in Section~\ref{sec:altBCs} the resulting horizon symmetries are precisely~\eqref{hongsymmetry}. This can be understood from the fact, restricted to radial gauge near the horizon, the conditions~\eqref{symcond1} and~\eqref{symcond2} precisely correspond to diffeomorphisms that preserve these boundary conditions.\footnote{Note~\eqref{diff2} does not itself preserve radial gauge. However $\delta g_{rr}(r_0) = \delta g_{\mu r}(r_0) = 0$ can be imposed through order $(r-r_0)$ corrections to~\eqref{diff2} Such corrections away from the horizon do not modify the conditions for a horizon symmetry in terms of $f, Z, \tilde{X}^{i}$.} In contrast, for the boundary conditions~\eqref{bcsintro} whose horizon symmetries were discussed in Section~\ref{sec:horizonsyms}, we found a distinct set of horizon symmetries. Specifically~\eqref{horizonchiA} has a different profile for the exponentially growing symmetry to~\eqref{hongsymmetry}. This is not inconsistent -- it is simply that in this case we are asking for diffeomorphisms that preserve a distinct set of horizon data to that demanded in~\cite{Knysh:2024asf}.

\section{An alternative approach in the literature} 
\label{app:bu}
\paragraph{}In this Appendix we discuss the alternative approach to obtaining a hydrodynamic action for AdS$_5$ gravity proposed in \cite{Bu:2025zad}, which works in a fixed gauge for metric perturbations. \cite{Bu:2025zad} applies the CGL analytic continuation prescription to these gauge-fixed solutions, which results in qualitatively different near-horizon boundary conditions than ours. \cite{Bu:2025zad} also has no explicit gravitational description of the hydrodynamic fields $\xi_{\mu}$ -- these are generated by acting on boundary sources with diffeomorphisms. 

\paragraph{} More importantly, by following the method of \cite{Bu:2025zad} we have been unable to reproduce their result for the quadratic partially on-shell action at first order in derivatives (equation (3.41) of \cite{Bu:2025zad}). Instead, with this method we obtain a result that is inconsistent as an SK action, as it contains terms involving products of symmetric fields on the SK contour. 

\paragraph{}For concreteness we will now present the results we obtain using the method of \cite{Bu:2025zad} for AdS$_5$ gravity.\footnote{We expect this method would also give an inconsistent action in other dimensions.} The gravitational action is $S = S_{\mathrm{EH}} + S_{\mathrm{GH}}$ where
\begin{equation}
S_{\mathrm{EH}} = \int d^{5} x \sqrt{-g} (R + 12), \hspace{2.0cm} S_{\mathrm{GH}} =  \int d^{4} x \sqrt{-\gamma} (2K- 6) + \dots,
\end{equation} 
and dots are additional counterterms that depend on curvatures of the boundary metric and can be ignored to first order in the boundary derivative expansion. The action has the equations of motion
\begin{equation}
E_{M N} = R_{M N} - \frac{1}{2} g_{M N} R - 6 g_{M N} = 0.
\end{equation}
We will consider metric perturbations on the CGL contour around the equilibrium spacetime
\begin{equation}
ds^2 = -D(r) dt^2 + 2 dt dr + r^2( dx^2 + dy^2 + dz^2),\quad\quad\quad\quad D(r) = r^2\bigg(1 - \frac{r_0^4}{r^4}\bigg).
\end{equation}
The method of~\cite{Bu:2025zad} is to work in the fixed gauge
\begin{equation}
\delta g^{\sigma}_{rr} = \frac{\delta g^{\sigma}_{tt}}{D(r)^2}, \hspace{2.0cm} \delta g^{\sigma}_{tr} = - \frac{\delta g^{\sigma}_{tt}}{D(r)}, \hspace{2.0cm} \delta g^{\sigma}_{ri} = - \frac{\delta g^{\sigma}_{ti}}{D(r)},
\label{Bugaugechoice}
\end{equation}
and then solve the linearised Einstein equations $\delta E^{\sigma}_{tt} = 0, \delta E^{\sigma}_{ti} = 0, \delta E^{\sigma}_{ij}=0$ (with $i, j \in (x, y, z)$) for the remaining metric perturbations, subject to the asymptotic boundary conditions\footnote{It is through this equation and the definition~\eqref{bmunu} that the hydrodynamic fields are introduced in the method of \cite{Bu:2025zad}. There is no explicit bulk description in terms of relative diffeomorphisms.}
\begin{equation}
\delta g^{\sigma}_{\mu \nu}(t,r,x^{i}) = r^2 {\mathcal B}^{\sigma}_{\mu \nu}(t, x^{i}) + {\mathcal O}(r).
\label{uvsourcesbu}
\end{equation}
One of the horizon boundary conditions of \cite{Bu:2025zad} is the same as ours: $\delta g^{\sigma}_{ti}(t, r_0, x^{i}) = 0$. But, unlike us, their other horizon boundary conditions are to relate the remaining fields on the two branches through analytic continuation around the horizon via the CGL contour.  

\paragraph{}For illustrating our point, it is sufficient to consider longitudinal metric perturbations with spatial dependence in the $x$ direction. The result for the quadratic action presented in equation (3.41) of \cite{Bu:2025zad} then reduces to 
\begin{eqnarray}
S_0 &=&r_0^4 \int d^{4} x \Biggl(\frac{3}{4} {\mathcal B}^{-}_{00} {\mathcal B}^{+}_{00} +  {\mathcal B}^{-}_{0x} {\mathcal B}^{+}_{0x} - \frac{1}{4} {\mathcal B}^{-}_{ii} {\mathcal B}^{+}_{00} + \frac{3}{4} {\mathcal B}^{-}_{00} {\mathcal B}^{+}_{ii} +  \frac{1}{4} {\mathcal B}^{-}_{ii} {\mathcal B}^{+}_{jj}- \frac{1}{2} ({\mathcal B}^{-}_{xx} {\mathcal B}^{+}_{xx}\nonumber\\
&+& {\mathcal B}^{-}_{yy} {\mathcal B}^{+}_{yy} + {\mathcal B}^{-}_{zz} {\mathcal B}^{+}_{zz})+ \frac{i}{6 \pi} (3 ({\mathcal B}^{-}_{xx})^2 + 3 ({\mathcal B}^{-}_{yy})^2 + 3 ({\mathcal B}^{-}_{zz})^2 - ({\mathcal B}^{-}_{xx} + {\mathcal B}^{-}_{yy} + {\mathcal B}^{-}_{zz})^2)\Biggr),\nonumber
\end{eqnarray}
at zeroth order in derivatives, and to 
\begin{equation}
S_1 =  \frac{r_0^3}{2}\int d^{4}x \Biggl(\frac{1}{3} {\mathcal B}^{-}_{ii} \partial_{t} {\mathcal B}^{+}_{jj} -  ({\mathcal B}^{-}_{xx} \partial_{t} {\mathcal B}^{+}_{xx} + {\mathcal B}^{-}_{yy} \partial_{t} {\mathcal B}^{+}_{yy} + {\mathcal B}^{-}_{zz} \partial_{t} {\mathcal B}^{+}_{zz})- \frac{i \pi}{4} {\mathcal B}^{-}_{00} \partial_{t} {\mathcal B}^{-}_{ii}\Biggr), 
\end{equation}
at first order in derivatives.

\paragraph{} Repeating this calculation ourselves, we have found discrepancies in the calculation at first order in derivatives. Specifically we find the first order calculation computes to
\begin{equation}
\begin{aligned}
S_1 =\frac{r_0^3}{2}  \int d^{4}x \Biggl(&\frac{1}{3} {\mathcal B}^{-}_{ii} \partial_{t} {\mathcal B}^{+}_{jj} -  ({\mathcal B}^{-}_{xx} \partial_{t} {\mathcal B}^{+}_{xx} + {\mathcal B}^{-}_{yy} \partial_{t} {\mathcal B}^{+}_{yy} + {\mathcal B}^{-}_{zz} \partial_{t} {\mathcal B}^{+}_{zz})+\frac{i \pi}{4} {\mathcal B}^{-}_{00} \partial_{t} {\mathcal B}^{-}_{ii} \\
&-i\pi {\mathcal B}^{+}_{00} \partial_{t} {\mathcal B}^{+}_{ii}\Biggr), 
\label{Bucorrected}
\end{aligned}
\end{equation}
with the discrepancies in the last two terms. The discrepancy is significant:~\eqref{Bucorrected} is not a consistent Schwinger-Keldysh action due to the presence of the ${\mathcal B}^{+}_{00} \partial_{t} \mathcal{B}^{+}_{ii}$ term. Such terms are not allowed in the Schwinger-Keldysh action, which by unitarity must vanish when ${\mathcal B}^{-}_{\mu \nu}$ is set to zero. We do not have a deep understanding of what produces such terms, but it is related to the fact that the solutions for $\delta g_{tt,1}^{\sigma}$ and $\delta g_{p,1}^{\sigma}$ obtained using the boundary conditions of~\cite{Bu:2025zad} are different on the two branches when antisymmetric sources vanish.

\paragraph{} We now present details of how we obtain~\eqref{Bucorrected}. The longitudinal metric perturbations, in the gauge~\eqref{Bugaugechoice}, can be parameterised in terms of $\delta g^{\sigma}_{tt},  \delta g^{\sigma}_{tx}, \delta g^{\sigma}_{xx}, \delta g^{\sigma}_{yy}, \delta g^{\sigma}_{zz}$ which are functions of $(t,r,x)$. We want to solve for these functions order by order in boundary derivatives by solving the equations $\delta E^{\sigma}_{tt} = 0, \delta E^{\sigma}_{tx} = 0, \delta E^{\sigma}_{xx} = 0, \delta E^{\sigma}_{yy} = 0, \delta E^{\sigma}_{zz} = 0$. To do so it is convenient to define
\begin{equation}
\delta g^{\sigma}_{p} = \delta g^{\sigma}_{xx} + \delta g^{\sigma}_{yy} + \delta g^{\sigma}_{zz},\quad\quad\quad
\delta g^{\sigma}_{my} = \delta g^{\sigma}_{xx} - \delta g^{\sigma}_{yy},\quad\quad\quad 
\delta g^{\sigma}_{mz} = \delta g^{\sigma}_{xx} - \delta g^{\sigma}_{zz}.
\label{decomposeads5pert}
\end{equation}
At zeroth order in derivatives the equations $\delta E^{\sigma}_{tx} = 0$, $\delta E^{\sigma} _{tt} = 0$, $\delta E^{\sigma}_{xx} - \delta E^{\sigma}_{yy} = 0$ and $\delta E^{\sigma}_{xx} - \delta E^{\sigma}_{zz} = 0$ diagonalise and can be solved for the zeroth order solutions $\delta g^{\sigma}_{tx, 0}$, $\delta g^{\sigma}_{p, 0}$, $\delta g^{\sigma}_{my, 0}$ and $\delta g^{\sigma}_{mz, 0}$ respectively. 
One can then insert the solution for $\delta g^{\sigma}_{p, 0}$ into the equation $\delta E^{\sigma}_{ii} = 0$ to obtain an equation that can be solved for $\delta g^{\sigma}_{tt, 0}$.

\paragraph{} With the boundary conditions described above we find the zeroth order solutions
\begin{eqnarray}
\delta g^{1}_{p, 0}  &=& r^2 {\mathcal B}^{+}_{p} + \frac{\sqrt{r^4 - r_0^4}}{2} {\mathcal B}_{p}^{-}, \quad\quad
\delta g^{2}_{p, 0}  = r^2 {\mathcal B}^{+}_{p} - \frac{\sqrt{r^4 - r_0^4}}{2}  {\mathcal B}_{p}^{-}, \nonumber \quad\quad \delta g^{\sigma}_{tx, 0} = D(r) {\mathcal B}^{\sigma}_{0x}, \nonumber \\
\delta g^{\sigma}_{my,0} &=&  {\mathcal B}_{my}^{\sigma} r^2  - \frac{i  {\mathcal B}_{my}^{-}}{2 \pi} r^2 \mathrm{log}\bigg(1 - \frac{r_0^4}{r^4}\bigg), \quad\quad\quad\;\;\;
\delta g^{\sigma}_{mz, 0} =  {\mathcal B}_{mz}^{\sigma} r^2  - \frac{i  {\mathcal B}_{mz}^{-}}{2 \pi} r^2 \mathrm{log}\bigg(1 - \frac{r_0^4}{r^4} \bigg),  \nonumber \\
\delta g^{1}_{tt,0} &=& \frac{\sqrt{r^4 - r_0^4}}{2}  {\mathcal B}_{00}^{-} + \frac{r^4 - r_0^4}{r^2} {\mathcal B}_{00}^{+} - \frac{r_0^4 \sqrt{r^4 - r_0^4}}{6 r^4} {\mathcal B}_{p}^{-},  \nonumber \\
\delta g^{2}_{tt,0} &=& -\frac{\sqrt{r^4 - r_0^4}}{2}  {\mathcal B}_{00}^{-} + \frac{r^4 - r_0^4}{r^2} {\mathcal B}_{00}^{+} + \frac{r_0^4 \sqrt{r^4 - r_0^4}}{6 r^4} {\mathcal B}_{p}^{-},  
\end{eqnarray}
where we note that analytic continuation flips the signs of the square root factors as we move from one branch of the contour to the other. These expressions agree with the results presented in equation (3.31) of~\cite{Bu:2025zad} (we are using $0$ to denote a $t$ index in order to more closely match the notation in~\cite{Bu:2025zad}, and the sources ${\mathcal B}^{\sigma}_{my}, {\mathcal B}^{\sigma}_{mz}, {\mathcal B}^{\sigma}_{p}$ are defined analogously to~\eqref{decomposeads5pert}).

\paragraph{} Proceeding to first order in derivatives we find the expressions (after imposing the same boundary conditions) 
\begin{eqnarray}
\delta g^{1}_{p,1}  &=& -\frac{i \pi \partial_{t} {\mathcal B}_{p}^{2}}{4 r_0} r^2 + \frac{i \pi \partial_{t} {\mathcal B}_{p}^{2}}{4 r_0} \sqrt{r^4 - r_0^4} - r_*(r) \partial_{t} g^{1}_{p, 0}, \nonumber \\
\delta g^{2}_{p,1}  &=&  \frac{i \pi \partial_{t} {\mathcal B}_{p}^{1}}{4 r_0} r^2 -  \frac{i \pi \partial_{t} {\mathcal B}_{p}^{1}}{4 r_0} \sqrt{r^4 - r_0^4} - r_*(r) \partial_{t} g^{2}_{p, 0}, \nonumber \\
\delta g^{\sigma}_{tx,1}  &=& - r_*(r) \partial_{t} g^{\sigma}_{tx, 0}, \nonumber \\
\delta g^{\sigma}_{my,1}  &=&  \frac{\partial_{t} {\mathcal B}_{my}^{\bar{\sigma}}}{4 r_0} r^2 \mathrm{log}\bigg(1 - \frac{r_0^4}{r^4} \bigg) - r_*(r)  \partial_{t} g^{\sigma}_{my, 0}, \nonumber \\
\delta g^{\sigma}_{mz,1} &=&  \frac{\partial_{t} {\mathcal B}_{mz}^{\bar{\sigma}}}{4 r_0} r^2 \mathrm{log}\bigg(1 - \frac{r_0^4}{r^4}\bigg) -  r_*(r) \partial_{t} g^{\sigma}_{mz, 0}, \nonumber \\
\delta g^{1}_{tt,1} &=&  -\frac{i \pi \partial_{t} {\mathcal B}^{2}_{p}}{4 r_0} \frac{r_0^4 \sqrt{r^4 - r_0^4}}{3 r^4} + \frac{i \pi \partial_{t} {\mathcal B}_{00}^{2}}{4 r_0} \sqrt{r^4 - r_0^4} -  \frac{i \pi \partial_{t} {\mathcal B}_{00}^{2}}{4 r_0 r^2} (r^4 - r_0^4) - r_*(r) \partial_{t} g^{1}_{tt, 0}, \nonumber \\
\delta g^{2}_{tt,1} &=&  \frac{i \pi \partial_{t} {\mathcal B}^{1}_{p}}{4 r_0} \frac{r_0^4 \sqrt{r^4 - r_0^4}}{3 r^4} - \frac{i \pi \partial_{t} {\mathcal B}_{00}^{1}}{4 r_0} \sqrt{r^4 - r_0^4} +  \frac{i \pi \partial_{t} {\mathcal B}_{00}^{1}}{4 r_0 r^2} (r^4 - r_0^4) - r_*(r) \partial_{t} g^{2}_{tt, 0}, \nonumber 
\end{eqnarray}
where 
\begin{equation}
r_*(r) = \int_{\infty}^{r} \frac{dy}{D(y)} = \frac{1}{4 r_0} ( 2 \tan^{-1}(r/r_0) + \mathrm{log}(r - r_0)  - \mathrm{log}(r + r_0) - \pi),
\end{equation}
is the tortoise coordinate. At this order we find discrepancies with the results presented in~\cite{Bu:2025zad}, including a factor of $2$ in the terms involving the zeroth order solution.

\paragraph{} We now proceed to compute the Einstein-Hilbert action on the solutions presented above. This can be done using the fact that in the gauge~\eqref{Bugaugechoice} the Einstein-Hilbert action, expanded to quadratic order in longitudinal perturbations, can be expressed as $S_{\mathrm{EH}} = S_{\mathrm{bulk}} + S_{\mathrm{bdy}}$
where 
\begin{equation}
 S_{\mathrm{bulk}}  = \int d^{5} x \bigg( - \frac{r^3}{D(r)^2} \delta E_{tt} \delta g_{tt}+ \frac{2r}{D(r)} \delta E_{tx} \delta g_{tx}  -  \frac{1}{r} \delta E_{ii} \delta g_{ii} \bigg),
\label{Bubulkaction}
\end{equation}
and 
\begin{equation}
\begin{aligned}
 &S_{\mathrm{bdy}}  = \int d^{4}x \frac{1}{r^4} \Biggl(r^5 (\delta g_{xx} + \delta g_{yy} + \delta g_{zz})(\partial_{r} \delta g_{tt} - \frac{2 r}{D(r)} \delta g_{tt})- 6 r^5 \delta g_{tx} \partial_{r} \delta g_{tx} \\
 &+ \frac{r^4 \delta g_{tt}( - 3 (r^4 + r_0^4) \delta g_{tt} + 2 r^3D(r) \partial_{r} \delta g_{tt})}{D(r)^2}+ r^5 \delta g_{tt} \partial_{r} (\delta g_{xx}+\delta g_{yy}+\delta g_{zz})\\
 &+ r^2D(r)\delta g_{xx}(-3 \delta g_{xx} + 2 r \partial_{r} \delta g_{xx}-r\partial_r\delta g_{yy}-r\partial_r\delta g_{zz}+\delta g_{yy}+\delta g_{zz})\\
 & + r^2D(r)\delta g_{yy}(-3 \delta g_{yy} + 2 r \partial_{r} \delta g_{yy}-r\partial_r \delta g_{xx}-r\partial_r \delta g_{zz}+\delta g_{xx}+\delta g_{zz})\\
 &+ r^2D(r)\delta g_{zz}(-3 \delta g_{zz} + 2 r \partial_{r} \delta g_{zz}-r\partial_r\delta g_{xx}-r\partial_r\delta g_{yy}+\delta g_{xx}+\delta g_{yy})+\frac{8 r^6 \delta g_{tx}^2}{D(r)}\Biggr).
      \label{buboundaryaction}
      \end{aligned}
      \end{equation}
In obtaining the above result we have integrated by parts in the $t,x$ directions and neglected boundary terms. The bulk action vanishes~\eqref{Bubulkaction} upon the imposed equations of motion, and so the Einstein-Hilbert action reduces to a boundary term. It is simple to check that for the solutions constructed above, the boundary terms coming from the horizon always cancel between the two branches of the CGL contour. This follows from the fact that the only part of the solution that is non-analytic on the CGL contour is $\delta g_{tx}$, and it is straightforward to verify that the $\delta g_{tx}$ dependent terms in \eqref{buboundaryaction} vanish at the horizon. The only contributions to the action therefore come from boundary terms in the UV and are of the form $S_{\mathrm{GH}} + S_{\mathrm{boundary}}$ on each contour. Evaluating the difference between these boundary terms on each branch of the contour gives the action presented above.

\bibliography{HydroActionPaper}
\bibliographystyle{JHEP}
\end{document}